%% file: TheiaWP.tex
\RequirePackage{fix-cm}
\documentclass[twocolumn,epjc3]{svjour3}  
\smartqed  % flush right qed marks, e.g. at end of proof
\RequirePackage{graphicx}

\usepackage[graphicx]{realboxes}
\usepackage{xspace}
\usepackage{isotope}
\usepackage{hyperref}

\usepackage{color}% Change font colors
\usepackage{gensymb}% Some nice math symbols...
\usepackage{wasysym}
\usepackage{amssymb}
\usepackage{enumerate}
\usepackage{comment}
\usepackage[utf8]{inputenc}	%UTF8 file encoding
\usepackage{mathrsfs}
\usepackage[T1]{fontenc}
\usepackage{booktabs}

\newcommand{\theia}[0]{\textsc{Theia}\xspace}

\journalname{Eur. Phys. J. C}

\begin{document}

\title{\theia: An Advanced Optical Neutrino Detector}

\input{authorlist}

\date{Received: date / Accepted: date}

\maketitle

\input{maintext}

\end{document}

%% file: authorlist.tex
\author{M.~Askins\thanksref{ucb,lbnl}
\and Z.~Bagdasarian\thanksref{jul}
\and N.~Barros\thanksref{penn, fcul, lip}
\and E.W.~Beier\thanksref{penn}
\and E.~Blucher\thanksref{chic}
\and R.~Bonventre\thanksref{lbnl}
\and E.~Bourret\thanksref{lbnl}
\and E.~J.~Callaghan\thanksref{ucb,lbnl}
\and J.~Caravaca\thanksref{ucb,lbnl}
\and M.~Diwan\thanksref{bnl}
\and S.T.~Dye\thanksref{uh}
\and J.~Eisch\thanksref{iowa}
\and A.~Elagin\thanksref{chic}
\and T.~Enqvist\thanksref{jyv}
\and V.~Fischer\thanksref{ucd}
\and K.~Frankiewicz\thanksref{bu}
\and C.~Grant\thanksref{bu}
\and D.~Guffanti\thanksref{mainz}
\and C.~Hagner\thanksref{ham}
\and A.~Hallin\thanksref{alb}
\and C.~M.~Jackson\thanksref{pnnl}
\and R.~Jiang\thanksref{chic}
\and T.~Kaptanoglu\thanksref{penn}
\and J.R.~Klein\thanksref{penn}
\and Yu.~G.~Kolomensky\thanksref{ucb,lbnl}
\and C.~Kraus\thanksref{laur}
\and F.~Krennrich\thanksref{iowa}
\and T.~Kutter\thanksref{lsu}
\and T.~Lachenmaier\thanksref{tub}
\and B.~Land\thanksref{ucb,lbnl,penn}
\and K.~Lande\thanksref{penn}
\and J.G.~Learned\thanksref{uh}
\and V.~Lozza\thanksref{fcul,lip}
\and L.~Ludhova\thanksref{jul}
\and M.~Malek\thanksref{sheff}
\and S.~Manecki\thanksref{laur,qu,snolab}
\and J.~Maneira\thanksref{fcul,lip}
\and J.~Maricic\thanksref{uh}
\and J.~Martyn\thanksref{mainz}
\and A.~Mastbaum\thanksref{rut}
\and C.~Mauger\thanksref{penn}
\and F.~Moretti\thanksref{lbnl}
\and J.~Napolitano\thanksref{temp}
\and B.~Naranjo\thanksref{ucla}
\and M.~Nieslony\thanksref{mainz}
\and L.~Oberauer\thanksref{mun}
\and G.~D.~Orebi Gann\thanksref{e1,ucb,lbnl}
\and J.~Ouellet\thanksref{mit}
\and T.~Pershing\thanksref{ucd}
\and S.T.~Petcov\thanksref{tri,kav}
\and L.~Pickard\thanksref{ucd}
\and R.~Rosero\thanksref{bnl}
\and M.~C.~Sanchez\thanksref{iowa}
\and J.~Sawatzki\thanksref{mun}
\and S.H.~Seo\thanksref{kor}
\and M.~Smiley\thanksref{ucb,lbnl}
\and M.~Smy\thanksref{uci}
\and A.~Stahl\thanksref{aach}
\and H.~Steiger\thanksref{mun}
\and M.~R.~Stock\thanksref{mun}
\and H.~Sunej\thanksref{bnl}
\and R.~Svoboda\thanksref{ucd}
\and E.~Tiras\thanksref{iowa}
\and W.~H.~Trzaska\thanksref{jyv}
\and M.~Tzanov\thanksref{lsu}
\and M.~Vagins\thanksref{uci}
\and C.~Vilela\thanksref{sbu}
\and Z.~Wang\thanksref{tsing}
\and J.~Wang\thanksref{ucd}
\and M.~Wetstein\thanksref{iowa}
\and M.J.~Wilking\thanksref{sbu}
\and L.~Winslow\thanksref{mit}
\and P.~Wittich\thanksref{corn}
\and B.~Wonsak\thanksref{ham}
\and E.~Worcester\thanksref{bnl,sbu}
\and M.~Wurm\thanksref{mainz}
\and G.~Yang\thanksref{sbu}
\and M.~Yeh\thanksref{bnl}
\and E.D.~Zimmerman\thanksref{boul}
\and S.~Zsoldos\thanksref{ucb, lbnl}
\and K.~Zuber\thanksref{dres}
}

\thankstext{e1}{e-mail: gorebigann@lbl.gov}

\institute{University of California, Berkeley, Department of Physics, CA 94720, Berkeley, USA \label{ucb}
           \and
           Lawrence Berkeley National Laboratory, 1 Cyclotron Road, Berkeley, CA 94720-8153, USA \label{lbnl}
\and 
Forschungszentrum J{\"u}lich, Institute for Nuclear Physics, Wilhelm-Johnen-Stra{\ss}e 52425 J{\"u}lich, Germany\label{jul} 
\and 
Department of Physics and Astronomy, University of Pennsylvania, Philadelphia, PA 19104-6396\label{penn}
\and 
Universidade de Lisboa, Faculdade de Ci{\^e}ncias (FCUL), Departamento de F{\'i}sica, Campo Grande, Edifício C8, 1749-016 Lisboa, Portugal\label{fcul}
\and 
Laborat{\'o}rio de Instrumenta{}{\c c}{\~a}o e F{\'i}sica Experimental de Part{\'i}culas (LIP), Av. Prof. Gama Pinto, 2, 1649-003, Lisboa, Portugal\label{lip}
\and 
The Enrico Fermi Institute and Department of Physics, The University of Chicago, Chicago, IL 60637, USA\label{chic}
\and 
Brookhaven National Laboratory, Upton, New York 11973, USA\label{bnl}
\and 
University of Hawai‘i at Manoa, Honolulu, Hawai‘i 96822, USA\label{uh}
\and 
Department of Physics and Astronomy, Iowa State University, Ames, IA 50011, USA\label{iowa}
\and 
Department of Physics, University of Jyv{\"a}skyl{\"a}, Finland\label{jyv}
\and 
University of California, Davis, 1 Shields Avenue, Davis, CA 95616, USA\label{ucd}
\and 
Boston University, Department of Physics, Boston, MA 02215, USA\label{bu}
\and 
Institute of Physics and Excellence Cluster PRISMA, Johannes Gutenberg-Universit{\"a}t Mainz, 55099 Mainz, Germany\label{mainz}
\and 
Institut f{\"u}r Experimentalphysik, Universit{\"a}t Hamburg, 22761 Hamburg, Germany\label{ham}
\and 
University of Alberta, Department of Physics, 4-181 CCIS, Edmonton, AB T6G 2E1, Canada\label{alb}
\and 
Pacific Northwest National Laboratory, Richland, WA 99352, USA\label{pnnl}
\and 
Laurentian University, Department of Physics, 935 Ramsey Lake Road, Sudbury, ON P3E 2C6, Canada\label{laur}
\and 
Department of Physics and Astronomy, Louisiana State University, Baton Rouge, LA 70803\label{lsu}
\and 
Kepler Center for Astro and Particle Physics, Universit{\"a}t T{\"u}bingen, 72076 T{\"u}bingen, Germany\label{tub}
\and 
University of Sheffield, Physics \& Astronomy, Western Bank, Sheffield S10 2TN, UK\label{sheff}
\and 
Queen's University, Department of Physics, Engineering Physics \& Astronomy, Kingston, ON K7L 3N6, Canada\label{qu}
\and 
SNOLAB, Creighton Mine 9, 1039 Regional Road 24, Sudbury, ON P3Y 1N2, Canada\label{snolab}
\and 
Department of Physics and Astronomy, Rutgers, The State University of New Jersey, 136 Frelinghuysen Road, Piscataway, NJ 08854-8019 USA\label{rut}
\and 
Department of Physics, Temple University, Philadelphia, PA, USA\label{temp}
\and 
University of California Los Angeles, Department of Physics \& Astronomy, 475 Portola Plaza, Los Angeles, CA 90095-1547, USA\label{ucla}
\and 
Physik-Department and Excellence Cluster Universe, Technische Universit{\"a}t M{\"u}nchen, 85748 Garching, Germany\label{mun}
\and 
Massachusetts Institute of Technology, Department of Physics and Laboratory for Nuclear Science, 77 Massachusetts Ave Cambridge, MA 02139, USA\label{mit}
\and 
SISSA/INFN, Via Bonomea 265, I-34136 Trieste, Italy\label{tri}
\and 
Kavli IPMU (WPI), University of Tokyo, 5-1-5 Kashiwanoha, 277-8583 Kashiwa, Japan\label{kav}
\and 
Center for Underground Physics, Institute for Basic Science, Daejeon 34126, Korea\label{kor}
\and 
University of California, Irvine, Department of Physics, CA 92697, Irvine, USA\label{uci}
\and 
Physikzentrum RWTH Aachen, Otto-Blumenthal-Stra{\ss}e, 52074 Aachen, Germany\label{aach}
\and 
State University of New York at Stony Brook, Department of Physics and Astronomy, Stony Brook, New York, USA\label{sbu}
\and 
Department of Engineering Physics, Tsinghua University, Beijing 100084, China.\label{tsing}
\and 
Cornell University, Ithaca, NY, USA\label{corn}
\and 
University of Colorado at Boulder, Department of Physics, Boulder, Colorado, USA\label{boul}
\and 
Institut f{\"u}r Kern und Teilchenphysik, TU Dresden, Zellescher Weg 19, 01069, Dresden, Germany \label{dres}
}

%% file: maintext.tex
\begin{abstract}
New developments in liquid scintillators, high-efficiency, fast photon detectors, and chromatic photon sorting have opened up the possibility for building a large-scale detector that can discriminate between Cherenkov and scintillation signals. Such a detector could reconstruct particle direction and species using Cherenkov light while also having the excellent energy resolution and low threshold of a scintillator detector. Situated deep underground, and utilizing new techniques in computing and reconstruction, this detector could achieve unprecedented levels of background rejection,  enabling a rich physics program  spanning topics in nuclear, high-energy, and astrophysics, and across a dynamic range from hundreds of keV to many GeV. The scientific program would include observations of low- and high-energy solar neutrinos, determination of neutrino mass ordering and measurement of the neutrino CP-violating phase $\delta$, observations of diffuse supernova neutrinos and neutrinos from a supernova burst, sensitive searches for nucleon decay and, ultimately, a search for neutrinoless double beta decay, with sensitivity reaching the normal ordering regime of neutrino mass phase space.
This paper describes \theia, a detector design that incorporates these new technologies in a practical and affordable way to accomplish the science goals described above. %We consider two scenarios, one in which \theia would reside in a cavern the size and shape of those intended to be excavated for the Deep Underground Neutrino Experiment, which we call \theia-25, and a larger 100-ktonne version (\theia-100) that could achieve an even broader and more sensitive scientific program.
\end{abstract}

%\newpage
\tableofcontents
\setcounter{tocdepth}{5}

\section{Introduction and \theia overview}\label{s:intro}

	Neutrinos are the fundamental particles we would most expect to be ignored: they interact too weakly and are too light to directly affect most microscopic processes.
	Yet
neutrinos access a breadth of science no other fundamental particle can: understanding the weak sector through direct measurements of neutrino properties; testing fundamental symmetries of Nature; probing near and distant astrophysical phenomena; peering into the interior of the Earth, and understanding the earliest moments of the Universe.  That scientific breadth has been mirrored by the broad array of technologies used to detect and study neutrinos, with the strength of each technology typically focused on a narrow slice of neutrino physics.  We discuss in this white paper a new kind of detector, called \theia (after the Titan Goddess of light), whose aim is to make world-leading measurements over as broad range of neutrino physics and astrophysics as possible.  We consider two scenarios, one in which \theia would reside in a cavern the size and shape of those intended to be excavated for the Deep Underground Neutrino Experiment (DUNE), which we call \theia-25, and a larger 100-ktonne version (\theia-100) that could achieve an even broader and more sensitive scientific program.

The broadband neutrino beam being built for the Long Baseline Neutrino Facility (LBNF)
~\cite{lbnf1,lbnf2} and DUNE~\cite{duneidr1,duneidr2,duneidr3} offer an opportunity for world-leading long-baseline neutrino oscillation measurements. Due to advances in Cherenkov ring reconstruction techniques, a \theia detector in the LBNF beam would have good sensitivity to neutrino oscillation parameters, including CP violation (CPV), with a relatively modestly sized detector. 
In addition to this long-baseline neutrino program, \theia will also contribute to atmospheric neutrino measurements and searches for nucleon decay, particularly in the difficult $p\rightarrow K^+ + \overline{\nu}$ and $N\rightarrow 3\nu$ modes~\cite{Mohapatra_6D,Araki:2005jt,Anderson:2018byx}.

	\theia will also make a definitive measurement of neutrinos from the Sun's Carbon-Nitrogen-Oxygen cycle (CNO neutrinos),
which to date have not been detected exclusively~\cite{sage,gallex,gno} but which would tell us
important details about how the Sun has evolved~\cite{CNO1}.
\theia will also provide a high-statistics, low-threshold (MeV-scale)
measurement of the shape of the $^8$B solar neutrinos and thus search for new
physics in the MSW-vacuum transition region~\cite{friedland2004,minakata2012}. Antineutrinos produced in the
crust and mantle of the Earth will be measured precisely by \theia with statistical uncertainty far exceeding all detectors to date. 

Should a supernova occur during
\theia  operations, a high-statistics detection of the $\bar{\nu}_e$ flux will be
made -- literally complementary to the  detection of the $\nu_e$ flux in the DUNE liquid argon
detectors. The simultaneous detection of both messengers and detection of an optical, x-ray, or gamma-ray component will enable a great wealth of neutrino physics
and supernova astrophysics. With a very deep location and with the detection of a combination of scintillation and Cherenkov light~\cite{Nakamura:2016kkl,branchesi2016}, \theia will have world-leading sensitivity to make a detection of the Diffuse Supernova Neutrino Background (DSNB) antineutrino flux~\cite{SK_DSNB:2015,priya2017,Sawatzki:2019}. The most ambitious goal,
which would likely come in a future phase, is a search for neutrinoless double beta decay (NLDBD), with a total isotopic mass of 30 tonnes or more, and with decay lifetime sensitivity in excess of $10^{28}$ years~\cite{biller2013,delloro2016}.

Table~\ref{t:physics} summarizes the physics reach of \theia.  This broad program would be addressed using a phased approach, as discussed in Sec.~\ref{s:detector}.

\begin{table*}
\centering
\caption{  \theia physics reach.  Exposure is listed in terms of the fiducial volume assumed for each analysis. %, except for those numbers marked $^*$, for which an optimized fiducial volume has yet to be evaluated and so the total detector mass is assumed.  
For NLDBD the target mass assumed is the mass of the candidate isotope within the fiducial volume.
\label{t:physics}}
\begin{tabular}{l l l} 
\hline\noalign{\smallskip}
 {\bf Primary Physics Goal} & {\bf  Reach} & {\bf  Exposure / assumptions } \\
\noalign{\smallskip}\hline\noalign{\smallskip}
    Long-baseline oscillations 	& 	$> 5 \sigma$ for 30\% of $\delta_{CP}$ values	&	524 kt-MW-yr \\
   Supernova burst			&	$<1(2)^\circ$ pointing accuracy	& 100(25)-kt detector, 10kpc	\\
 								& 20,000 (5,000) events  &\\
   DSNB 	&	$ 5 \sigma$ discovery	&	125 kton-yr \\
    CNO neutrino flux	&	$< $ 5 (10)\%	& 300 (62.5) kton-yr	\\
     Reactor neutrino detection	&	2000 events	&	100 kton-yr \\
     Geo neutrino detection 	&	2650 events	& 100 kton-yr	\\
     NLDBD               	&    	T$_{1/2} > 1.1\times10^{28}$~yr	 &     211 ton-yr $^{130}$Te	\\
     Nucleon decay $p\rightarrow \overline{\nu}K^{+}$ 	&	$T>3.80\times10^{34}$~yr (90\% CL)	& 800 kton-yr	\\
\noalign{\smallskip}\hline
\end{tabular}
\end{table*}

\theia is able to achieve this broad range of physics by exploiting new
technologies to act simultaneously as a (low-energy) scintillation detector and
a (high-energy) Cherenkov detector. Scintillation light provides the energy
resolution necessary to get above the majority of radioactive backgrounds and
provides the ability to see slow-moving recoils; Cherenkov light enables event
direction reconstruction which provides particle ID at high energies and
background discrimination at low-energies. Thus, the scientific program benefits in many cases on
the ability of \theia to discriminate efficiently and precisely between the scintillation
and Cherenkov photons. 

Discrimination between Cherenkov and scintillation photons can be achieved in several ways.
The use of a cocktail like water-based liquid scintillator (WbLS) provides a
favorable ratio of Cherenkov/scintillation light~\cite{WBLS:2011}.
Combining angular
and timing information allows discrimination between Cherenkov and
scintillation light for high-energy events even in a standard scintillator like
LAB-PPO~\cite{CHESS2:2017}.  Slowing scintillator emission time down by using slow
secondary fluors can also provide excellent separation~\cite{slowscin,slowls}.
Recent R\&D with dichroic filters to sort photons by wavelength has shown
separation of long-wavelength Cherenkov light from the typically shorter-wavelength
scintillation light, even in LAB-PPO, with only small reductions in the total scintillation
light.   This could be realized in a large detector by using Winston light
concentrators built from dichroic filters, termed ``dichroicons''~\cite{dichroic}.  In principle, all of these techniques could be
deployed together if needed to achieve the full \theia physics program.  New
reconstruction techniques, to leverage the multi-component light detection, are
being developed and with the fast timing of newly available PMTs and the
ultrafast timing of LAPPDs (Large Area Picosecond Photo-Detectors), allow effective tracking for high-energy
events and excellent background rejection at low energies.

\subsection{Detector configuration}\label{s:detector}

\begin{figure*}[htp!]
\centering
\includegraphics[width=0.99\textwidth]{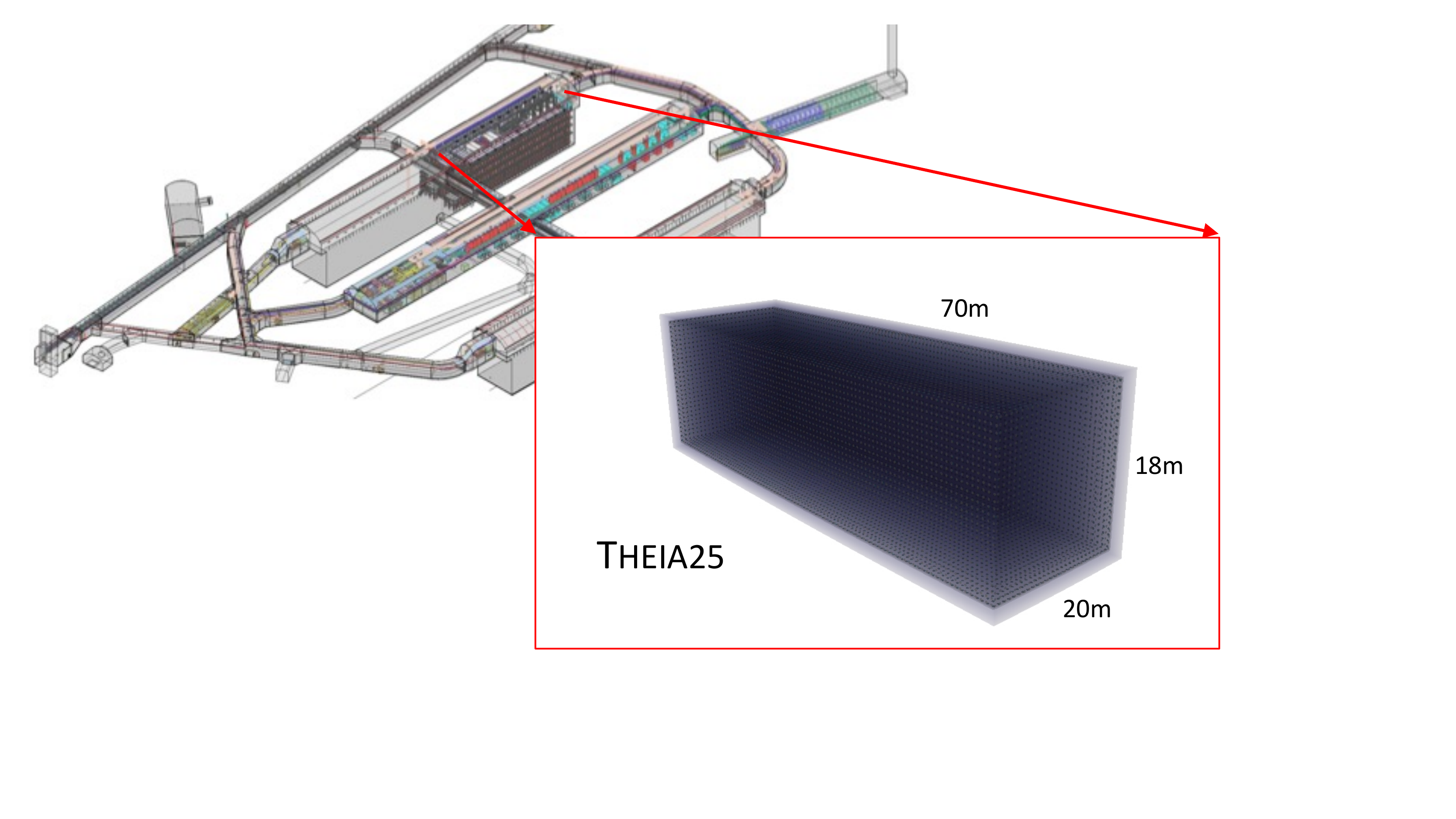}
\includegraphics[width=0.36\textwidth]{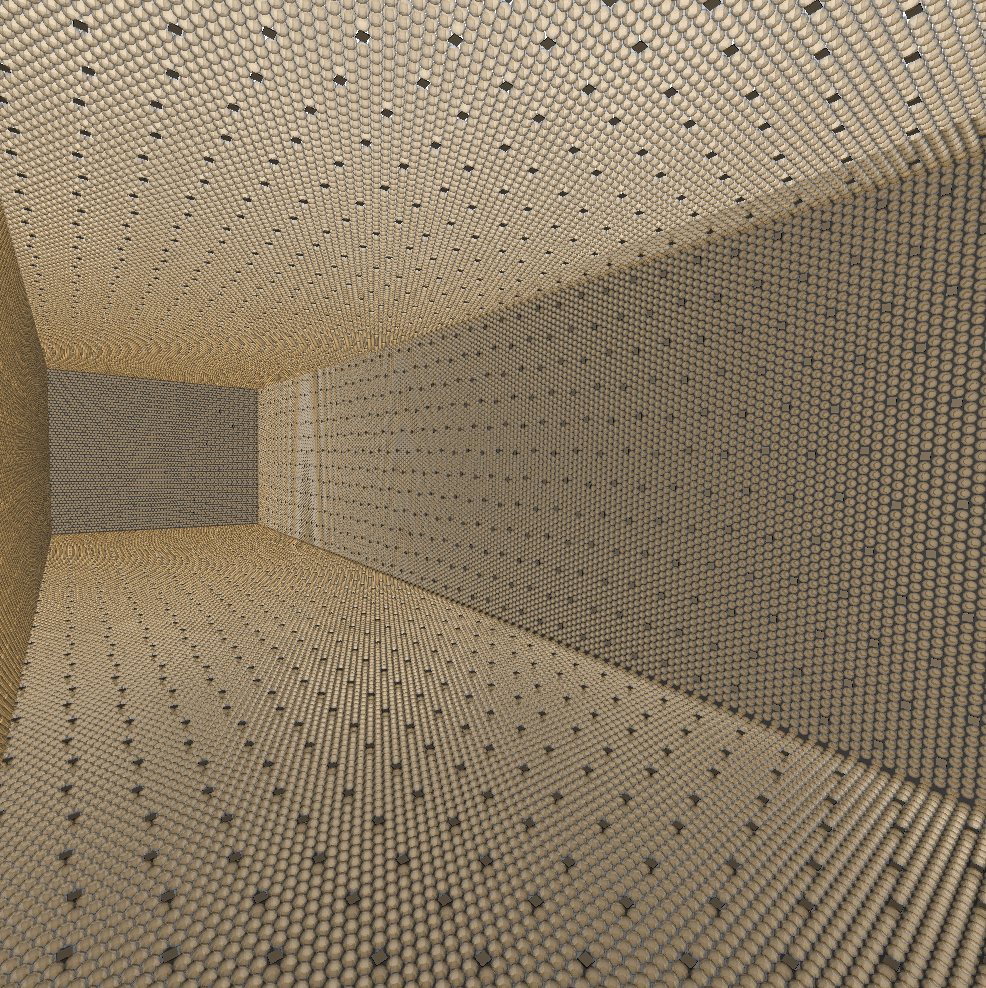}
\includegraphics[width=0.26\textwidth]{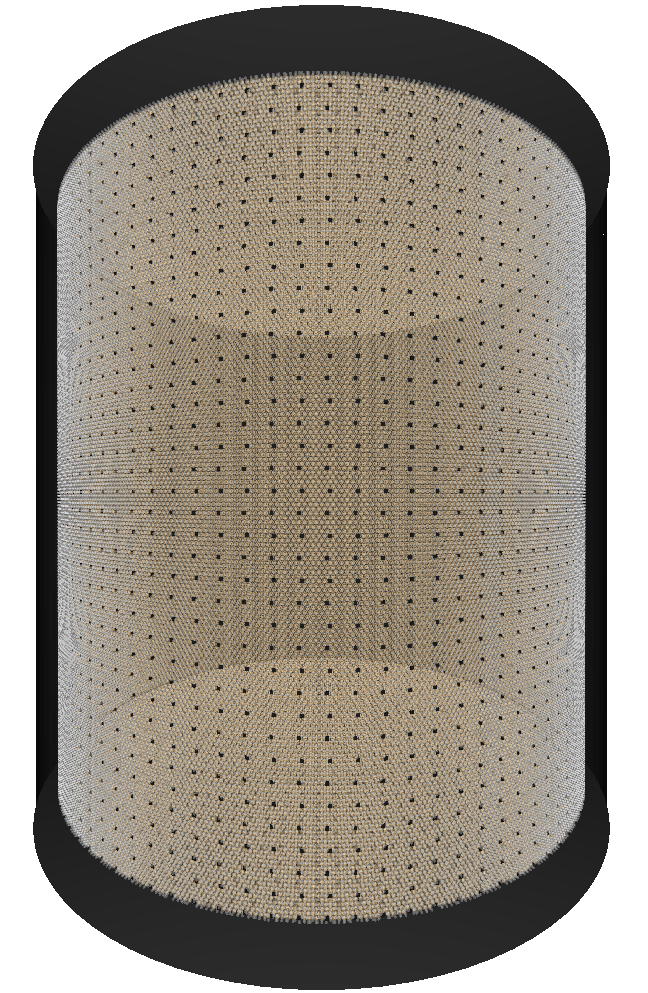}
\includegraphics[width=0.36\textwidth]{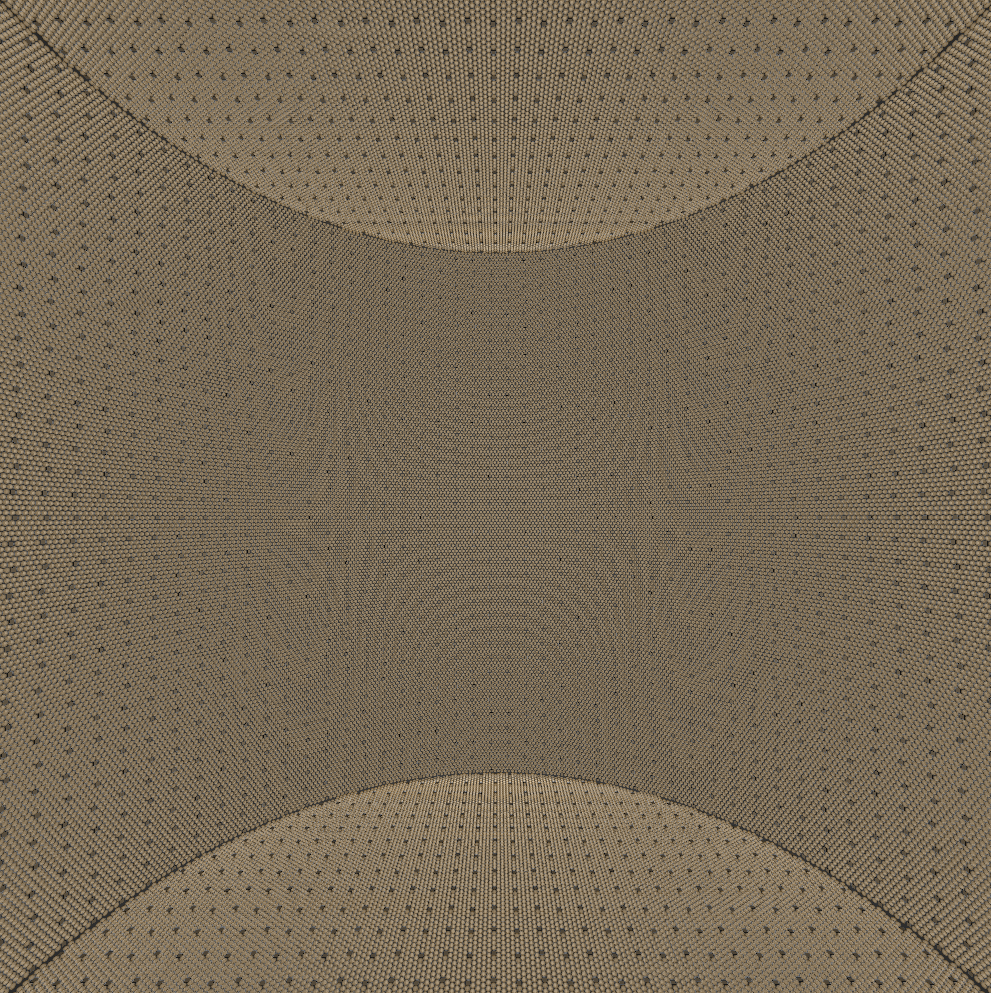}
\caption{The \theia detector.  
{\it Top  panel:} \theia-25 sited in the planned fourth DUNE cavern; 
{\it Lower left panel:} an interior view of \theia-25 modeled using the Chroma optical simulation package~\cite{chroma}; 
{\it Lower middle panel:} exterior view of \theia-100 in Chroma;
{\it Lower right panel:} an interior view of \theia-100 in Chroma.  In all cases, \theia has been modelled with 86\% coverage using standard 10-inch PMTs, and 4\% coverage with LAPPDs, uniformly distributed, for  illustrative purposes.}
\label{f:detector}
\end{figure*}

    The requirements for each of \theia's physics goals are different, although in nearly all cases increased detector mass unsurprisingly provides better sensitivity.  We consider for our sensitivity studies  two distinct size configurations:  a 25-kt total mass detector with a geometry consistent with one of the planned DUNE caverns and which could be deployed relatively quickly (\theia-25); and a detector with 100-kt total mass in a right-cylinder geometry (\theia-100).  Fig.~\ref{f:detector} shows simulation-derived images of each detector.  Ultimately, the limitation on size is likely driven by the attenuation lengths of the scintillator mixture in the detector. 
    
    The variation in requirements in some cases is inclusive: the need for good energy resolution at low energies only improves the background rejection and reconstruction capabilities at high energies, and the same is true for requirements on direction reconstruction via Cherenkov photons. Other requirements may be exclusive or at least partially so: the presence of inner containment to hold a $0\nu\beta\beta$ isotope-loaded scintillator would make a long-baseline analysis more complex from an optical standpoint, or reduce fiducial mass.
    
    A major advantage of \theia is that the target can be modified in a phased program to address the science priorities. In addition, since a major cost of \theia is expected to be photosensors, investments in \theia-25 instrumentation can be transferred directly over to \theia-100. Thus, \theia can be realized in phases, with an initial phase consisting of lightly-doped scintillator and very fast photosensors, followed by a second phase with enhanced photon detection to enable a very low energy solar neutrino program, followed by a third phase that could include doping with a $0\nu\beta\beta$ isotope and perhaps an internal containment vessel. Table~\ref{tbl:phases} lists the primary physics targets and the general configuration required to achieve those physics goals for each phase.

\begin{table*}
\centering
\caption{\theia physics goals and phased program.  Each successive phase adds to the breadth of the physics program.   The configuration column lists potential approaches to each phase, rather than a finalized detector design.
\label{tbl:phases}}
\begin{tabular}{c l l l l} 
\hline\noalign{\smallskip}
{\bf Phase} & {\bf Primary Physics Goals} & {\bf Detector capabilities} & {\bf Configuration options} \\
\noalign{\smallskip}\hline\noalign{\smallskip}
  I         &  Long-baseline oscillations 			& High-precision ring imaging & Low-yield WbLS  & \\
            &  $^{8}$B flux  					&&  Low photosensor coverage \\ 
            & Supernova burst, DSNB 			&& Fast timing     & \\ \hline
  II        & Long-baseline oscillations  			& Low threshold &  High-yield WbLS or slow LS & \\         
            &  $^8$B MSW transition    			& Cherenkov/scintillation separation &  Potential $^7$Li loading  \\
            &  CNO, $pep$ solar 				& High light yield & High photosensor coverage     \\
            &  Reactor and geo $\bar{\nu}$  		&&   Potential dichroicon deployment \\ 
            & Supernova burst ($\bar{\nu}_e$ and $\nu_e$), & &  & \\ 
            & DSNB ($\nu_e$ and $\bar{\nu}_e$) \\ \hline
 III        & $0\nu\beta\beta$  				& Low threshold&  Inner vessel with LAB+PPO+isotope \\               
        & $^8$B MSW transition  				&Cherenkov/scintillation separation& High photosensor coverage \\  
            &  Reactor and geo $\bar{\nu}$  		&High light yield&   Potential dichroicon deployment \\ 
            &  Supernova burst and DSNB ($\bar{\nu}_e$) && \\ 
\noalign{\smallskip}\hline
\end{tabular}
\end{table*}
  
    We discuss the various detector possibilities in terms of their total mass (25 kt vs. 100 kt) because {\it fiducial mass} depends upon the individual physics goals.  Low-energy, single-event physics (solar neutrinos, double-beta decay) can be sensitive to backgrounds from low-energy gamma rays coming from the photon sensors or containment vessel and thus a strict fiducial cut is needed; low-energy time-correlated events such as reactor antineutrinos or supernova burst neutrinos are more easily distinguished from such ``external'' backgrounds and thus a larger fiducial volume can be supported. High-energy physics, such as beam neutrino events, atmospheric neutrinos, and nucleon decay, are insensitive to such backgrounds but because they include extended tracks and showers they also require a fiducial cut (a ``distance-to-nearest-wall'') to ensure enough information is available for precise energy reconstruction and particle ID.  Thus, in the following sections, we include the relevant fiducial volumes for each physics topic, but keep the total mass fixed to our two defined configurations.

 \section{Detector capabilities}\label{s:cap}

The \theia detector is made possible by the development of new technology, mainly in the areas of fast photosensors, novel scintillating liquids, spectral photon sorting, and advanced image recognition techniques, especially those utilizing machine learning (ML) and other new techniques to discern underlying patterns from complex image data. In this section we present the status of these new technologies, and in addition discuss plans to further develop and incorporate them into the \theia design.  

\subsection{Water-based liquid scintillator}

\theia will be a unique detector, designed as the first large realization of the Advanced Scintillation Detector Concept (ASDC)~\cite{ASDC:2014}. Thus far, large Water Cherenkov (WC) detectors such as Super-Kamiokande (SK) have suffered in sensitivity due to the inability to detect particles with energy below the Cherenkov threshold.  For example, this limits sensitivity to the DSNB~\cite{SK_DSNB:2015} due to enhanced backgrounds from low-energy atmospheric neutrino interactions, and reduced signal from the inability to detect positron annihilation, which enhances the prompt signal from the leading reaction $\overline{\nu}_e + p\rightarrow e^{+} + n$. In the area of proton decay, the kaon from $p\rightarrow \overline{\nu}K^{+}$ is below the Cherenkov threshold, and in the area of solar neutrinos the $^{7}$Be and CNO neutrinos are practically undetectable as much of the energy from the neutrino electron scattering reaction is invisible.

Organic liquid scintillators (LS) have been used to enhance sensitivity for below Cherenkov threshold particles. LS is currently being used in the KamLAND, Borexino, and SNO+ detectors, and is planned for use in the JUNO detector now under construction. While this is very effective at increasing sensitivity at low energies, it comes at the loss of the directional sensitivity and multi-track resolution that is a hallmark of WC detectors. Use of organic LS also introduces issues of high cost, short optical transmission lengths, and undesirable environmental and safety problems. 

The recent development of water-based liquid scintillator (WbLS)~\cite{WBLS:2011} has the potential to alter this situation. By introducing a small amount (typically 1\%-10\%) of liquid scintillator into water, the liquid yield can be adjusted to allow detection of particles below Cherenkov threshold  while not sacrificing directional capability, cost, or environmental friendliness. First developed at Brookhaven National Lab (BNL), WbLS is a leading candidate for the main target medium for \theia, and will enhance the proposed scientific program significantly, as described in subsequent sections.

There is an active R\&D effort to realize the novel WbLS liquid target being considered for \theia. These efforts include a precision measurement of attenuation at long distances, demonstration of material compatibility with detector components, and accurate costs and production capabilities. Examples include WbLS development at BNL~\cite{BNL_WbLS_Dev}, compatibility studies at UC Davis, characterization and optimization with the CHESS detector at UC Berkeley and LBNL~\cite{CHESS:2017,CHESS2:2017}, fast photon sensor development at Chicago and Iowa State~\cite{lappd1}, spectral photon sorting at Penn~\cite{dichroic}, development of reconstruction and  particle identification algorithms~\cite{Wonsak:2018uby,Missert:2017qdz,Jiang:2019xwn,Wonsak_Dresden,Dunger:2019dfo,Aberle:2014,Elagin:2016zgp,Jiang:2019cnb,Li_2019}, and potential nanoparticle loading in NuDot at MIT~\cite{NUDOT}.   A practical purification system is being developed at UC Davis~\cite{UCD_WbLS_Dev}. This purification system, based on separating the scintillator component from the water component using nanofiltration (NF) techniques, has been shown to work well at small scales. After separation, standard water purification techniques can be used.  Since NF is widely used in the food industry, systems of the size needed for \theia scale detectors are commercially available.  The WbLS R\&D program for \theia also strongly leverages existing efforts and synergy with other programs, such as ANNIE~\cite{ANNIE}, SNO+~\cite{SNOPLUS_WP}, WATCHMAN~\cite{WM_WP}, and others.

For the purposes of the studies presented in here we have made the following reasonable assumptions for WbLS performance:
(1) absorption and scattering are simply weighted averages of pure water and LAB-based LS, and (2) a 10\% of LS light yield can be achieved with good stability and reasonable costs.

%\begin{enumerate}
%\item Demonstrate sufficiently high intrinsic light yield and long attenuation length to meet minimal light collection requirements.   %(Note: This  requirement can be offset by high efficiency, high coverage photon detection).
%\item Successful separation of  Cherenkov and scintillation signals, with sufficiently high Cherenkov light yield to maintain direction resolution and ring imaging capability.  This can be achieved by ultra-fast timing photon detection, such as LAPPDs, tuning of the WbLS cocktail, or a combination of the two.
%\item Stability of the above properties over long timescales, and with respect to possible isotope loading ({\it e.g.} Gd, Li, Te).
%\item Demonstrate chemical compatibility with common detector materials, such as PMT glass, PVC, PVDF, stainless steel, and various epoxies. Many of these studies are now in progress.
%\item Demonstrate reconstruction \& particle ID capability in the high energy and low energy ranges. The ability of WbLS to produce both Cherenkov and scintillation light necessitates the development of new pattern recognition and reconstruction techniques to take advantage of this new capability.
%\end{enumerate}

For the studies of \theia performance for CP violation, the advantages of WbLS have not been incorporated into the analysis. For example, it is expected that this will likely provide better vertex resolution and enable detection of below Cherenkov threshold charged hadrons, but given the high light levels and a reasonable Cherenkov/scintillation photon separation, the tracking performance already achieved in existing WC detectors will not be degraded.

\subsection{Photodetection}

Complementary to the development of new chemical loadings and WbLS is the development of new advanced  photodetection capabilities. Progress in photosensor technology will enable significant improvements in time resolution, improved light collection,  spatial granularity at different scales, and even the ability to separate photons by production process, production point, and wavelength. Many of the technologies, once speculative, are now reaching maturity. The specific combination of these technologies that will optimize the physics reach of  \theia is currently being explored. These are summarized below.

Since the construction of the last generation of large water optical neutrino detectors, significant progress has also been made in the advancement of conventional vacuum PMTs. High Quantum Efficiency (QE) PMTs with QE greater than 35\% are now readily available from multiple suppliers. Direct comparisons in the laboratory between these new devices and the 20-inch Hamamatsu PMTs installed in Super-Kamiokande have shown that they have a factor of 1.5 better photon collection efficiency per square centimeter~\cite{NIM12in}. Thus, a wall coverage of only 27\% is needed to be equivalent to the 40\% in Super-Kamiokande in terms of photon collection capabilities. In addition, timing is significantly better (e.g. 1.3 ns FWHM~\cite{NIM12in} versus 5.1 ns~\cite{sknim}), and other performance characteristics are also much improved~\cite{NIM11in,NIM12in}. In some studies below, PMTs with modern performance characteristics were used (e.g. solar neutrinos), while in others older performance characteristics have not yet been updated (e.g. CP violation search) and thus results are expected to improve.

%Hybrid Avalanche Photodiodes (HPDs) combine conventional large-area PMT vacuum tubes with a solid state gain stage and have been rigorously tested by the Hyper-K collaboration~\cite{hPDs}.

Large Area Picosecond Photo-Detectors (LAPPDs) are 20 cm x 20 cm imaging photosensors with single photoelectron (SPE) time resolutions below 100 picoseconds and sub-cm spatial resolutions~\cite{lappd0,lappd1}. The combination of these capabilities  makes it possible to better separate individual photons and develop reconstruction tools that fully capture correlations between the time and spatial patterns of light, rather than treating them independently. LAPPDs are now commercially available through Incom, Inc and will soon be deployed in their first neutrino application in the ANNIE experiment~\cite{ANNIE}. Studies done by ANNIE show a significant increase in high-energy reconstruction capabilities~\cite{ANNIE}, albeit at rather short distances. Note that ANNIE has only one LAPPD for every twenty-five PMT's and it was determined from simulations that this was sufficient to meet the initial goals of ANNIE to measure neutron production from single-track quasi-elastic events. The addition of LAPPDs was shown to improve the vertex and tracking resolution by a factor of two over just using PMTs  in this experiment. The addition of more LAPPDs to ANNIE to handle multi-track events and further improve performance is still under study.   Studies of the impact of LAPPD deployment on \theia are underway.  Preliminary results suggest substantial improvement in vertex resolution, but a smaller effect on Cherenkov/scintillation separation at this scale, since dispersion has a substantial effect over the distances in question.  The use of LAPPDs is not assumed in any of the following physics discussions.

LAPPD costs are expected to drop with increasing production yields and market extent. Further work in developing new production techniques, such as the use of ceramic bodies, and the development of {\it in situ} photocathodes, once mature, could have a significant impact in further reducing these prices.
%~\cite{gen2LAPPDs}. No reference for this?

Another advancement in photon detection is the development of multi-PMT modules and mixed PMT coverage schemes. IceCube, KM3Net, and Hyper-Kamiokande are developing Digital Optical Modules (DOMs)~\cite{DOMs1,DOMs2,DOMs3} in place of conventional large area photomultipliers. These transparent modules each contain an array of smaller PMTs along with the readout electronics. These DOMs provide similar area coverage to more conventional large-area PMTs, but with finer spatial and time resolution, as well as the ability to resolve directionality. The JUNO collaboration is pursuing a detector design with a mixture of large-area and small-area PMTs to achieve increased coverage,  and to provide different scales of spatial granularity~\cite{JUNOPMTs}.

%These new photosensors require commensurate electronic readout systems. The development of low-cost, ultrafast CMOS waveform sampling chips such as the PSEC4 and DRS can bring down the per channel cost of detailed pulse reconstruction. Another improvement is the digitization of pulses at the photodetector, rather than transmitting signals over long cables. DOMs…small digitizers at the base… 

Many of the key advances in photodetection look beyond the photosensors themselves and consider optics for using photodetectors more efficiently. Traditional light collectors such as Winston Cones and scintillating light guides have been rigorously pursued for the Long Baseline Neutrino Experiment (LBNE)~\cite{lbne_opt} and Hyper-K~\cite{Abe:2018uyc}. Using arrays of lenses, plenoptic imaging would add directional information to detected light~\cite{plenoptic}. Novel designs using specular reflection off  mirrors could also be used to enhance coverage. One particularly promising optical concept is the application of dichroic filters to separate light by wavelength. This would provide a strong additional handle for discriminating between the largely monochromatic scintillation light and broad-band Cherenkov light, as well as enabling better correction for chromatic dispersion of Cherenkov photons~\cite{dichroic}.  

%Retroreflectors redirect light back along its initial trajectory, increasing effective coverage with a well defined ray geometry~\cite{retrorelfection}.

Moving forward, the \theia collaboration will leverage detailed simulations and reconstruction tools to evaluate the optimal suite of photosensors, light collection and sorting optics, and readout electronics. The challenge is to enable improved physics over a wide range of energies and to co-develop the photodetector systems with the optimization of the particular WbLS cocktail. The deployment of photosensors in \theia will likely be different than in other existing applications, so another key task will be the development of application-ready modules, in parallel with the process of technology down selection.

\subsection{Reconstruction techniques}
\label{sec:Reconstruction}
While  Cherenkov detectors have been very successful in reconstructing various properties of the particles involved
in a neutrino event, liquid scintillation detectors have long been considered a source for calorimetric information only. 
However, in recent years it has become clear that the time information of the light in liquid scintillators can be used to 
access a wide range of information, similar or even superior to what a pure Cherenkov detector can deliver~\cite{Learned:2009}. 

\begin{figure*}[!h]
  \centering
    \includegraphics[trim=0.1cm 0.1cm 0.0cm 0.1cm,clip=true,width=\textwidth]{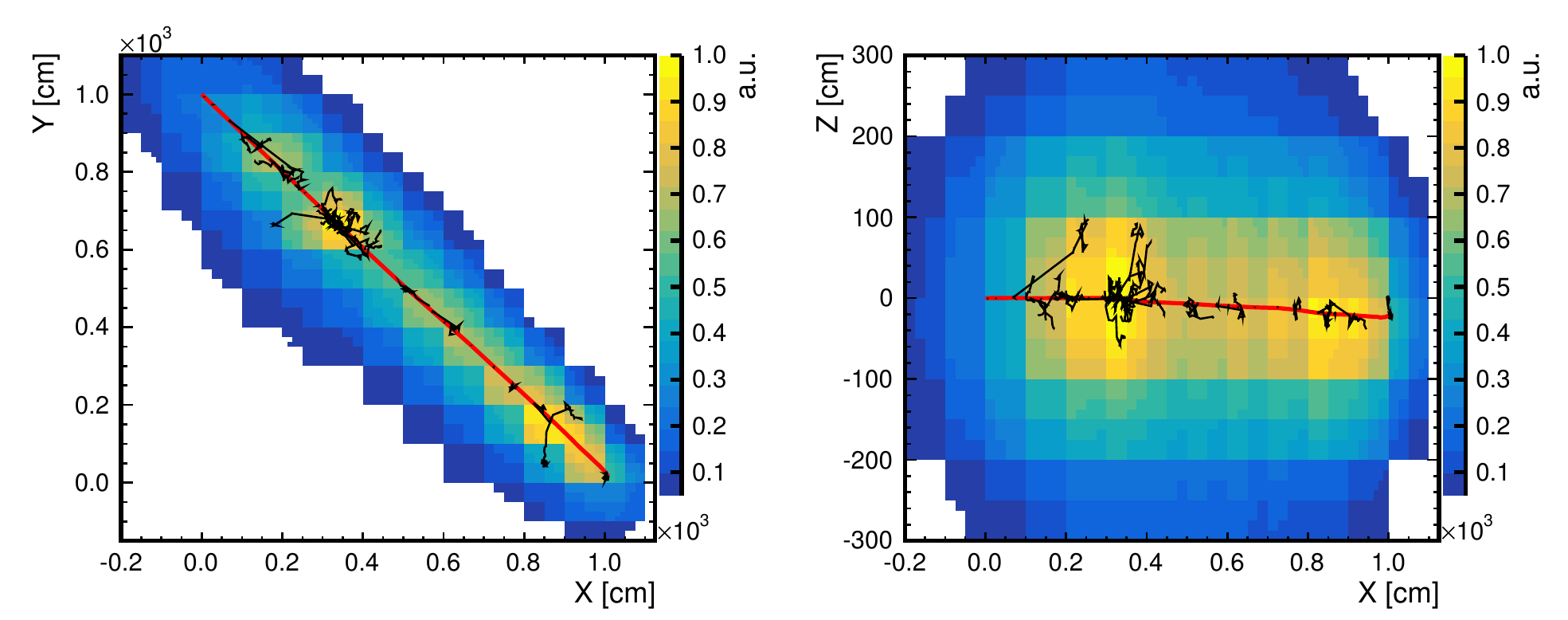}
  \caption{Reconstruction results for a 
simulated muon with $3$~GeV initial kinetic energy in the cylindrical LENA detector projected 
along the symmetry axis (left) or a radial $y$-axis (right). The primary particle started at 
$(0,\,1000,\,0)$~cm in the direction $(1,\, -1,\, 0)$. Both the projected tracks of the 
primary particle (red) and of secondary particles (black) are shown. The cell content is given in a.u.
and rescaled such that the maximum content is $1$. The plots are taken from \cite{Wonsak:2018uby}, which provides details on the reconstruction procedure.
}
  \label{fig:RecoResults}
\end{figure*}

There are two complementary approaches to reconstruction in both detector types and, consequently, also in WbLS 
detectors. The first approach developed in MiniBooNE~\cite{patt09} and extended to the more complicated event topologies of Super-Kamiokande~\cite{Missert:2017qdz} follows a likelihood
ansatz to 
find the optimal track parameters and compare different hypotheses. This technique can naturally accommodate Cherenkov and scintillation light, as was required for MiniBooNE, by combining Cherenkov and scintillation light predictions for each photosensor in the calculation of the likelihood.
In contrast to this,  three-dimensional topological 
reconstruction \cite{Wonsak:2018uby} tries to picture the spatial distribution of the energy deposition within the detector without using a 
specific hypothesis. This technique has been developed for the LENA~\cite{Wurm:2011zn} detector and also been implemented for the JUNO detector
\cite{An:2015jdp}.
The application of these algorithms to Cherenkov detectors is straightforward. An example for a reconstructed
stopping muon in LENA (a liquid scintillator detector)
clearly showing the accessibility of the energy loss per unit length in shown in Fig.~\ref{fig:RecoResults}.

Both methods have been improved considerably over the last couple of years. For example fiTQun \cite{Missert:2017qdz}, the reconstruction
software 
used by Super-Kamiokande/T2K, is now able to reconstruct up to 6 Cherenkov rings produced by electron, muon, or pion particle hypotheses. This allows for a simultaneous determination of the identity and number of particles.
%This alone will increase the expected sensitivity for example for CP-violation in the LBNF beam significantly in comparison to previous studies like \cite{Goon:2012if}.
Recently, the Super-Kamiokande collaboration also published results using fiTQun \cite{Jiang:2019xwn} with improved kinematic and particle identification capabilities. This allowed them to increase the volume accessible to the analysis by 32\%.

\begin{figure*}[b!t]
  \centering  
  \includegraphics[width = 0.48\textwidth]{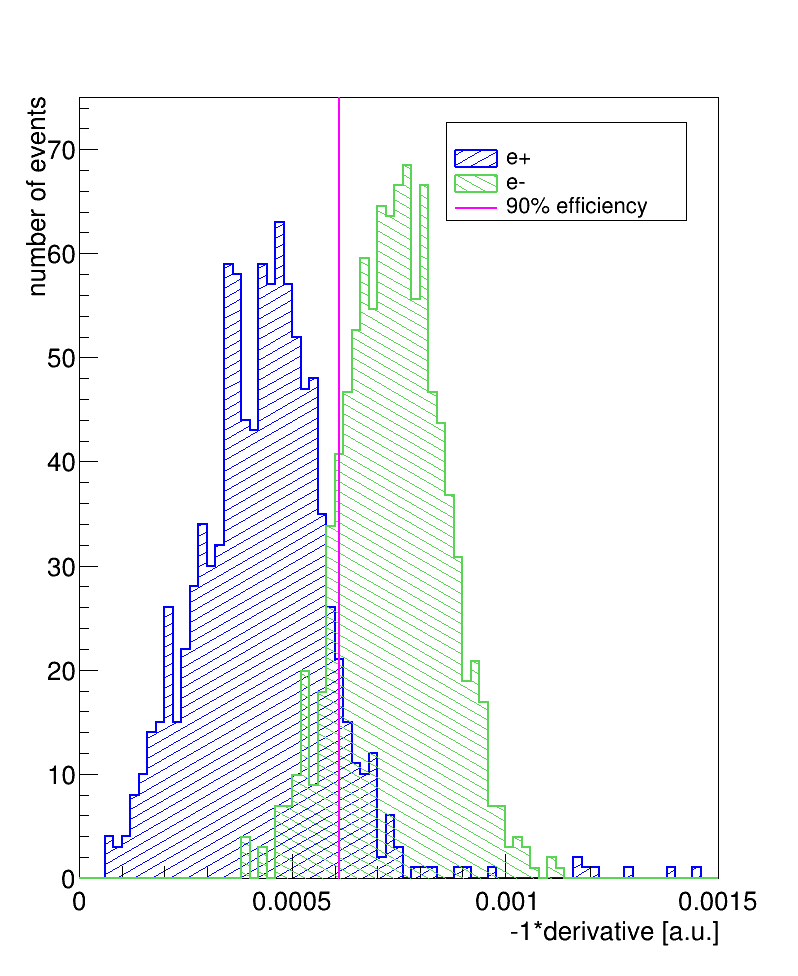}
  \includegraphics[width = 0.48\textwidth]{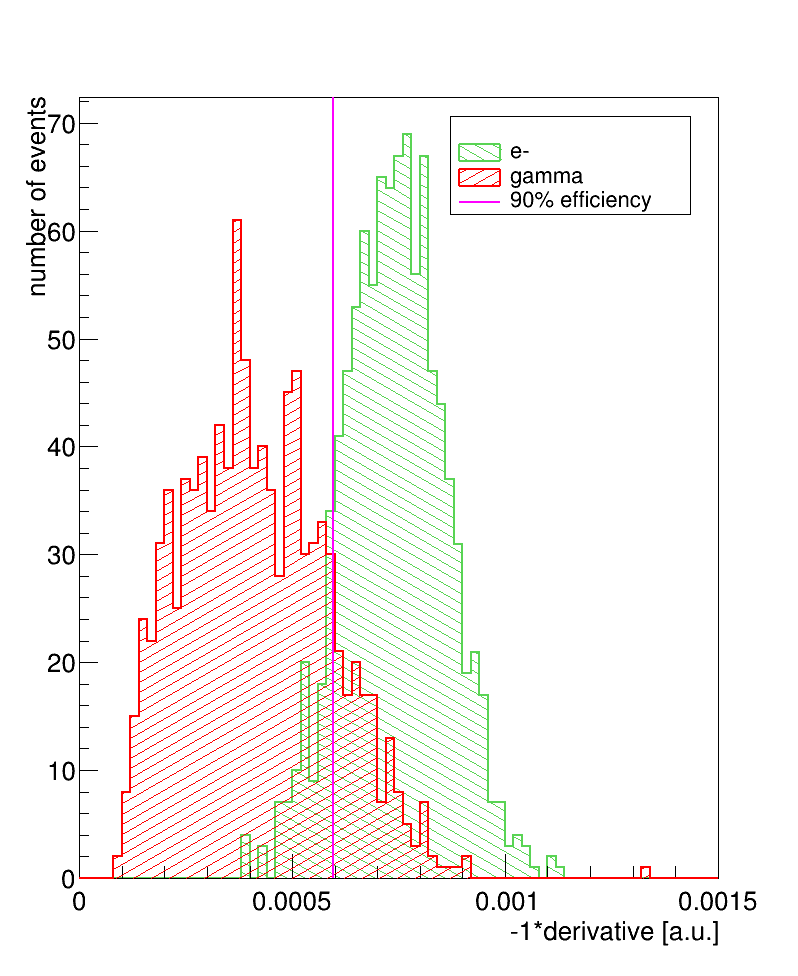}
  \caption{Maximum derivative of the radial profile of the 3D-topological reconstruction at 2~MeV in JUNO for (left) electrons  and positrons  and (right)  electrons 
  and gammas  ~\cite{bwon18}.}
  \label{fig:ParticleDiscrimination}
\end{figure*}

Topological reconstruction offers large volume liquid detectors the same capabilities as  highly segmented detectors (with all the resulting implications). This includes possibilities for particle identification at energies as low as
a few MeV based on topological information. For example it is now possible to distinguish point-like events from multi-site events in liquid
scintillator \cite{Wonsak_Dresden} using various techniques, including 
likelihood-based pulse shape discrimination methods \cite{Dunger:2019dfo}. 
Fig.~\ref{fig:ParticleDiscrimination} shows an example of the separation of electrons from gammas, critical for separation of neutrino scatters, which produce electrons, from common gamma emitters in the uranium and thorium decay chains. 

\begin{figure*}[!h]
    \centering
    \includegraphics[width=0.48\textwidth]{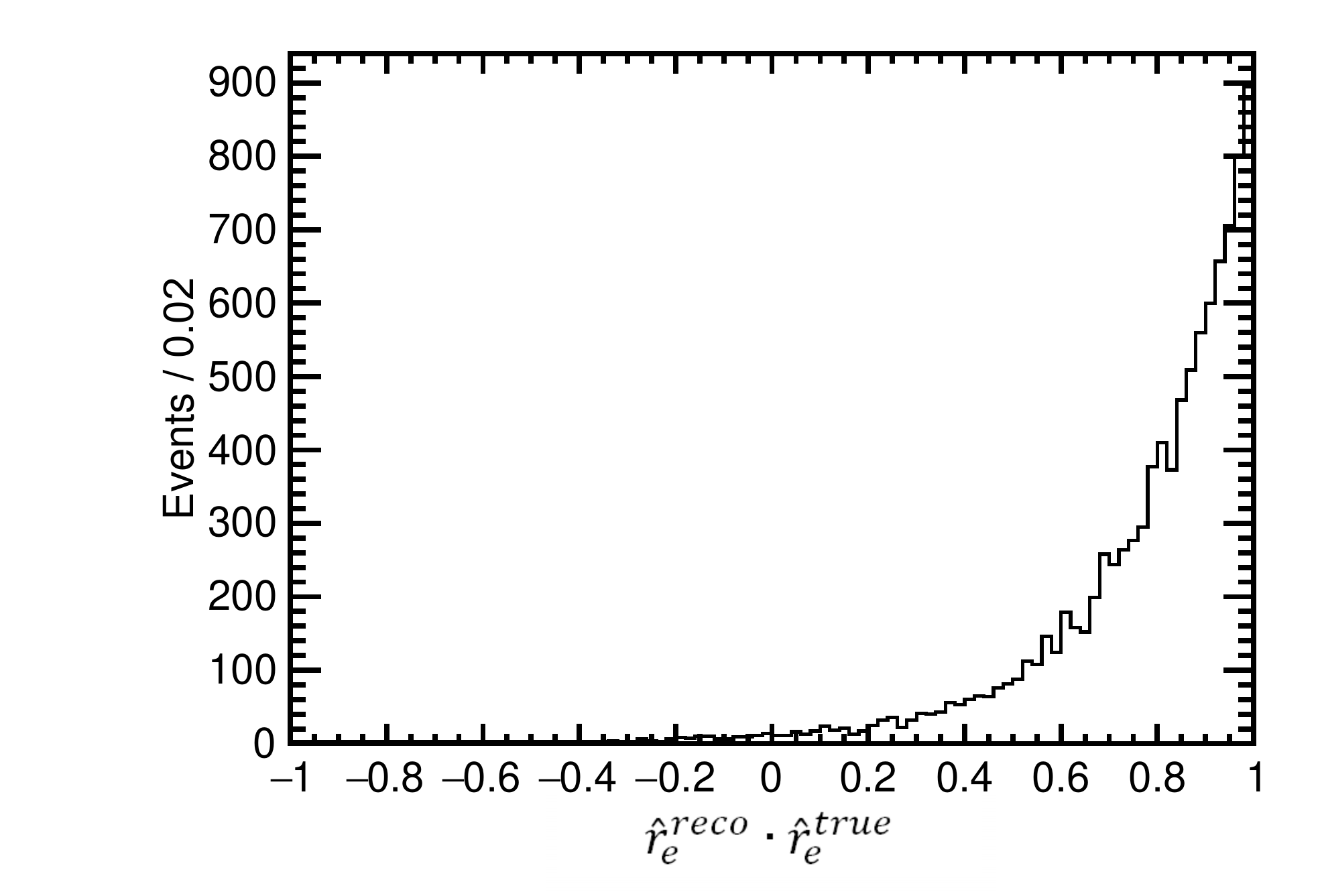}
    \includegraphics[width=0.48\textwidth]{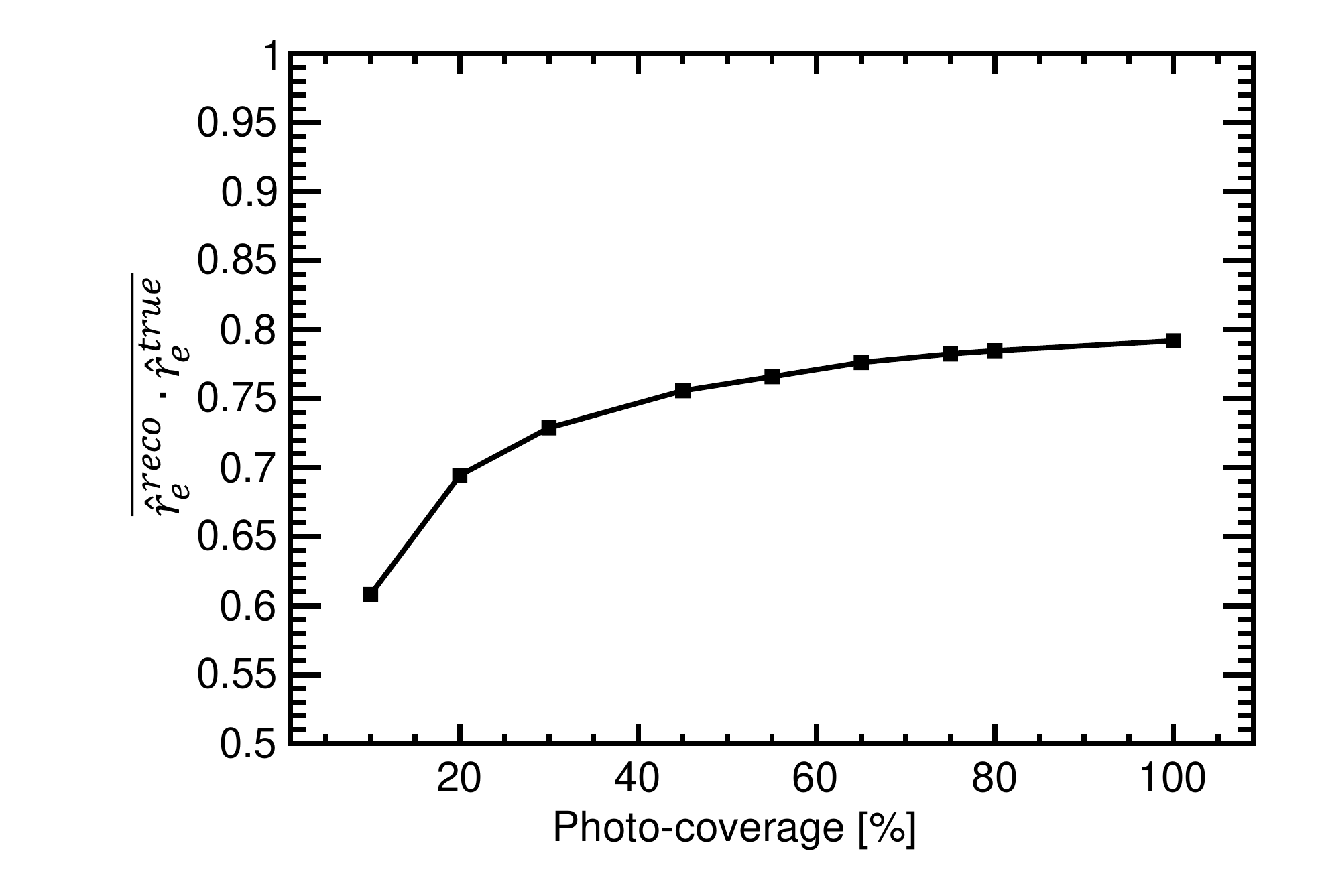}
    \caption{
    Simulation of a liquid scintillator detector similar to KamLAND-Zen, but equipped with 100 ps time resolution photo-detectors. Details about the simulation can be found in~\cite{Jiang:2019cnb}
    (Left): The inner product between the reconstructed direction and the true electron direction for simulation of 10,000 one-track background events.
    (Right): The mean value of the inner product distribution as a function of photo-coverage.
    }
    \label{fig:inner_product_coverage}
\end{figure*}

At low energies, reconstruction techniques have also improved remarkably. Reconstruction of events with energies down to 3-5 MeV using Cherenkov light in
pure water has been done by both the Super-Kamiokande and SNO experiments~\cite{LETA,Smy_bibtex_from_inspire,Smy_Method_in_PRD}. The ability to separate Cherenkov from scintillation light allows the use of additional information for event reconstruction. This was pioneered by the LSND experiment, which successfully used Cherenkov light in a diluted liquid scintillator to reconstruct electron tracks in the energy range of about 45 MeV~\cite{LSND_diluted_scintillator}. More recently, the feasibility of directional reconstruction of few-MeV electrons in a conventional high-light yield liquid scintillator has also been demonstrated in simulations relevant to NLDBD experiments~\cite{Aberle:2014}. In both these examples, the use of timing information was crucial for separating directional Cherenkov light from the isotropic scintillation light.

Direction reconstruction based on Cherenkov-scintillation light separation continues to make rapid progress~\cite{Wonsak_Dresden,Jiang:2019cnb}. At their core, these new algorithms rely on selecting a sample of photoelectrons (PEs) with a favourable ratio of Cherenkov to scintillation light. Then the
`center-of-mass' of the PE distribution on the detector surface becomes aligned with the direction of the parent particle. An angular resolution of 0.5-0.8 radians can be achieved for a large spherical detector similar to KamLAND-Zen, but equipped with 100-ps time resolution photo-detectors. An example of  direction reconstruction for 2.5-MeV electrons is shown in Fig.~\ref{fig:inner_product_coverage}. 
 We note that direction reconstruction of MeV-scale electrons relies on the ability to separate Cherenkov and scintillation light. Fig.~\ref{fig:inner_product_coverage} shows the performance of a reconstruction algorithm that achieves Cherenkov/scintillation separation by using timing information in a detector much smaller than \theia.  It can  be expected that the effects of dispersion will be inherently greater in the larger \theia detector and, thus, the primarily blue scintillation light should be further delayed relative to the longer-wavelength Cherenkov tail.  Work is ongoing to develop direction reconstruction for the \theia detector. The required Cherenkov detection efficiency, and methods for achieving it, are still being developed, as discussed in Secs.~\ref{s:intro} and~\ref{s:cap}.

Beyond directionality, topological features of events in the few-MeV energy range can also be extracted. A spherical harmonics analysis of PE
distributions has been used to separate NLDBD from $^{8}$B solar neutrino events~\cite{Elagin:2016zgp}. This has been improved by introducing a Cherenkov-scintillation space-time boundary that allows for reliable
and more general Cherenkov-scintillation light separation~\cite{Jiang:2019cnb}. The Cherenkov-scintillation space-time boundary is defined as the light cone in the 2-dimensional space of the arrival time and the polar angle of each PE with respect to the axis from the center of the detector to the vertex. The PEs located near the boundary correspond to photons that were emitted early and contain a high fraction of directional Cherenkov PEs. In a consequent analysis each individual PE can be assigned a weight based on its distance from the boundary, thus maximizing the
contribution from Cherenkov PEs. 

Reconstruction of various characteristics of candidate events in detectors with scintillation and Cherenkov light can further benefit from the  widely spreading use of machine learning techniques. For example, separation of NLDBD from $^{10}$C events using a convolutional neural network (CNN) has been explored in~\cite{Li:2018}. In a simulation using detector parameters similar to KamLAND-Zen, including photo-coverage, time and position resolution, 60\% rejection of $^{10}$C events has been achieved with a 90\% signal efficiency for NLDBD candidate events.

\section{Physics sensitivities and detector requirements}\label{s:physics}

\theia would address a broad program of physics, including: geo-neutrinos, supernova neutrinos, nucleon decay, measurement of the neutrino mass hierarchy and CP-violating phase, and even a next-generation neutrinoless double beta decay search. The sections below discuss the progress in estimating sensitivities and detector requirements in the context of \theia-25 and \theia-100. 

\subsection{Long baseline}
Neutrino oscillations arise from mixing among the flavor and mass states of the neutrino which can be described
by a complex unitary matrix that depends on three mixing angles and a potentially CP-violating phase. The
parameters of this mixing matrix determine the probability amplitudes of neutrino oscillation and the
differences between the neutrino masses determine the frequency of oscillation. These parameters have
all been measured, with the exception of the value of the CP-violating phase, $\delta_{CP}$~\cite{nufit4.1}, although the ordering
of the mass states is also not definitively determined. Long-baseline neutrino oscillation experiments have significant sensitivity to the
mixing parameters $\theta_{23}$, $\theta_{13}$, and $\delta_{CP}$, as well as to the mass splitting
$\Delta m^{2}_{32}$, and the neutrino mass ordering via matter effects. The atmospheric parameters, $\theta_{23}$ and $\Delta m^{2}_{32}$, have been measured by the existing long-baseline oscillation experiments T2K~\cite{Abe:2018} and NOvA~\cite{nova2019}. These experiments are also beginning to have sensitivity to $\delta_{CP}$. DUNE~\cite{duneidr2,duneidr3} is a next-generation long-baseline neutrino oscillation experiment being built at the Long Baseline Neutrino Facility (LBNF)~\cite{duneidr1} that will make a definitive determination of the neutrino mass ordering, will have sensitivity for a definitive discovery of CP-violation for much of the possible parameter space, and will make precise measurements of all the oscillation parameters governing long-baseline oscillation in a single experiment. DUNE plans to use liquid argon Time Projection Chambers (TPC) for the large detectors at the LBNF far site in the United States. Hyper-Kamiokande~\cite{Abe:2018uyc} is a next-generation long-baseline experiment using a water Cherenkov detector (WCD) at a far site in Japan. Both are anticipated to run on a similar timescale.

The long-baseline oscillation sensitivities of two potential configurations of \theia positioned at the LBNF far site have been considered. 
 \theia sensitivity is compared to that of a DUNE-like liquid-argon TPC, using the same flux, location, and true oscillation parameters, such that the impact of detector reconstruction and selection efficiency may be directly compared. 
The full realization of \theia is a 100-kt right cylindrical volume (\theia-100), similar to the geometry of Super-Kamiokande and Hyper-Kamiokande, which corresponds approximately to 70-kt fiducial volume. A smaller 25-kt realization (\theia-25) with a 17-kt fiducial volume in one of the four LBNF caverns would be able to supplement the DUNE measurements with CP-violation sensitivity comparable to a 10-kt (fiducial) liquid argon detector~\cite{Acciarri:2015uup}. 
%We also consider a 50-kt detector between the two scales (\theia50) which would have a 35-kt fiducial volume.

The ability to measure long-baseline neutrino oscillations with a distinct set of detector systematic uncertainties and neutrino interaction uncertainties relative to the liquid argon detectors, would provide an important independent cross-check of the extracted oscillation parameter values.

For the studies presented here, we use GLoBES~\cite{Huber:2004ka,Huber:2007ji} to calculate predicted spectra for different
oscillation parameter hypotheses and compare these to quantify experimental sensitivity. We make use of
the publicly available LBNF beam flux and DUNE detector performance description~\cite{Alion:2016uaj}. For the DUNE
sensitivity we assume a 10-kt fiducial mass, corresponding to a single DUNE far detector module. For the
\theia sensitivity, we use \theia's expected 70- or 17-kton fiducial mass and assume the detector
can be designed to perform as well as and no better than a conventional WCD, using
Monte Carlo simulations from Super-Kamiokande to define this performance.
%More detailed simulation of efficiency and reconstruction capabilities near the detector walls may allow for a more precise estimate of the impact of fiducialization.
Detailed simulations of improved
performance from using LAPPDs, WbLS, and advanced image recognition algorithms are planned and expected to
demonstrate improved performance. Consistent with DUNE, we assume seven years exposure with equal running
in neutrino and
antineutrino mode for both detectors, where the running time in each year assumes a typical Fermilab accelerator uptime of 56\%. We use oscillation parameter central values and uncertainties
from NuFit 4.0~\cite{Esteban:2018azc,nufitweb}.

Previous studies of a water Cherenkov detector in the LBNF beam were performed in the context of the
predecessor experiment to DUNE: LBNE~\cite{Goon:2012if}. These studies were based on Super-Kamiokande event reconstruction techniques
developed within the first several years of Super-Kamiokande data taking, and were restricted to single-ring events with
no Michel electrons from stopped pion and muon decay. In the decade since, important advancements have been made in
Cherenkov reconstruction that have substantially improved particle identification and ring counting. As described in Section~\ref{sec:Reconstruction}, the fiTQun event reconstruction package used for \theia sensitivity studies has now been fully implemented in the most recent T2K analyses~\cite{Abe:2018}. These improvements, when applied to the LBNF beam, enhance the sensitivity to neutrino oscillations in three important ways:
\begin{enumerate}
\item The improved ring counting has removed 75\% of the neutral current background, relative to the previous analysis,
due to improvements in the detection of the faint second ring in boosted $\pi^{0}$ decays;
\item The improved electron/muon particle identification has allowed for an additional sample of 1-ring, one-Michel-electron events from $\nu_{e}$-CC$\pi^{+}$ interactions, without significant contamination from $\nu_{\mu}$ backgrounds
\item Multi-ring $\nu_{e}$ event samples can now be selected with sufficient purity to further enhance sensitivity to
neutrino oscillation parameters.
\end{enumerate}

\begin{figure*}[h!]
  \centering
  \includegraphics[width=0.45\linewidth]{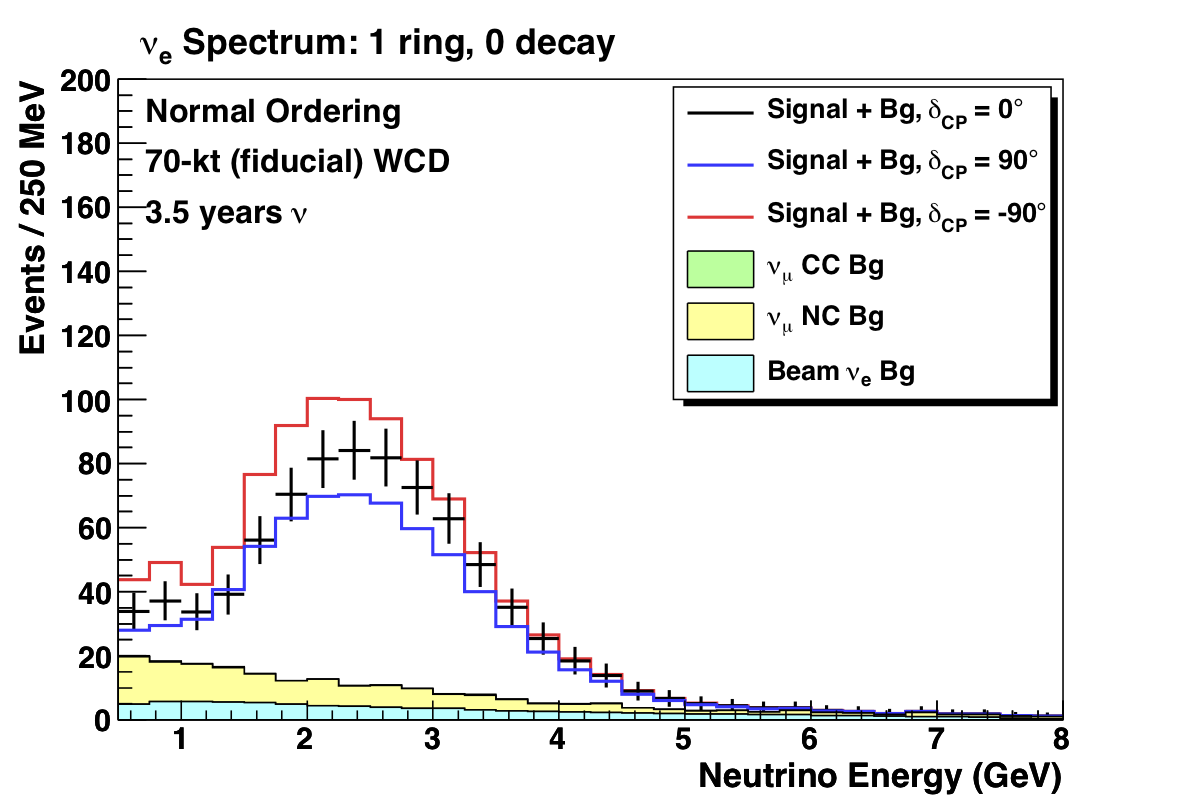}
  \includegraphics[width=0.45\linewidth]{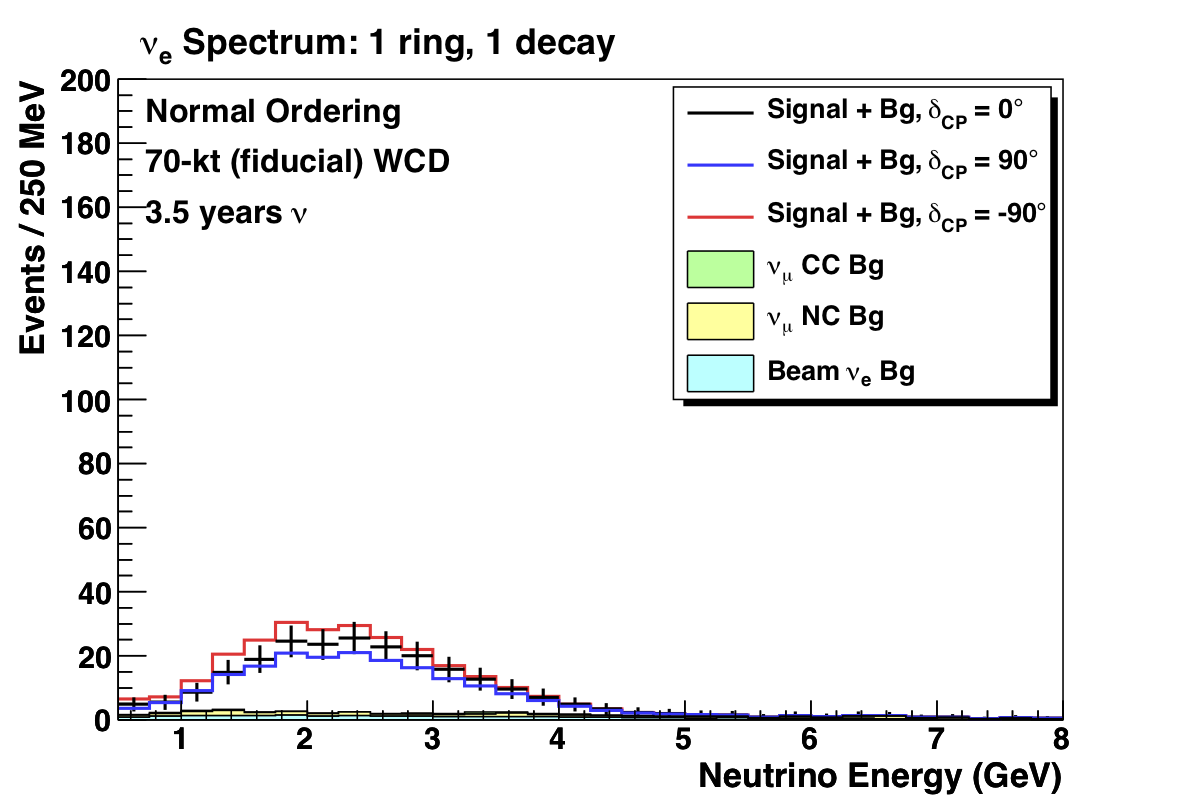}
  \includegraphics[width=0.45\linewidth]{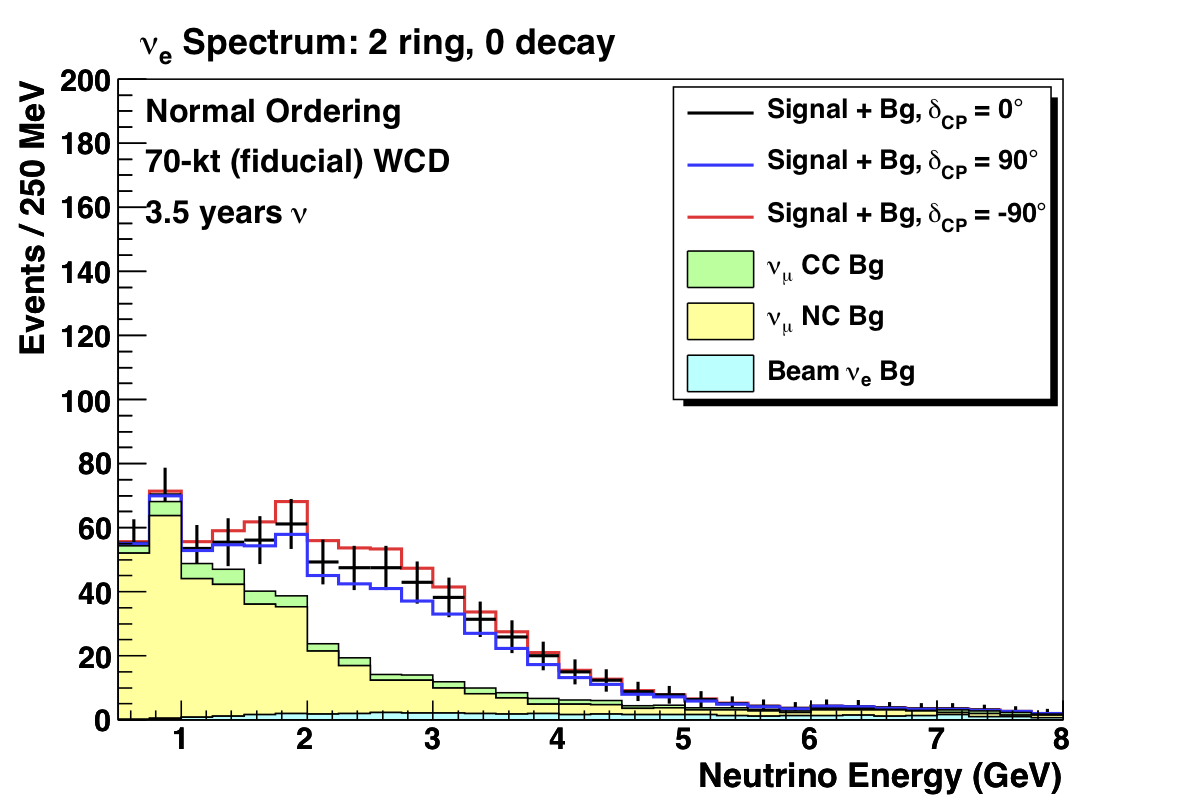}
  \includegraphics[width=0.45\linewidth]{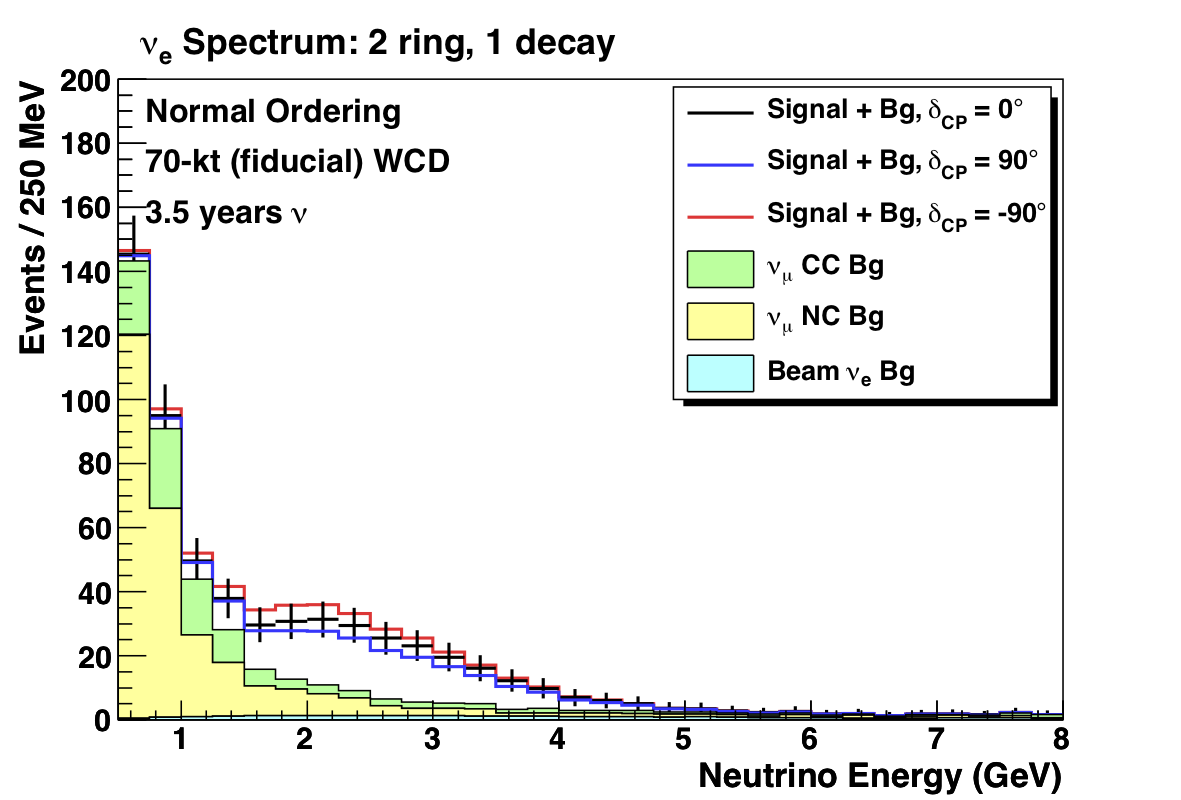}
  \includegraphics[width=0.45\linewidth]{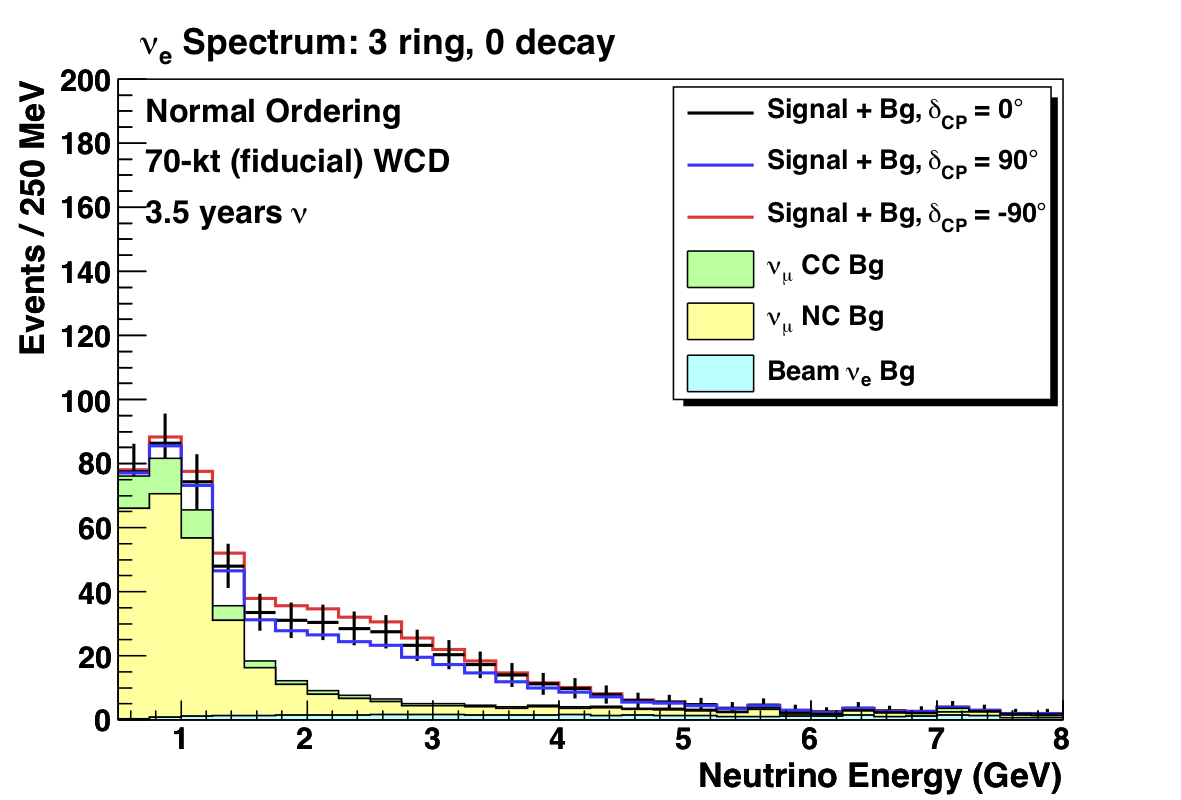}
  \includegraphics[width=0.45\linewidth]{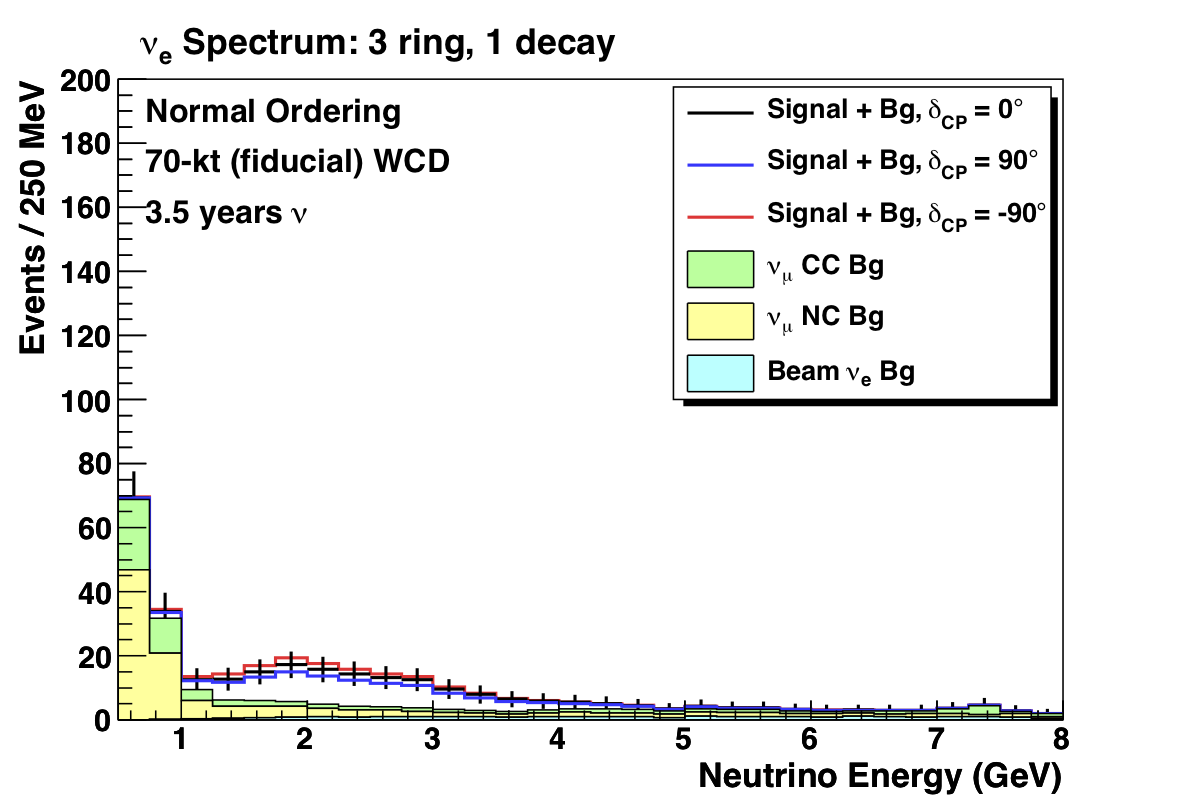}
  \caption{Expected event rates in neutrino mode as a function of reconstructed neutrino energy for \theia-100 after 3.5 years in the LBNF beam for the six selected neutrino-mode samples: one ring (top), two ring (middle), three ring (bottom) with zero Michel electrons (left) or one Michel electron (right).}
  \label{fig:lblspectra_nu}
\end{figure*}

\begin{figure*}[h!]
  \centering
  \includegraphics[width=0.45\linewidth]{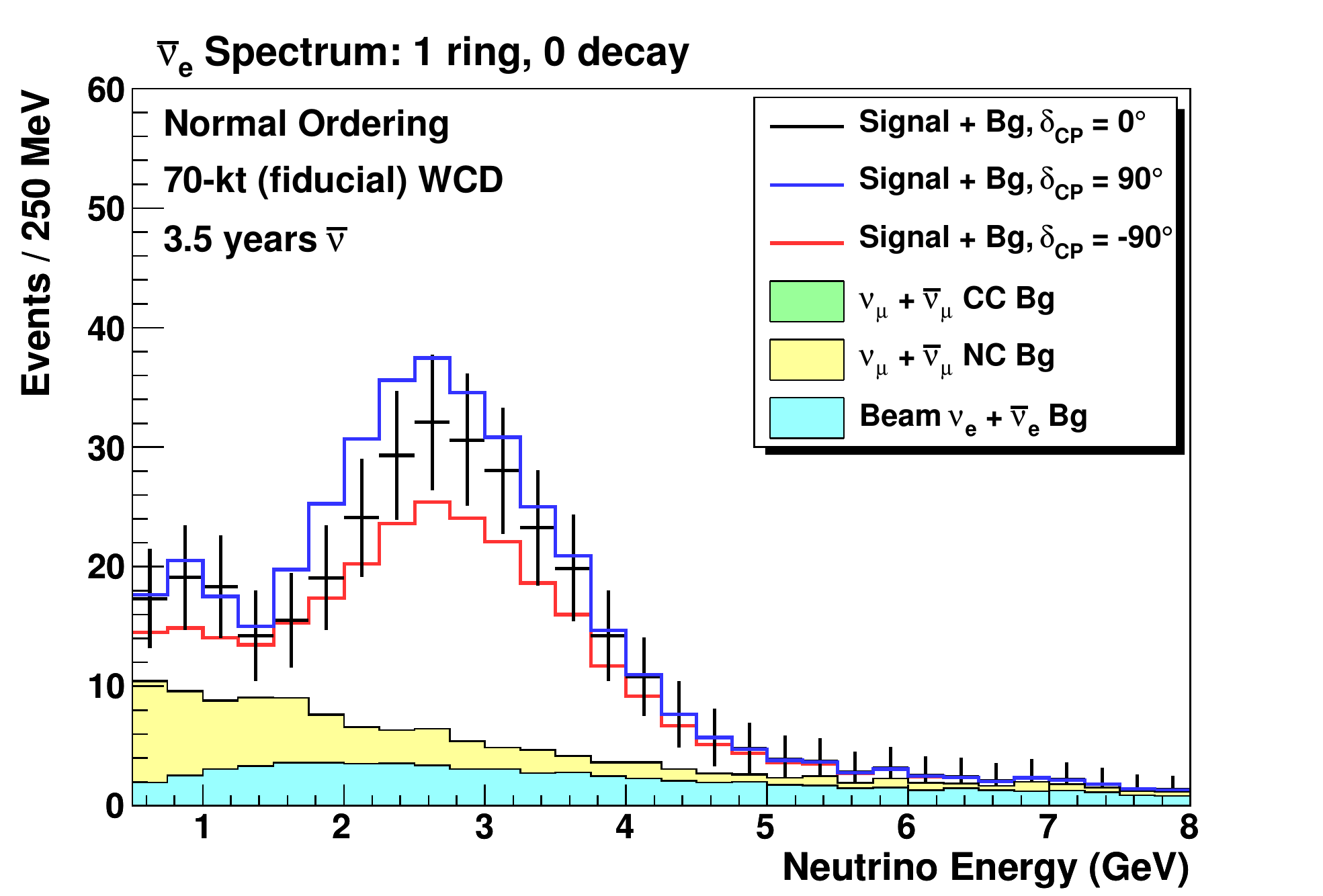}
  \includegraphics[width=0.45\linewidth]{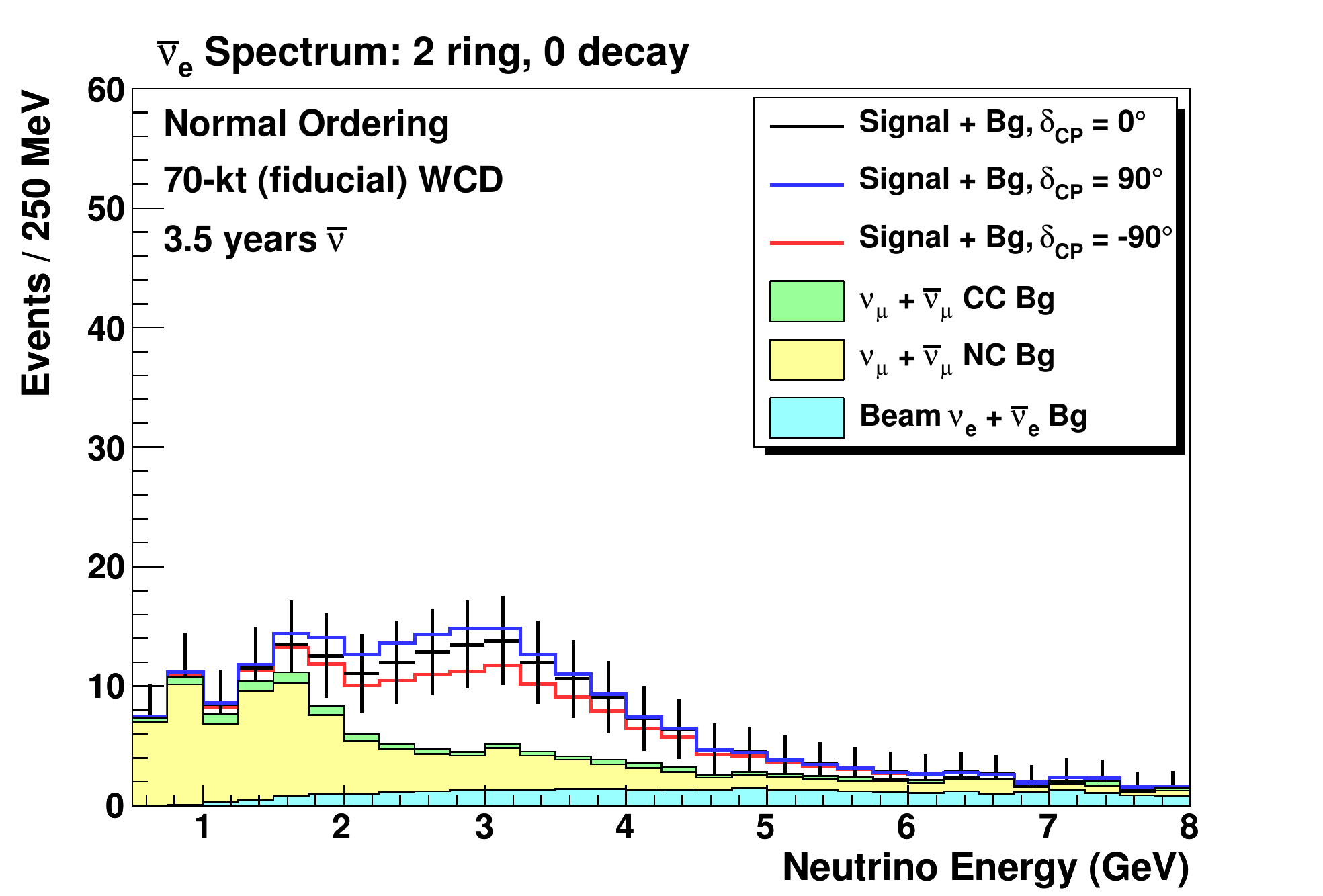}
  \includegraphics[width=0.45\linewidth]{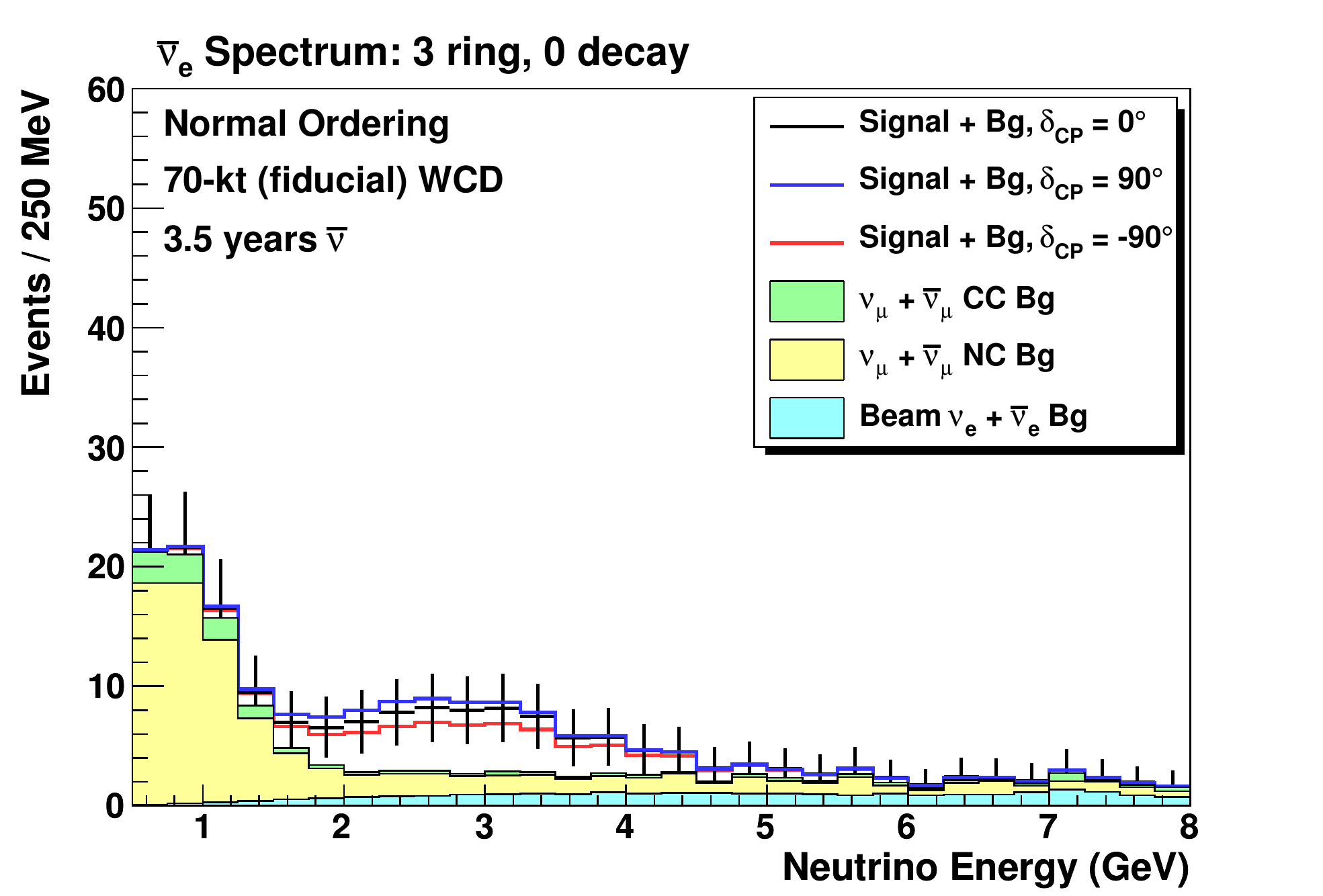}
  \caption{Expected event rates in antineutrino mode as a function of reconstructed neutrino energy for \theia-100 after 3.5 years in the LBNF beam for the three selected antineutrino-mode samples: one ring (top left), two ring (top right), three ring (bottom) with zero Michel electrons. Note that the vertical scale is different compared to Fig.~\ref{fig:lblspectra_nu}. Samples with Michel electrons are not considered for antineutrino mode.}
  \label{fig:lblspectra_anu}
\end{figure*}

The long-baseline oscillation analysis now includes nine samples that are analyzed with independent systematic
uncertainties within a single fit: one-, two-, and three-ring events with either zero or one Michel electron
in neutrino mode, and the corresponding zero Michel electron samples for antineutrino mode.  A boosted decision tree is employed to reduce the neutral current background, which uses the best-fit likelihoods of all one-, two-, and three-ring hypothesis fits, and the lowest reconstructed particle momentum in each fit. The resulting neutrino-mode samples are shown in Fig.~\ref{fig:lblspectra_nu} and the antineutrino-mode samples are shown in Fig.~\ref{fig:lblspectra_anu}. The two-
and three-ring samples tend to have higher background than the single-ring samples, but do make significant
contributions to the overall sensitivity.

%Updated to Theia100

\begin{figure*}[h!]
  \centering
  \includegraphics[width=0.45\textwidth]{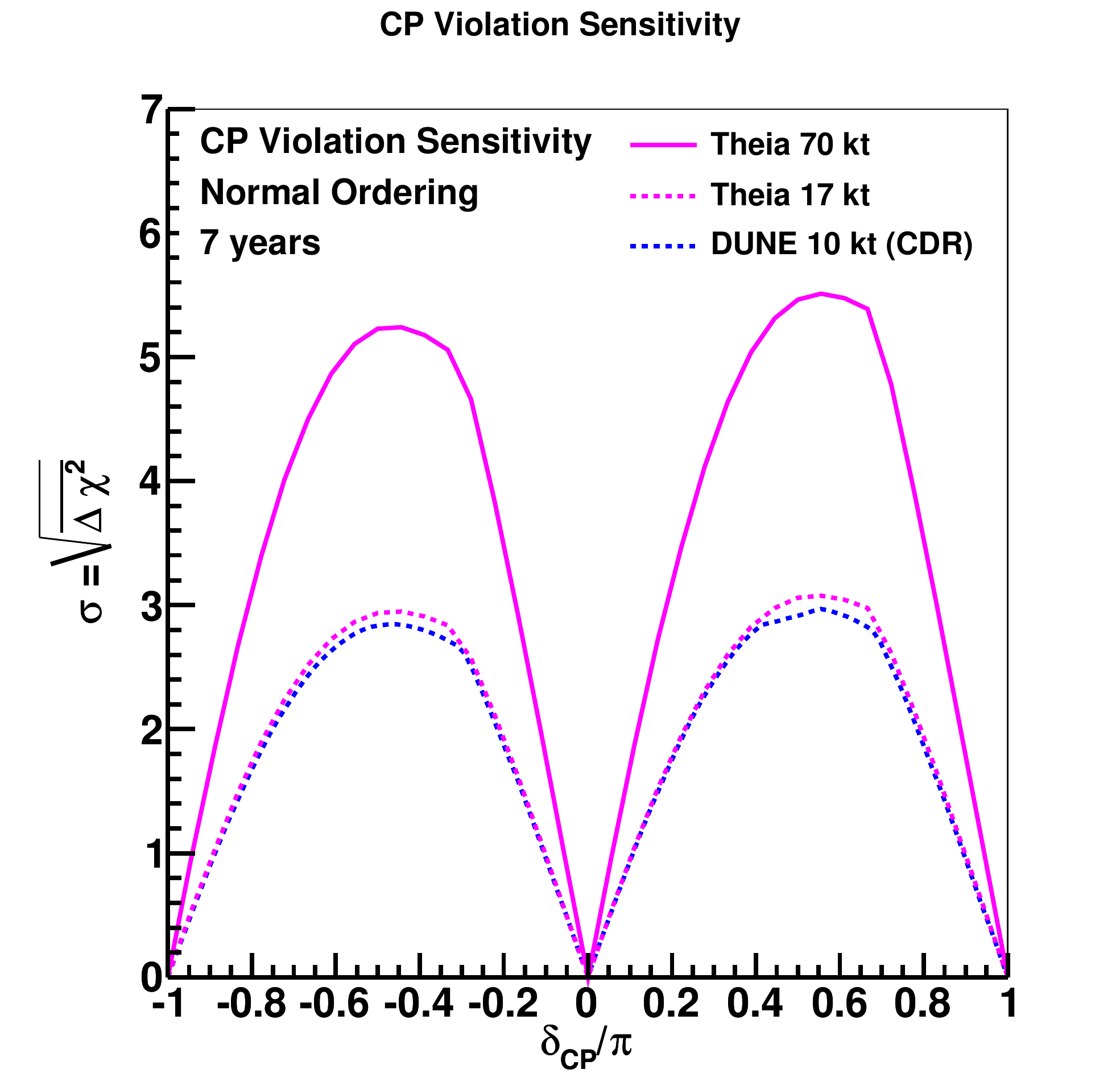}
  \includegraphics[width=0.45\textwidth]{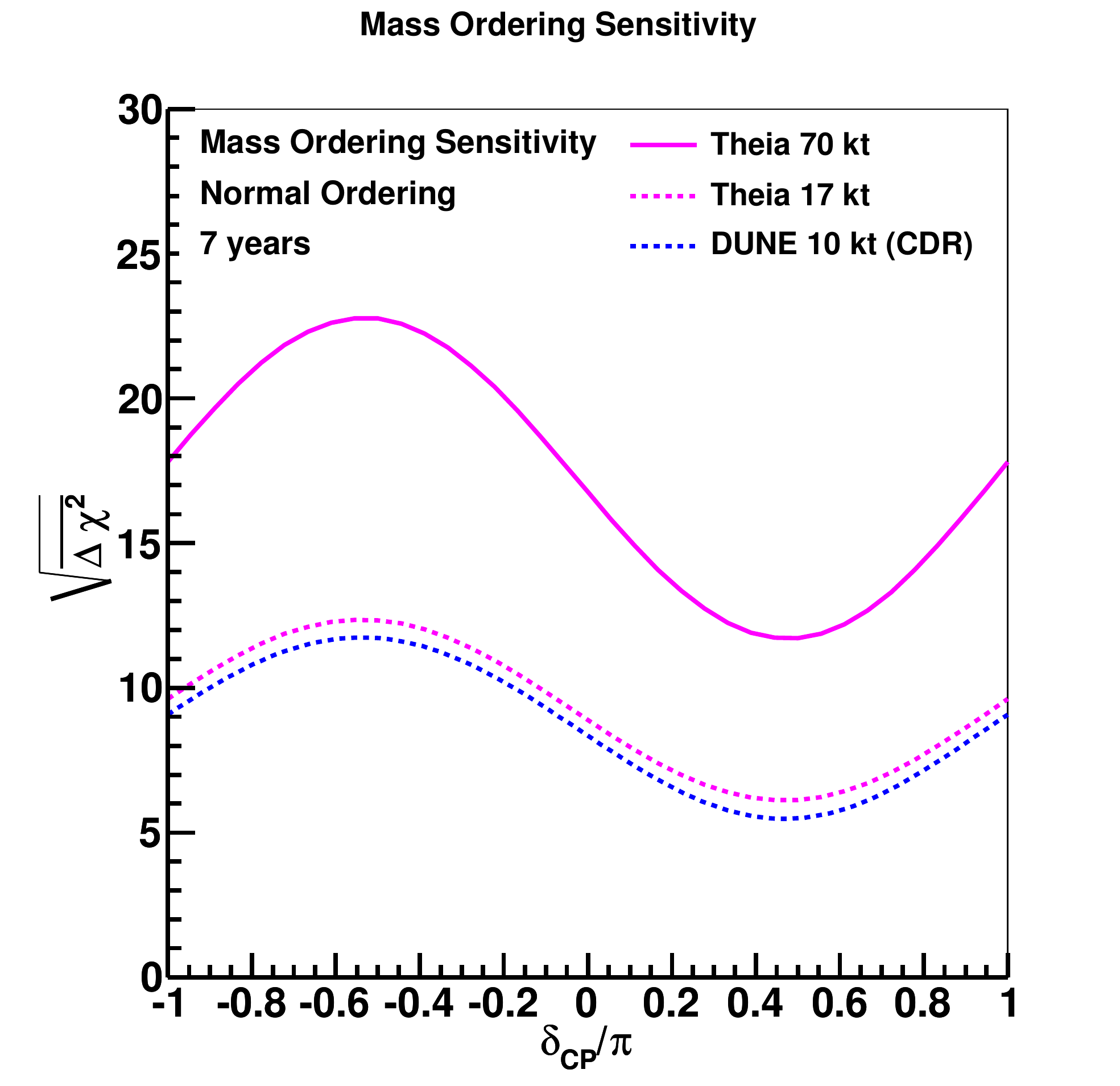}
  \caption{Sensitivity to CP violation (i.e.: determination that $\delta_{CP} \ne$ 0 or $\pi$) (left) and sensitivity
    to determination of the neutrino mass ordering (right), as a function of the true value of $\delta_{CP}$, for the \theia 70-kt fiducial volume detector (pink). Also shown are sensitivity curves for a 10-kt (fiducial) LArTPC (blue dashed) compared to a 17-kt (fiducial) \theia-25 WCD detector (pink dashed). Seven years of exposure to the LBNF beam
    with equal running in neutrino and antineutrino mode is assumed. LArTPC sensitivity is based on
    detector performance described by \cite{Alion:2016uaj}.}
  \label{fig:sens_lbl}
\end{figure*}

We assign independent normalization uncertainties of 2\%(5\%) on each of the $\nu_{e}$ and $\overline{\nu}_{e}$ appearance
signal(background) modes. We do not explicitly include the $\nu_{\mu}$ disappearance samples, but the choice of
uncertainty for the appearance samples assumes some systematics constraint from the disappearance samples.
This treatment of systematic uncertainty is comparable to that in the DUNE Conceptual Design Report (CDR) analysis, and assumes significant constraint of systematic uncertainty from the DUNE near detector. Under these assumptions, we find better than 3$\sigma$ sensitivity to CP violation for much of parameter space for \theia-100, with the sensitivity of the \theia-25 WCD being comparable that of the DUNE 10-kt LArTPC, as shown in Fig.~\ref{fig:sens_lbl}.
% <MW comment: this is redundant with the same statement above.> It is expected that this may improve when the modern fast HQE PMTs, WbLS, and LAPPDs are considered. 

Both the \theia $\delta_{CP}$ and mass hierarchy sensitivities, and the corresponding LAr sensitivities from the DUNE CDR are based on similar assumptions about the eventual systematic uncertainty levels that will be achieved. In practice, achieving these targets will require a high-precision near detector program. A conceptual design for the DUNE near detector complex is presently under development, and is expected to include 3 major detector systems: a LArTPC, a downstream multipurpose detector, consisting of a high-pressure argon gas TPC and surrounding energy calorimeter (ECAL), and a 3-dimensional scintillator tracker (3DST) consisting of 1~cm$^3$ scintillator cubes surrounded by tracking detectors, all installed within the KLOE ECAL and magnet. The argon detectors are designed to move off-axis to take data with many different neutrino energy spectra (DUNE-PRISM) in order to provide additional constraints on neutrino-nucleus interactions.

The 3DST uses the same scintillator cubes as the Super-FGD detector, which is the new detector technology being implemented in the upgrade of the ND280 near detector of the T2K experiment. T2K has a detector configuration similar to a potential \theia detector located at SURF, with a water-based far detector and a scintillator-cube-based near detector. The T2K near detector upgrade is targeted at a more detailed understanding of neutrino interaction modeling by observing short proton and pion tracks, as well as an ability to detect neutrons, and then applying this improved understanding on carbon to oxygen. A similar strategy is currently envisioned for \theia based on the existing DUNE near detector configuration.

\subsection{Solar neutrinos}

Solar neutrinos have a long history of driving development and discovery in the field, using a range of technologies. 
There is a variety of untapped physics potential related to the Sun's production of neutrinos. Among these is the opportunity to detect solar neutrinos from the subdominant CNO production mechanism and discriminate between Standard Solar Models to resolve the ``solar metallicity problem'', emerging from tensions between theory and observation~\cite{Serenelli:2009yc}. A precision measurement of the neutrino flux from the CNO cycle could prove extremely useful in resolving the discrepancy, as the flux depends linearly on solar metallicity~\cite{Bahcall:2005va}. 
A sensitive search for new physics can be performed by probing  the transition region between low-energy vacuum oscillation, below 1~MeV, and the matter-dominated oscillation above 5~MeV, which is best achieved by a measurement of the shape of the low-energy region of the spectrum of $^8$B neutrinos.
Other exciting physics opportunities related to solar neutrinos include tests of solar luminosity through precision measurements of $pep$ and $pp$ neutrinos; tests of the solar temperature; and, potentially, separation of the different components of the CNO flux to probe the extent to which this cycle is in equilibrium in the Sun's core~\cite{HRS}.  
The LENA collaboration have explored in detail the power of a large-scale scintillator detector for resolving open questions in solar neutrino physics~\cite{Wurm:2011zn}.  \theia would have similar capability,  with the advantage of being able to distinguish ES events from backgrounds (such as $^{210}$Bi) using directionality.
For this study, we focus on the prospects for CNO detection.

\subsubsection{CNO neutrino flux}

The world-leading limit on the CNO flux comes from the Borexino experiment~\cite{Collaboration:2011nga}. A primary obstacle to  observation is the challenge of distinguishing the elastic scattering (ES) of CNO neutrinos from radioactive backgrounds, particularly $^{210}$Bi beta decays, using only the energy spectrum.  Sensitivity to event direction  could substantially add to the discrimination power, as the neutrino-scattered electron direction should be strongly correlated with the direction of the Sun, whereas the background should appear isotropic.  The combined detection of both Cherenkov and scintillation signals in \theia should allow for both direction reconstruction and the required low threshold.

Such an approach has been explored for 25- and 50-kton WbLS detectors, with promising results in~\cite{bonventre:2018}. The approach here is based on that of \cite{bonventre:2018}, which is also presented in ~\cite{OrebiGann:2019ncf} and ~\cite{Gann:2015fba}, with minor departures, including focusing on the specific \theia-25 and \theia-100 configurations. 
Sensitivity to the neutrino flux normalizations is evaluated by performing a 2D binned maximum likelihood fit  in reconstructed electron energy and $\cos\theta_{\odot}$. The same simulation packages, generators and energy reconstruction methods are used as in \cite{bonventre:2018}. Simulation of the neutrino interactions and radioactive decays is performed using RAT-PAC~\cite{seib14}, a specialized simulation and analysis package employing Geant4 as the primary physics simulation software. Monte Carlo generators used include the neutrino-electron elastic scattering developed by Joe Formaggio and Jason Detwiler, the radioactive decay generator developed by Joe Formaggio, and Decay0~\cite{decay0}.  WbLS is simulated using a multi-component model of water, linear alkylbenzene (LAB) and 2,5-Diphenyloxazole (PPO), which allows flexible combination of the optical properties of the separate materials. The assumed Standard Solar Model (SSM) is the BS05OP model, and survival probabilities are calculated using~\cite{Bonventre:2013wia}.

The energy reconstruction method is presented and evaluated in \cite{bonventre:2018} and is shown to be quite robust for the energy scale considered. While the method is not a full reconstruction, it more realistically allows the folding in of geometric effects of the detector than would be achieved by a simple smearing to the Monte Carlo information.  %We do not perform an explicit energy reconstruction for any configuration. Rather for direction reconstruction, 
The direction of events due to radioactive decay is assumed to be perfectly isotropic.  For neutrino interactions, the direction of the electron relative to the neutrino is determined using the full differential cross section, and this is then convolved with  a chosen angular resolution using the functional form developed in \cite{Jillings}. This is done to study the impact of a range of potential direction resolutions that might be achieved. Alpha and coincidence rejection are applied as scalings to the relevant assumed background rate normalizations. A judicious fiducial volume cut of 60\% in \theia-100 and 50\% in \theia-25 is used as a proxy for removing external backgrounds such as ${}^{208}$Tl gammas from the PMT glass and support structure. 
This corresponds to roughly a 4-m (3-m) standoff from the PMTs for \theia-100 (\theia-25).
As in \cite{bonventre:2018}, the fits are found to be unbiased by observing the pull distributions.

In order to save on computation power, Monte Carlo signals for the fit are generated for a lower coverage, kton-scale detector. The coverage is scaled up in the reconstruction process by  adjusting the number of PMT hits accordingly, while normalizations are recalculated for the larger exposure of the 25- or 100-kton detector. This approach assumes that effects of absorption and scattering in WbLS on energy reconstruction are minimal.  Absorption is expected to be low and, while scattering may be more significant due to the presence of micelles, this should have only a small impact on energy resolution, which is dominated by photon counting.  A more significant impact can be expected on the achievable angular resolution, but this is an input to this study -- we investigate the sensitivity that can be achieved for a range of angular resolution values.  An ongoing study is exploring the angular resolution that can be achieved for different detector configurations. This method of scaling the results has been verified for a 50-kton detector against the full simulation approach taken in \cite{bonventre:2018}, and the results are found to be in good agreement, thus validating the assumptions described. The signals considered in the fit are slightly altered from those in \cite{bonventre:2018}, with the addition of a water component of the cosmogenic $^{11}$C background from spallation, as well as the addition of the cosmogenic $^{15}$O background in water, with normalizations derived from \cite{Li:2014sea}. The inclusion of these added components is found to be minimal, though they are kept in place for completeness. The depth of the LBNF location at SURF in the Homestake mine is around 4300 mwe, and the muon rate at this location ($5,3 \pm 0.17 \times 10^{-5}$ m$^{-2}$ s$^{-1}$ \cite{Abgrall}) is reduced compared to that measured at Borexino ($3.41 \pm 0.01 \times 10^{-4}$ m$^{-2}$ s$^{-1}$ \cite{Bellini:2012te}).  The latter was used in the previous studies, so the nominal cosmogenic rate is adjusted from the previous work using the formulas developed in \cite{mei06}. The remaining components are the same as in \cite{bonventre:2018}. The normalizations over which the fit is performed are:
  ${}^8$B $\nu_e$ and $\nu_\mu$;
  ${}^7$Be $\nu_e$ and $\nu_\mu$;
  CNO $\nu_e$ and $\nu_\mu$;
  $pep$ $\nu_e$ and $\nu_\mu$;
  ${}^{39}$Ar and ${}^{210}$Po, floated together;
  ${}^{210}$Bi;
  ${}^{11}$C;
  ${}^{40}$K;
  ${}^{85}$Kr;
  ${}^{15}$O;
the  ${}^{232}$ Th chain;
and the  ${}^{238}$ U chain.            

%shown in Table \ref{tab:params}. Components on the same line are floated together.
%\begin{table}[]
%\begin{tabular}{c}
%\hline
%Signals                        \\ \hline
%${}^{39}$Ar and ${}^{210}$Po   \\
%${}^8$B $\nu_e$ and $\nu_\mu$  \\
%${}^7$Be $\nu_e$ and $\nu_\mu$ \\
%${}^{210}$Bi                   \\
%${}^{11}$C                     \\
%CNO $\nu_e$ and $\nu_\mu$      \\
%${}^{40}$K                     \\
%${}^{85}$Kr                    \\
%${}^{15}$O                     \\
%$pep$ $\nu_e$ and $\nu_\mu$    \\
%${}^{232}$ Th chain            \\
%${}^{238}$ U chain            
%\end{tabular}
%\caption{The normalization parameters floated in the fit to extract the sensitivity to the CNO flux}
%\label{tab:params}
%\end{table}

%\begin{itemize}
%\item ${}^8$B $\nu_e$ and $\nu_\mu$  \\
%\item ${}^7$Be $\nu_e$ and $\nu_\mu$ \\
%\item CNO $\nu_e$ and $\nu_\mu$      \\
%\item $pep$ $\nu_e$ and $\nu_\mu$    \\
%\item ${}^{39}$Ar and ${}^{210}$Po, floated together   \\
%\item ${}^{210}$Bi                   \\
%\item ${}^{11}$C                     \\
%\item ${}^{40}$K                     \\
%\item ${}^{85}$Kr                    \\
%\item ${}^{15}$O                     \\
%\item ${}^{232}$ Th chain            \\
%\item ${}^{238}$ U chain            
%\end{itemize} 

The sensitivity is studied for a range of detector configuration parameters including WbLS scintillator fraction, angular resolution, background levels, energy threshold, energy scale and energy resolution. Five years of data taking assumed.

We take as our baseline result for the 100-kt (25-kt) detector a 5\% WbLS-loaded detector with 90\% photocathode coverage, a 5 year runtime, a 60\% (50\%) fiducial volume corresponding to roughly a 4-m (3-m) standoff from the PMTs and $25^\circ$ angular resolution. The energy region of interest is between 0.6 and 6.5 MeV. The cosmogenic levels are as discussed above, appropriately scaled to LBNF from the Borexino-measured rate.  The levels of radioactive background are based on previous experiments: in the scintillator component, we assume Borexino-level backgrounds.  In the water component, we consider the U-chain and Th-chain backgrounds to be at the level of SNO water, the ${}^{40}$K to be one tenth of the Borexino CTF water-level, and remaining backgrounds to be at Borexino levels~\cite{balata:1996}\cite{bxo09}\cite{alimonti:2009}\cite{okeefe:2008}. Table \ref{tab:bgrate} shows the corresponding levels. Alpha rejection is taken to be 95\% across all energies; coincidence rejection of BiPo events falling in distinct trigger windows is taken to be 100\%, and 95\% for events that fall in the same trigger window.

\begin{table*}[]
\centering
\begin{tabular}{c|cc}
                 & H$_2$O level (g/gH$_2$O)   & LS level (g/gLS)                        \\ \hline
$^{238}$U chain  & $6.6 \times 10^{-15}$ \cite{okeefe:2008}      & $1.6 \times 10^{-17}$ \cite{bxo09}                  \\
$^{234}$Th chain & $8.8 \times 10^{-16}$  \cite{okeefe:2008}    & $6.8 \times 10^{-18}$  \cite{bxo09}                  \\
$^{40}$K         & $6.1 \times 10^{-16}$   \cite{balata:1996}   & $1.3 \times 10^{-18}$     \cite{alimonti:2009}              \\
$^{85}$Kr        & $2.4 \times 10^{-25}$      & $2.4 \times 10^{-25}$   \cite{alimonti:2009}                  \\
$^{39}$Ar        & $2.8 \times 10^{-24}$      & $2.8 \times 10^{-24}$       \cite{alimonti:2009}              \\
$^{210}$Bi       & $3.8 \times 10^{-28}$      & $3.8 \times 10^{-28}$              \cite{alimonti:2009}       \\
$^{210}$Po       & $4.2 \times 10^{-24}$      & $4.2 \times 10^{-24}$     \cite{alimonti:2009}                 \\
$^{11}$C         & 100 (event per kt per yr)  \cite{Li:2014sea}\cite{Abgrall}  \cite{mei06}& $2.0 \times 10^4$ (event per kt per yr)  \cite{Abgrall} \cite{Bellini:2012te}  \cite{mei06} \\
$^{15}$O         & 3000 (event per kt per yr) \cite{Li:2014sea}\cite{Abgrall}\cite{mei06} & 0                                      
\end{tabular}
\caption{Natural radioactive and cosmogenic background levels for the  baseline detector configuration.}
\label{tab:bgrate}
\end{table*}

\begin{figure}[!t]
	\includegraphics[width=0.49\textwidth]{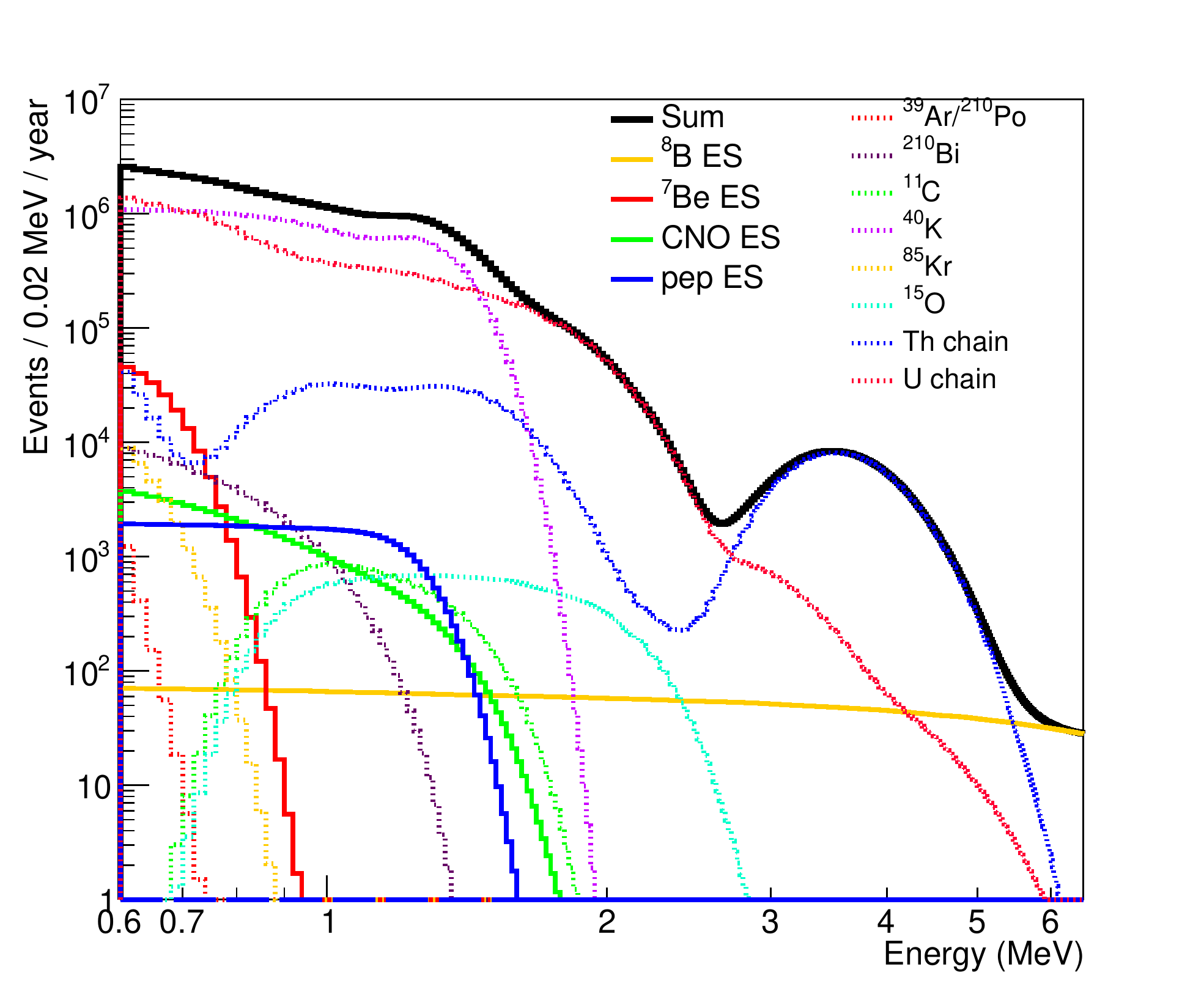}
	\caption{Energy spectrum of expected signals for a 25 kiloton, 5\% WbLS loaded detector} 
	\label{fig:spec}
\end{figure}

\begin{table*}[!h]
\centering
\begin{tabular}{cc|cccccc}
\hline
Target mass & WbLS  & 25$^\circ$ & 35$^\circ$ & 45$^\circ$ & 55$^\circ$ & 60$^\circ$ & 65$^\circ$ \\ \hline
100 kt      & 0.5\% & 4.7\%   & 6.8\%   & 8.6\%  & 10.5\%  & 12.1\%  & 13.6\%  \\
100 kt      & 1\%   & 4.5\%   & 6.4\%   & 8.0\%  & 10.0\%  & 11.5\%  & 12.9\%  \\
100 kt      & 2\%   & 4.1\%   & 5.9\%   & 7.5\%   & 9.3\%  & 10.6\%  & 12.3\%  \\
100 kt      & 3\%   & 3.7\%   & 5.3\%   & 6.9\%   & 8.4\%   & 9.8\%  & 11.3\%  \\
100 kt      & 4\%   & 3.6\%   & 5.2\%   & 6.6\%   & 8.0\%   & 9.5\%   & 10.9\%  \\
100 kt      & 5\%   & 3.4\%   & 4.9\%   & 6.3\%   & 7.4\%   & 8.7\%   & 9.9\%  \\ \hline
25 kt       & 0.5\% & 11.1\%  & 16.2\%  & 20.6\%  & 24.1    & 28.4\%  & 32.8\%   \\
25 kt       & 1\%   & 10.0\%  & 14.1\%  & 18.1\%  & 22.1\%  & 25.8\%  & 29.6\%  \\
25 kt       & 2\%   & 8.7\%   & 12.6\%  & 16.1\%  & 19.7\%  & 23.1\%  & 26.4\%  \\
25 kt       & 3\%   & 8.0\%   & 11.6\%  & 14.9\%  & 18.0\%  & 21.5\%  & 24.2\%  \\
25 kt       & 4\%   & 7.7\%   & 11.1\%  & 14.3\%  & 17.4\%  & 20.4\%  & 23.0\%  \\
25 kt       & 5\%   & 7.2\%   & 10.2\%  & 12.9\%  & 15.5\%  & 18.0\%  & 20.3\%  \\ \hline
\end{tabular}
\caption{Fractional uncertainty, as a percentage, on the fitted CNO normalization parameter for various detector configurations scanned over size, WbLS scintillator fraction and assumed angular resolution. The quoted results are the average of 100 such fits.}
\label{tab:cnomoney}
\end{table*}

\begin{figure}[!t]
	\includegraphics[width=0.49\textwidth]{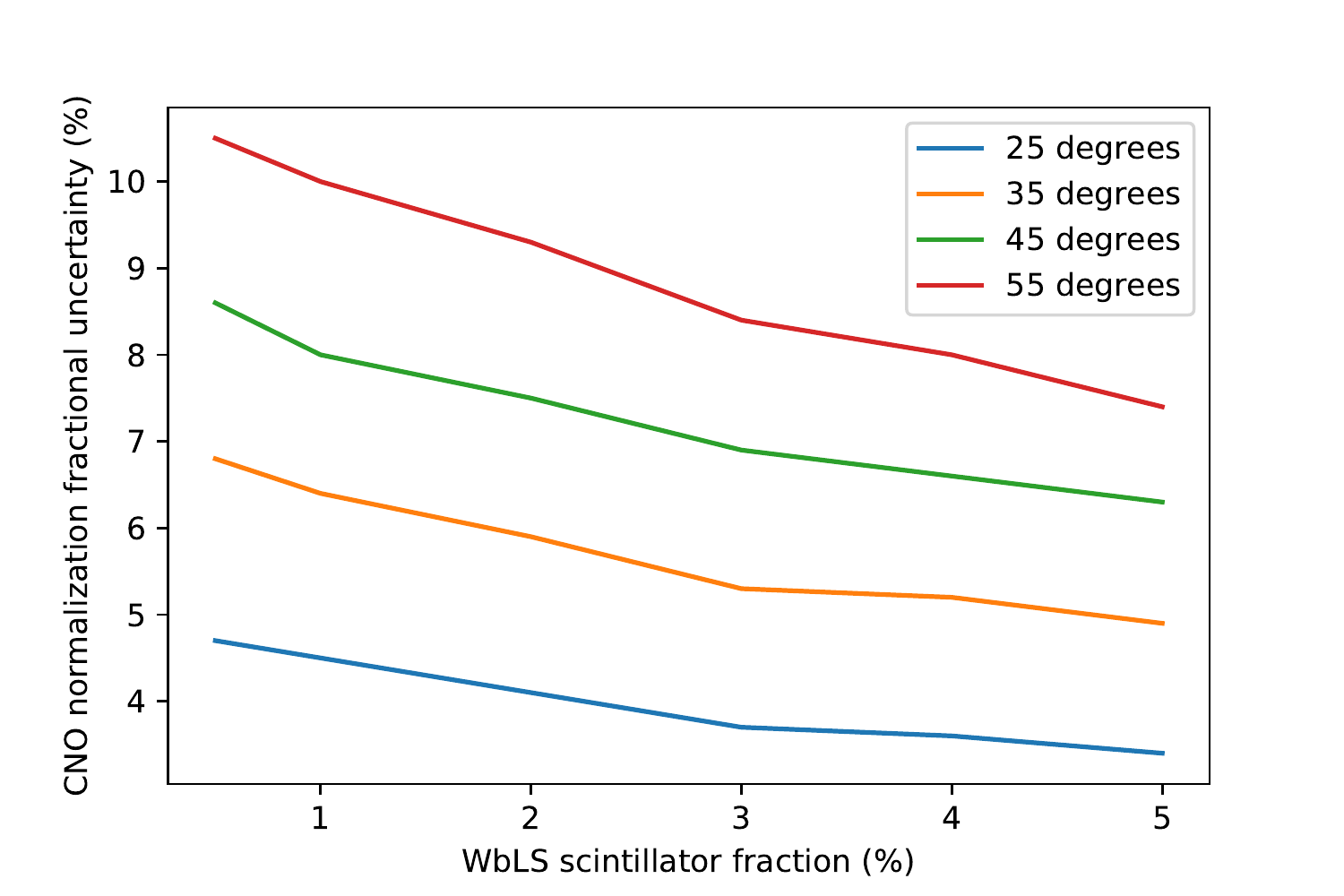}
	\caption{Fractional uncertainty on the CNO normalization parameter as a function of angular resolution and WbLS scintillator fraction, for \theia-100.} 
	\label{fig:tableplot}
\end{figure}

The resulting energy spectrum plot for the 25-kt, 5\% WbLS detector can be seen in Fig.~\ref{fig:spec}.  Table~\ref{tab:cnomoney} shows the fractional uncertainty on the CNO normalization parameter as a function of the detector volume, assumed angular resolution, and WbLS fraction, holding all other configurations from the baseline fit constant. Fig.~\ref{fig:tableplot} highlights the dependence on angular resolution and WbLS fraction in the 100-kton detector. A  strong dependence on angular resolution is seen, underlining the important of good direction reconstruction for discriminating signal from radioactive background.
%as improving this differentiates the neutrino signals from the radioactive backgrounds the most. 
High photocoverage is essential  for such a requirement, particularly at low energy.

The dependence of the CNO sensitivity on various background assumptions has been explored in~\cite{bonventre:2018}, including
the cosmogenic rate, alpha rejection, BiPo coincidence rejection, and the level of background in water for the low energy backgrounds ${}^{210}$Bi, ${}^{39}$Ar, ${}^{85}$Kr  and ${}^{40}$K. As the dominant low-energy background, additional attention was paid here to ${}^{40}$K, beyond the previous work.  This isotope  has a somewhat complicated decay scheme involving $\beta^-$ decay to ${}^{40}$Ca, and electron captures to the ground and $J^\pi = 2^+$ states of ${}^{40}$Ar, as well as $\beta^+$ decay to the ground state of ${}^{40}$Ar \cite{Chen}. In order to evaluate any impact of  additional uncertainties on this decay, we consider altering the branching fractions consistent with tabulated uncertainties, as well as changing the spectral shape associated with the $\beta^-$ decay  to an experimentally determined one \cite{Mougeot}. We find that propagating these effects changes the final results to the optimal cases for \theia-25 and \theia-100 at the 0.1\% level or less. The impact of uncertainties on energy scale and resolution were considered in~\cite{bonventre:2018} and found to be small.

Varying the energy threshold  is found to have an understandably large effect on the sensitivity. For the 5\% WbLS detector with 25$^\circ$ degree angular resolution, the CNO uncertainty climbs to 34\% (15\%) for the 25-kton (100-kton) detector if the threshold is raised from 0.6~MeV to 1~MeV.  

%Reduction of the K, U and Th contamination in the water might also be considered in order to increase sensitivity.

%Raising the threshold will cut off more and more of the spectrum and reduce statistics, so it is desirable for the target material to have a reasonably high light yield to push the threshold down. 
%Energy scale and resolution effects are determined by appropriately shifting by a fixed percentage (scale) or smearing by a Gaussian of fixed width (resolution), and then redoing the fit and it is found that the effects are minor. %Additionally, the impact of rejection of alphas and BiPo coincidences is found to be fairly minimal.

\subsubsection{Additional possibilities for solar neutrinos}

\begin{figure}[!t]
	\includegraphics[width=0.49\textwidth]{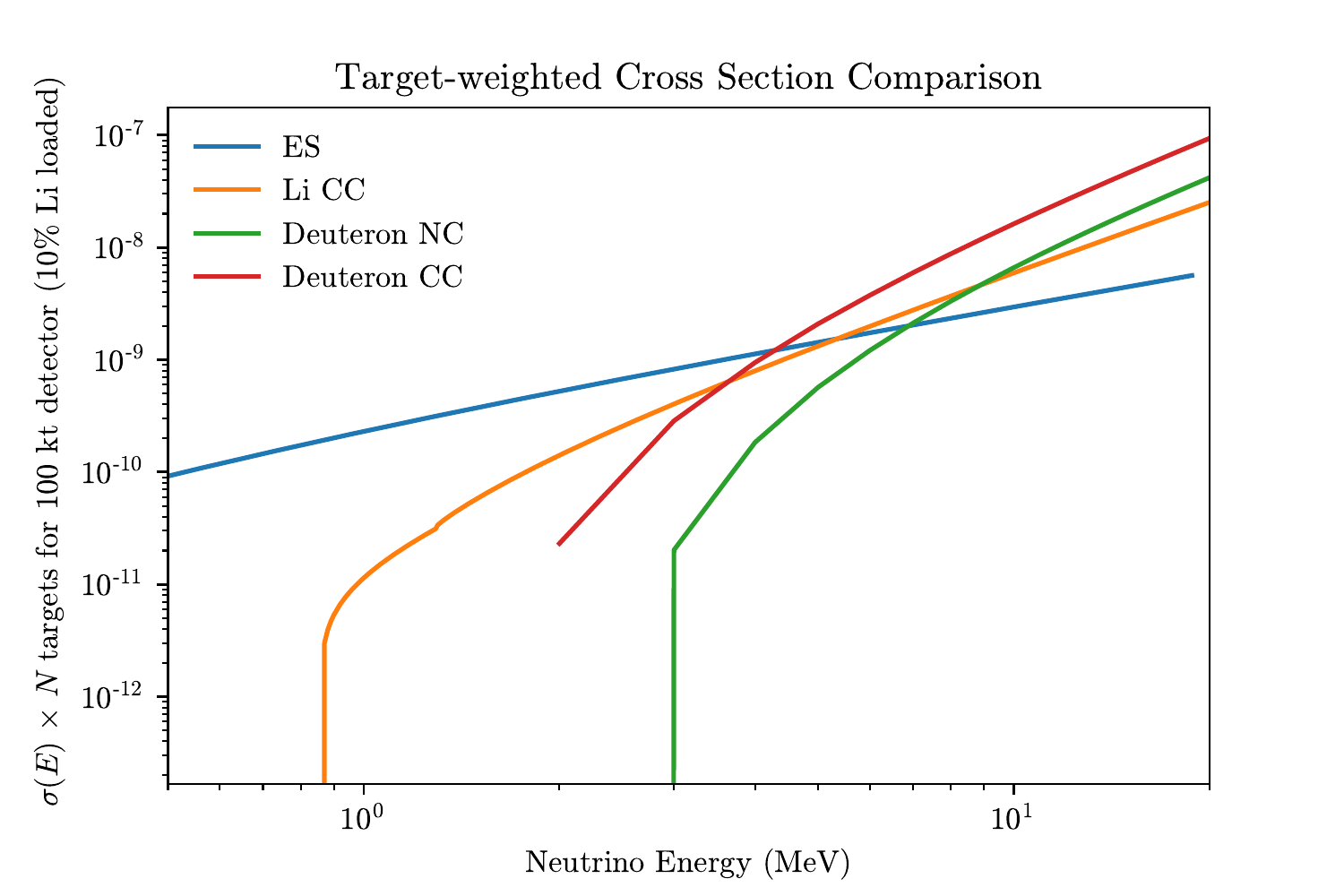}
	\caption{Comparison of various interaction cross sections, weighted by the number of available targets for that interaction in a 100 kiloton detector. The elastic scattering (ES) and $^{}7$Li charged current (Li CC) interactions are weighted for a 100 kiloton detector of 5\% WbLS, loaded with 10\% Lithium. The Li CC cross section is taken from~\cite{sigli}. The deuteron neutral current (Deuteron NC) and charged current (Deuteron CC) are weighted for a 100 kiloton detector entirely of heavy water, D$_2$O.  Deuteron cross section are taken from~\cite{Kubodera:1993rk}.} 
	\label{fig:targetweight}
	\vspace{-0.2in}
\end{figure}

Other studies of interest include probing the ${}^8$B transition region between 1 and 5 MeV for potential non-standard effects, as well as precision measurements of the $pp$, $pep$ and the individual CNO constituent fluxes to refine stellar modeling. The prospects for these studies could potentially be improved by loading the target material with a favorable isotope that enhances the signal through other interaction channels, such as ${}^7$Li, which has a high cross section for the neutrino-nucleus charged current (CC) interaction \cite{haxton:1996} \cite{ASDC:2014}. Fig.~\ref{fig:targetweight} compares the neutrino-electron elastic scattering ${}^7$Li CC interactions as a function of energy, weighted by target in a 100-kt detector loaded with 5\% WbLS and 10\% ${}^7$Li. Overlaid are the target-weighted cross sections for the charged and neutral current interactions on the deuteron, again for a 100-kt detector. It can be seen that the ${}^7$Li cross section becomes dominant around 5~MeV, which is of particular interest for a study of the ${}^8$B transition region. An additional benefit from this interaction is that the cross section highly correlates the incoming neutrino energy and outgoing electron energy, which provides an additional lever for studies in which knowledge of the neutrino energy is key, whereas the breadth of the elastic scattering differential cross section washes this out. Studies are ongoing to quantitatively  evaluate the potential for these additional physics topics, and the impact of isotope loading.

\subsection{Supernova neutrinos}

%\paragraph{Motivation}

The neutrino burst detected from the next galactic supernova (SN) will provide us with a wealth of information on the dynamics of the core collapse (neutronization, reheating, proto-neutron star cooling) and the properties of the neutrinos themselves (mass hierarchy, absolute mass scale, collective oscillations) (e.g.~\cite{Scholberg:2018meq}). Since the first detection of SN neutrinos in 1987, there has been a continuous stream of new features predicted for the SN neutrino signal, hinting at new stellar or particle physics \cite{Raffelt:2007nv}. So while we are uncertain what (superposition of) signatures to expect from the next event, it is beyond doubt that only a concerted effort of all neutrino observatories available will enable us to extract the full information contained in the burst signal (e.g.~\cite{Scholberg:2017czd}), then to be combined with electromagnetic and gravitational wave observations (e.g.~\cite{Nakamura:2016kkl}).

If a SN neutrino burst would pass by the Earth today, the largest event statistics would be collected by the two large Cherenkov detectors, SK and IceCube \cite{Abe:2016waf,Kopke:2017req}. Ten years from now, we may expect that additional information will be added by JUNO's liquid scintillator and DUNE's liquid argon neutrino targets \cite{An:2015jdp,Ankowski:2016lab}. In a simplified picture, SK, JUNO and IceCube will dominate the information on $\bar\nu_e$ flux and energies, while DUNE has the potential for a high-statistics $\nu_e$ measurement. JUNO will provide information on the combined flux of $\nu_\mu$ and $\nu_\tau$ and antineutrinos (denoted commonly as $\nu_x$).

%\subsubsection{Detector Configuration}

What will \theia add to the global picture of SN neutrino observations? To answer this, we assume \theia-100 with a WbLS target of 10\,\% organic fraction and 90\,\% optical coverage. The resulting photoelectron yield of $\sim$200\,p.e./MeV (75\,\% scintillation) provides a 7\,\% energy resolution comparable to present-day organic scintillator detectors and a sufficiently low threshold for high-efficiency neutron tagging.

\begin{figure}[!ht]
\begin{center}
\includegraphics[width=0.42\textwidth]{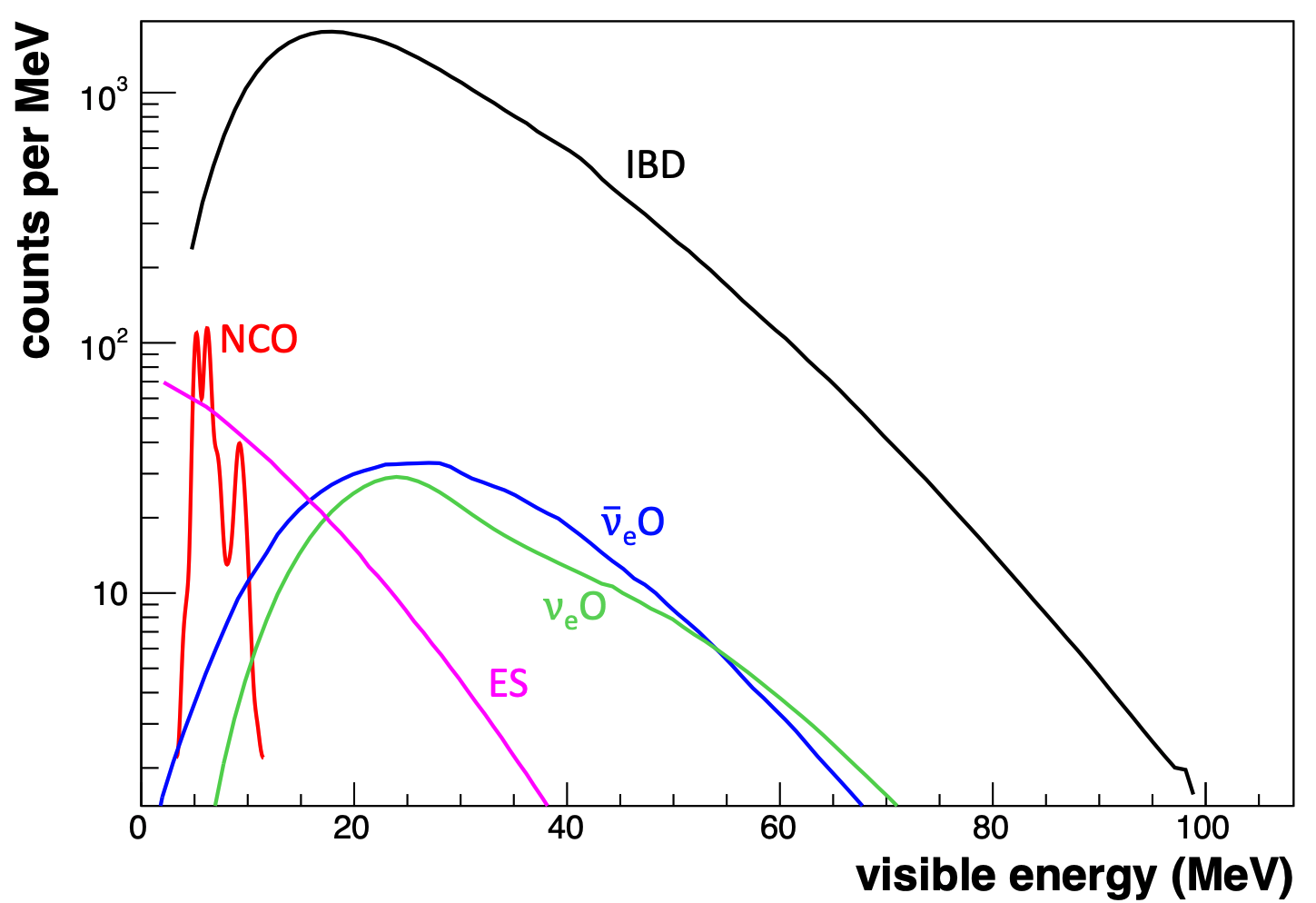}
\caption{Visible energy spectra of the prompt events, corresponding to the event rates of Table~\ref{tab:snrates} (GVKM model \cite{Gava:2009pj}). A Gaussian energy resolution of 7\,\% at 1\,MeV is applied.}
\label{fig:snspectra}
\end{center}
\vspace{-1.5\baselineskip}
\end{figure}

\begin{figure}[!t]
\centering
\vspace{-1.5\baselineskip}
\includegraphics[width=0.42\textwidth]{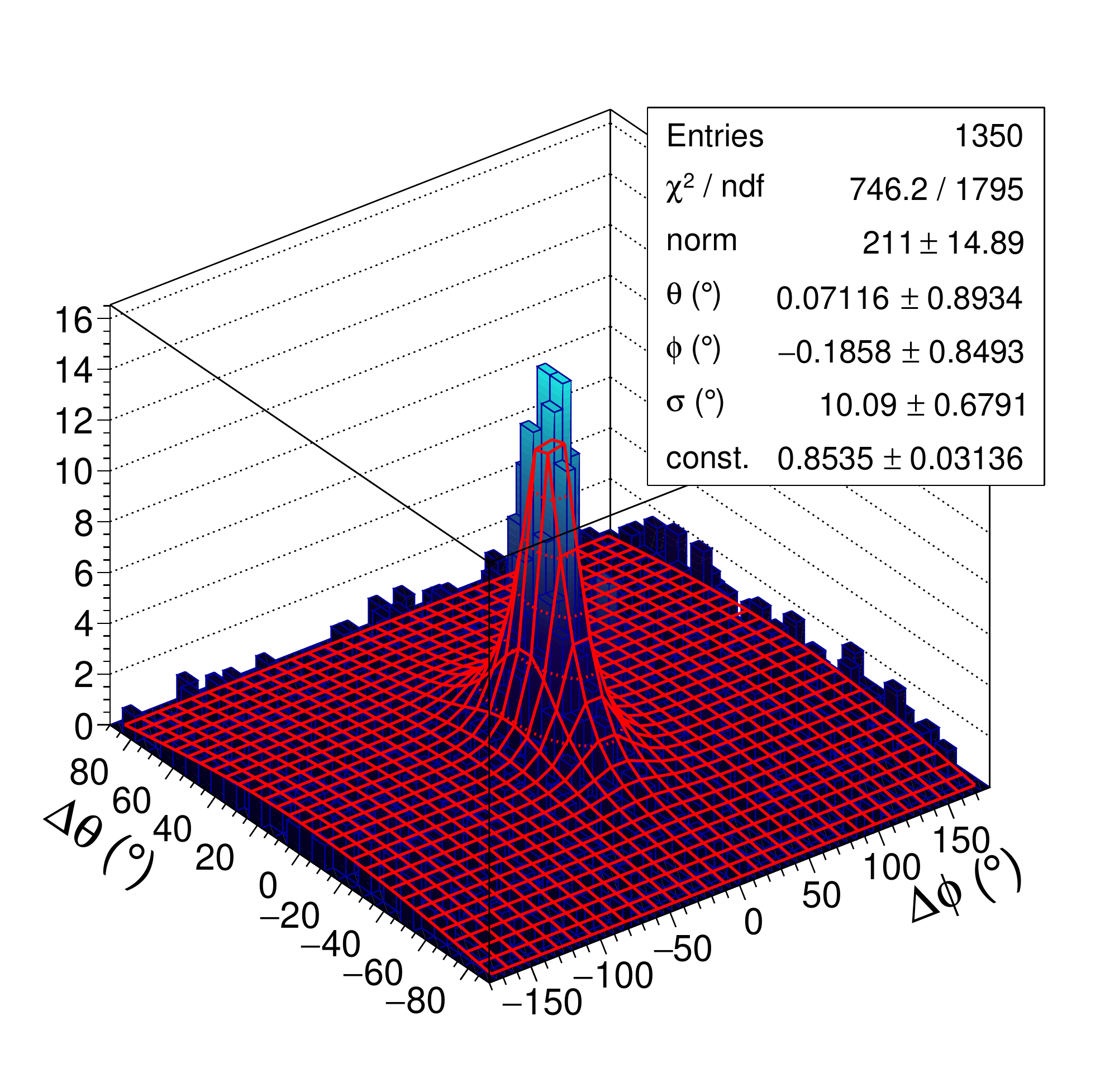}
\hfill
\includegraphics[width=0.42\textwidth]{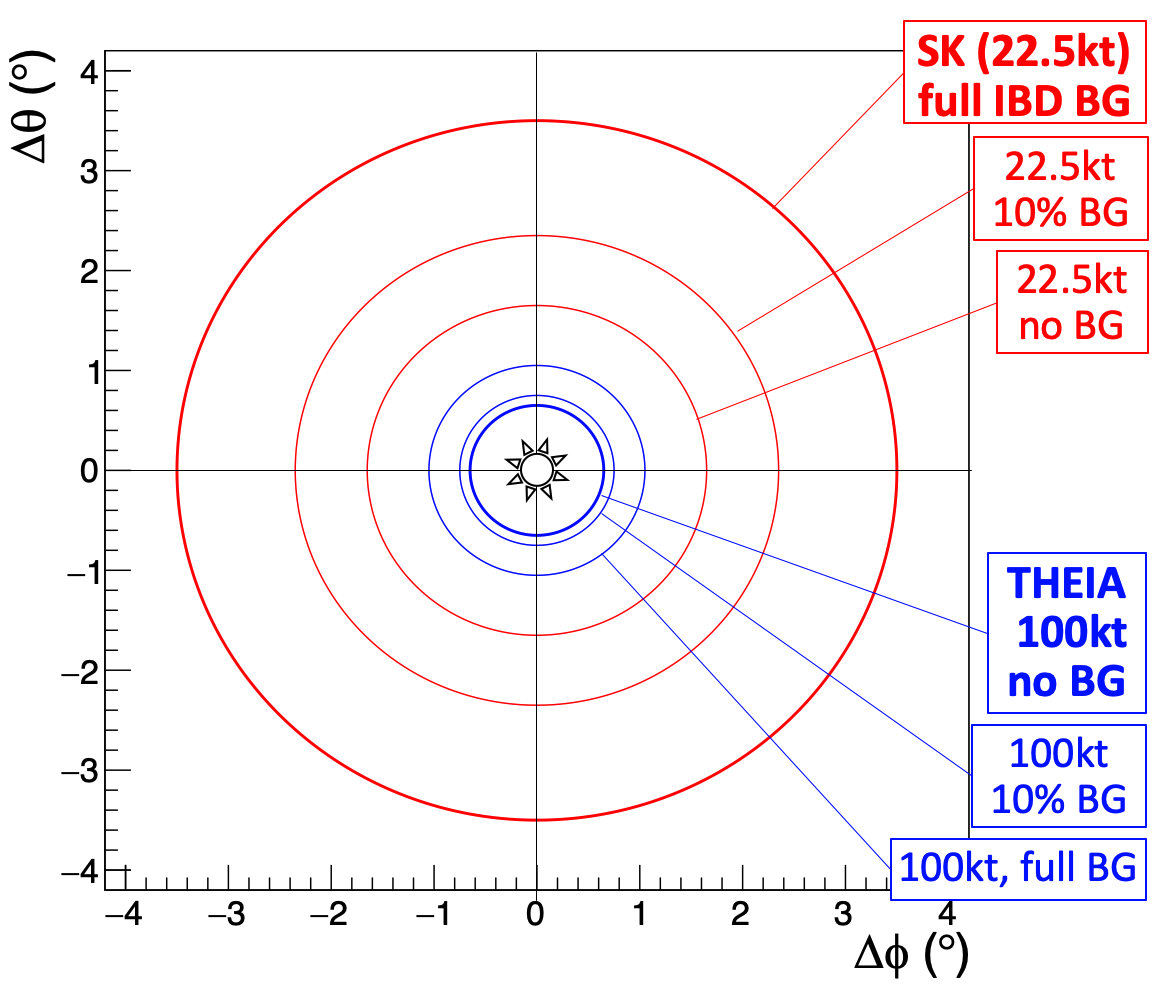}
\caption{SN pointing capability of \theia, based on the reconstruction of the ES directional signal. {\it Top panel:} Example angular distribution, assuming 90\,\% in the flat IBD spectrum. Based on a fit to this and similar distributions (red net), the {\it Bottom panel} depicts the pointing accuracy for \theia, assuming different IBD background levels for 100\,kt as well as 22.5\,kt target mass (comparable to SK).}
\label{fig:snpointing}
\end{figure}

\begin{figure}[h!]
\centering
\includegraphics[width=0.48\textwidth]{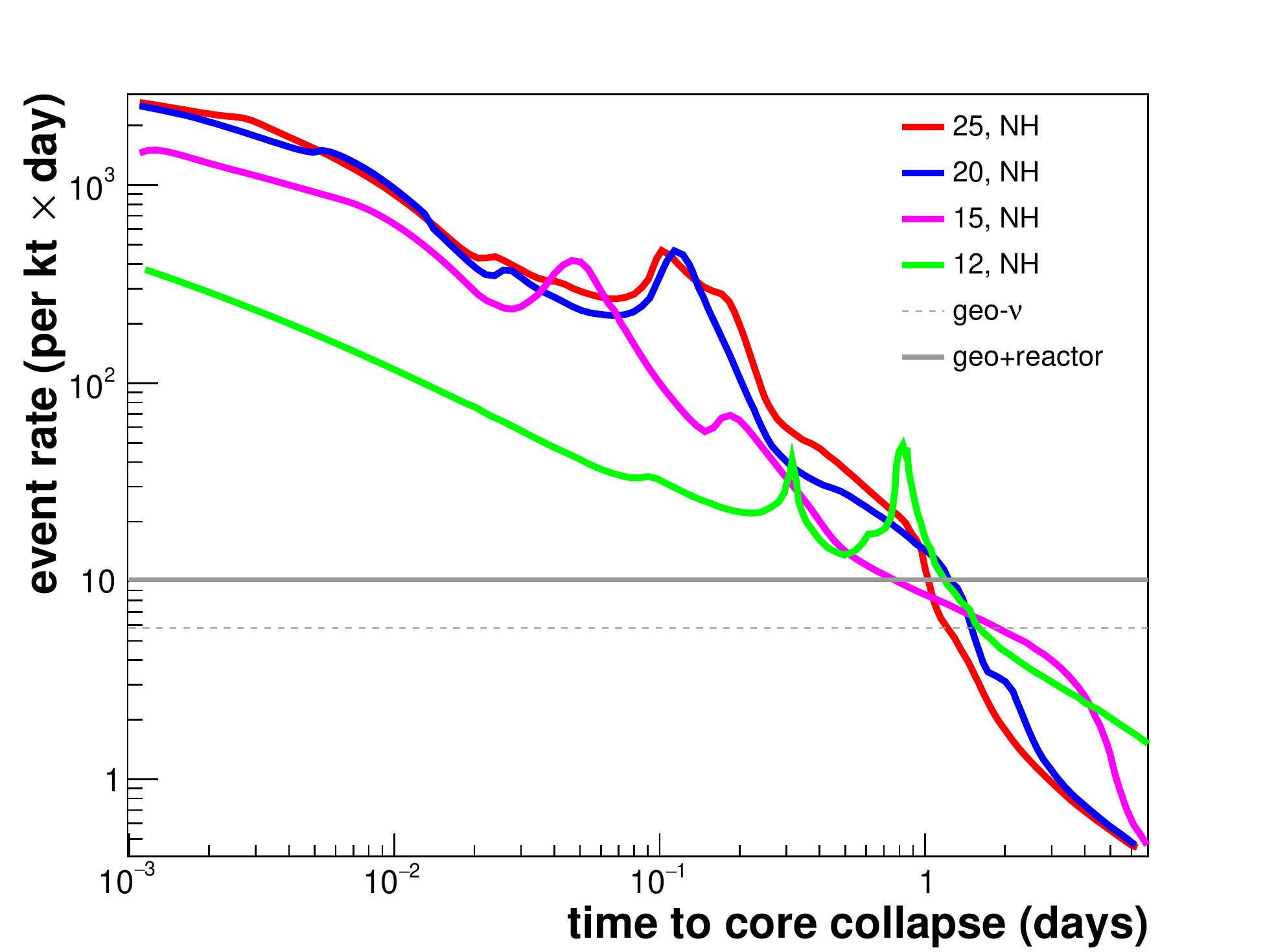}
\caption{Time-dependent pre-Supernova neutrino events expected in Theia-100: In total, more than 100 events are expected for a heavy-mass progenitor ($25\,M_\odot$) at 1\,kpc distance and NH. In all cases, the $\bar\nu_e$ event rate exceeds the relatively low background from geo- and reactor neutrinos $\sim$1 day before the onset of the core-collapse. For IH, expected event rates are about a factor of 3 lower. Light curves adopted from \cite{Guo:2019orq}. }
\label{fig:presn}
\end{figure}

\begin{enumerate}
\item {\it A high-statistics and low-threshold signal:} At 100\,kt, \theia will more than double the statistics expected for both SK and JUNO combined in $\bar\nu_e$-induced IBD signals and add hundreds of events for $\nu_e$'s and $\nu_x$'s (Table~\ref{tab:snrates}). Together with a good energy resolution, this will be very useful for correlation of time-dependent spectral features with other observation techniques, e.g.~with gravitational wave emission in the early accretion phase (SASI), or when looking for energy-dependent oscillation patterns (e.g.~the spectral swaps induced by collective oscillations) \cite{Kopke:2017req}.
\item {\it Flavor-resolved neutrino spectra:} The presence of delayed tags from neutron capture (IBD) and re-decays of ${^{16}{\rm N}}$, the presence of $\gamma$-lines from NCO reactions  as well as the directional signature for ES will allow to resolve the integrated SN neutrino signal into its individual spectral components (Fig.~\ref{fig:snspectra}). This will enable unambiguous spectroscopy of the $\nu_e$ (ES+$\nu_e$O) and $\bar\nu_e$ (IBD+$\bar\nu_e$O) signals as well as a measurement of the combined $\nu_e+\bar\nu_e+\nu_x$ flux via NCO.
\item {\it Supernova pointing:} The presence of a high-efficiency neutron tag greatly simplifies the selection of a clean ES sample from an otherwise overwhelming IBD background \cite{Tomas:2003xn}, providing pointing accuracy on the sub-$1^\circ$ level and thus extremely valuable information for multi-messenger observation of the early SN phases. The top panel of Fig.~\ref{fig:snpointing} exemplifies an angular distribution of directional ES and nearly isotropic IBD prompt events, assuming a tagging efficiency of 90\,\% and an intrinsic angular resolution of 10$^\circ$ ($1\sigma$). The lower panel compares the pointing capabilities of \theia and SK for varying assumptions on the tagging efficiency, including the upcoming SK-Gd phase.
\item {\it Neutronization burst:} While the ES signal induced by the $\nu_e$ burst from the initial phase of the core-collapse is comparatively weak, the large mass of \theia provides $\cal O$(20) events for an SN at 10\,kpc. For a close-by SN (e.g.~1\,kpc), statistics will become sufficient to look as well for the $\nu_e$ spectrum and potential oscillation effects impacting on the burst.
\item {\it Complementarity of \theia-25 and DUNE LAr:} The combination of WbLS and LAr detectors at the same site will provide for a co-detection of $\nu_{e}$ and $\bar\nu_e$ signals, offering a unique opportunity to search for potential differences in flavor/antiflavor oscillations for neutrinos traversing the Earth. Moreover, even the comparatively low volume of \theia-25 will provide ${{\cal O}(10^2)}$ 
events for an SN as far away as the Large Magellanic Cloud and thus a reliable and fast trigger signal for the DUNE LArTPCs.
\item {\it Complementarity to other observatories:} In relation to SK and JUNO, \theia will be a further high-statistics $\bar\nu_e$ detector but on the opposite side of the Earth, allowing investigation of Earth matter effects in a direct spectral comparison and superior pointing capabilities.
\item  {\it Pre-Supernova neutrinos.} Theia will be a sensitive observatory for the $\bar\nu_e$ signal emitted by the Si burning phase of a close-by SN progenitor star. Fig.~\ref{fig:presn} scales the time-dependent event rates calculated in \cite{Guo:2019orq} to the conditions of Theia-100: depending on the progenitor mass and the neutrino mass ordering, between 10 and 110 pre-SN $\bar\nu_e$ events can be expected for a star at 1~kpc remove (for $12M_{\odot}$ (IH) and $25M_{\odot}$ (NH), respectively). Given the relatively low background from geo- and reactor neutrinos of $\sim$10 events per day (see below), a $3\sigma$ detection of the signal will be possible out to 3.3\,kpc in the most favourable conditions (compared to $\sim$1.5\,kpc for JUNO).
\end{enumerate}
 
 \begin{table}[h!]
\caption{Event rates expected in 100\,kt of WbLS (10\,\% scintillator) for an SN at 10\,kpc distance (GVKM model \cite{Gava:2009pj} and SNOwGLoBES). We list inverse beta decays (IBDs), elastic scattering off electrons (ES) as well as charged-current ($\nu_e$O,$\bar\nu_e$O) and neutral-current (NCO) interactions on oxygen. Comparatively small event rates on carbon are not listed.}
\label{tab:snrates}
\begin{minipage}[b]{0.4\textwidth}
\begin{tabular}{llr}
\hline\noalign{\smallskip}
\multicolumn{2}{l}{Reaction} & Rate \\
\noalign{\smallskip}\hline\noalign{\smallskip}
(IBD) & $\bar\nu_e+p\to n+e^+$ & 19,800 \\
(ES) & $\nu+e \to e+\nu$ & 960 \\
($\nu_e$O) & ${^{16}\rm{O}}(\nu_e,e^-){^{16}{\rm F}}$ & 340 \\
($\bar\nu_e$O) & ${^{16}\rm{O}}(\bar\nu_e,e^+){^{16}{\rm N}}$ & 440 \\
(NCO) & ${^{16}\rm{O}}(\nu,\nu){^{16}{\rm O}^*}$  & 1,100 \\
\hline
\noalign{\smallskip}\hline
\end{tabular}
\end{minipage}\hfill
\end{table}

%\begin{itemize}
%\item a high-statistics and low-threshold IBD ($\bar\nu_e$) signal, offering considerably better energy resolution than SK and about twice its statistics; both aspects will prove very useful when trying to correlate spectral features changing over time with other signals, e.g.~gravitational wave emission during the SASI phase, or when looking for energy-dependent oscillation patterns (e.g.~the spectral swaps induced by collective oscillations)
%\item improved pointing accuracy for the $\nu_e$ elastic scattering signal, improving the current day $\sim$3$^\circ$ resolution of SK to the level of 1$^\circ$ or better. The key is the virtually complete subtraction of IBDs from the highly directional ES event sample that is facilitated by a high-efficiency neutron capture tag \cite{Tomas:2003xn}; SK+Gd can expect a similar improvement but offers a neutron tagging efficiency of only $\sim$70\,\%
%\item the chance to glimpse the initial $\nu_e$ neutronization burst \cite{Kachelriess:2004ds}: however, as $\cal O$(10) events are expected for a SN at 10\,kpc, detailed information can only be expected for a relatively nearby Supernova
%\item 
%\item relative to DUNE, the co-detection of $\nu_e$ and $\bar\nu_e$ signals in the very same location, allowing a direct comparison 
%\end{itemize}

\subsection{Diffuse supernova neutrino background}

%\paragraph{Motivation}

The Diffuse Supernova Neutrino Background (DSNB) consists of neutrinos emitted by all core-collapse supernovae (SNe) throughout the Universe \cite{Ando:2004hc,Beacom:2010kk}. Travelling over vast distances and red-shifted by cosmic expansion, these neutrinos constitute a faint isotropic background flux. The discovery and subsequent spectroscopy of the DSNB will provide unique information on the redshift-dependent SN rate, the relative frequency of neutron star and black hole formation and the equation of state of the emerging neutron stars (e.g.~\cite{KresseMSc}).

The primary detection reaction for the $\bar\nu_e$ component of the DSNB flux is the inverse beta decay (IBD). With an expected event rate of 0.1 per year per kiloton of detector material, overwhelming backgrounds have to be faced. However, with SK-Gd and JUNO there are now two contenders with a realistic chance to obtain a first ($3\sigma$) evidence of the DSNB within the next 5$-$10 years \cite{Beacom:2003nk,An:2015jdp}.

\theia may play a pivotal role in the discovery and exploration of the DSNB signal. With a target mass several times the size of SK or JUNO, \theia-100 will obtain a $\sim$5$\sigma$ discovery of the DSNB in less than 1 year of data taking and reach ${\cal O}(10^2)$ DSNB events within $\sim$5 years. Even the smaller \theia-25 will profit considerably from the dual detection of Cherenkov and scintillation signals that offer a background discrimination capability unparalleled by Gd-doped water or pure organic scintillator: For instance, a signal efficiency of 95\,\% can be maintained while reducing the most crucial background from atmospheric neutrino NC interactions to a residual $\sim$1.7\,\% \cite{Sawatzki:2019}. 

\subsubsection{DSNB signal and background levels}

\begin{figure*}
\centering
\includegraphics[width=0.9\textwidth]{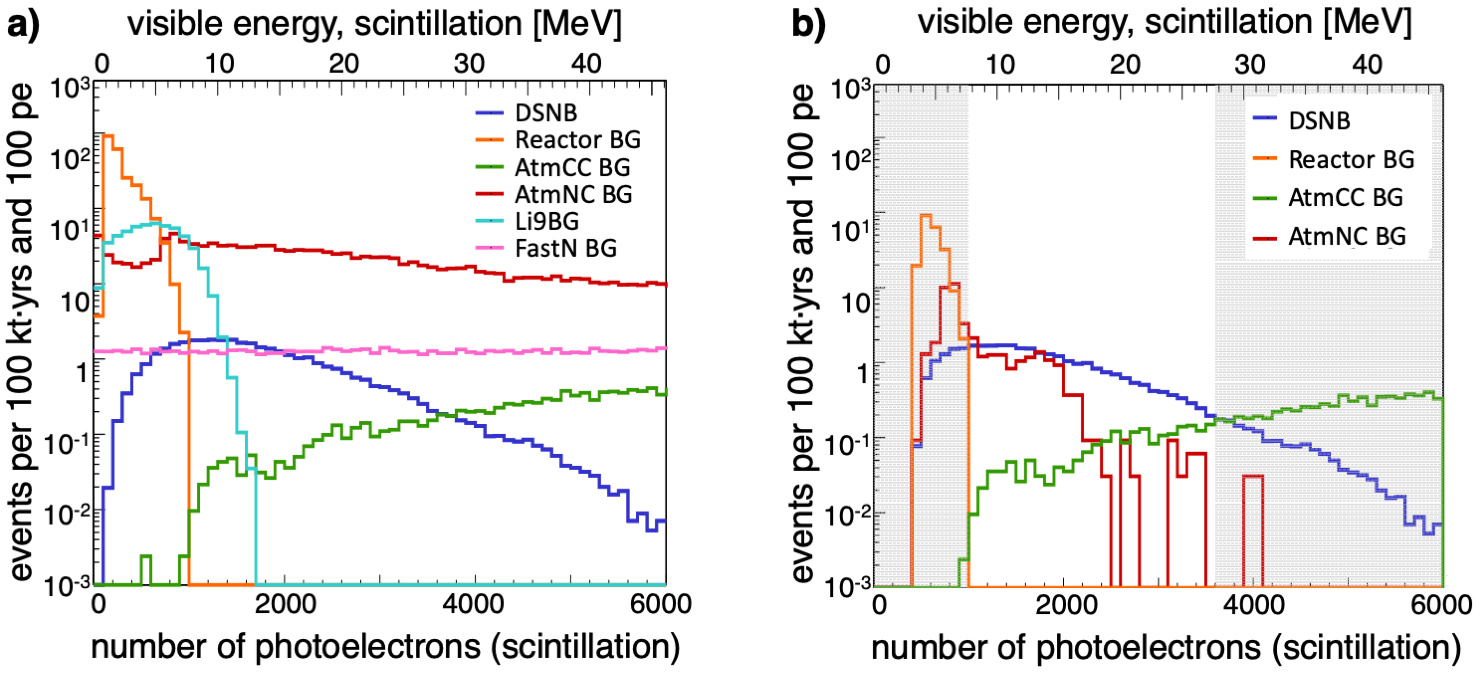}
\caption{The visible energy spectrum from scintillation expected for the DSNB signal and its ample backgrounds: While panel (a) includes reactor neutrinos, cosmogenic Li-9, fast neutrons as well as atmospheric neutrino CC and NC interaction rates before application of discrimination techniques, panel (b) illustrates that $-$ within the observational window from $\sim$8 to 30\,MeV $-$ all backgrounds can be greatly reduced by ring counting, Cherenkov/scintillation ratio and delayed decay cuts to obtain a signal-to-background ratio $>$1.}
\label{fig:dsnb_spectrum}
\end{figure*}

For IBD events, the prompt energy of the positron signal translates almost directly to the incident neutrino energy, while the delayed neutron capture provides a fast coincidence tag to reduce the ample single-event backgrounds. Fig.~\ref{fig:dsnb_spectrum}a) depicts the visible energy spectrum (scintillation only) for the prompt positrons. The chosen detector configuration corresponds to a WbLS of 10\,\% organic fraction and 70\,\% photocoverage, resulting in a photoelectron yield of 120 (80) for scintillation (Cherenkov) component. The DSNB model is based on work by the Garching group \cite{KresseMSc}. 

Fig.~\ref{fig:dsnb_spectrum}a) shows as well the relevant background spectra: IBDs from {\it reactor and atmospheric} $\bar\nu_e$'s constitute an indistinguishable background that overwhelms the signal at low and high energies, effectively limiting the detection to an energy window of 10--30\,MeV.

Even within this window, several further background sources contribute, all of cosmogenic origin: {\it cosmogenic $\beta n$-emitters}, primarily $^9$Li, are created by muon spallation on the oxygen (and carbon) nuclei of the target; {\it fast neutrons} are induced by muons in the rock surrounding the detector and are able to enter the detector unnoticed. The combination of a prompt signal created by elastic scattering off protons and the subsequent neutron capture may mimic the IBD signature. Finally,  {\it NC reactions of atmospheric neutrinos (atm-NC)} resemble the IBD coincidence in case a prompt signal is generated due to the recoils and possible de-excitation of the fragments of the target nucleus and a delayed signal in case a neutron is released from the nuclear break-up. First recognized by the KamLAND experiment for its special importance in organic scintillators \cite{Collaboration:2011jza}, it dominates the DSNB signal by more than one order of magnitude. For scintillator detectors like JUNO, the quenched nuclear signal constitutes a major challenge to be overcome by pulse shape discrimination \cite{An:2015jdp}. SK-Gd (or HK-Gd) will be much less affected but features this background as well below $\sim$16\,MeV \cite{Moller:2018kpn}.

\subsubsection{Background discrimination in WbLS} 

\begin{figure}[htp!]
\centering
\includegraphics[width=0.49\textwidth]{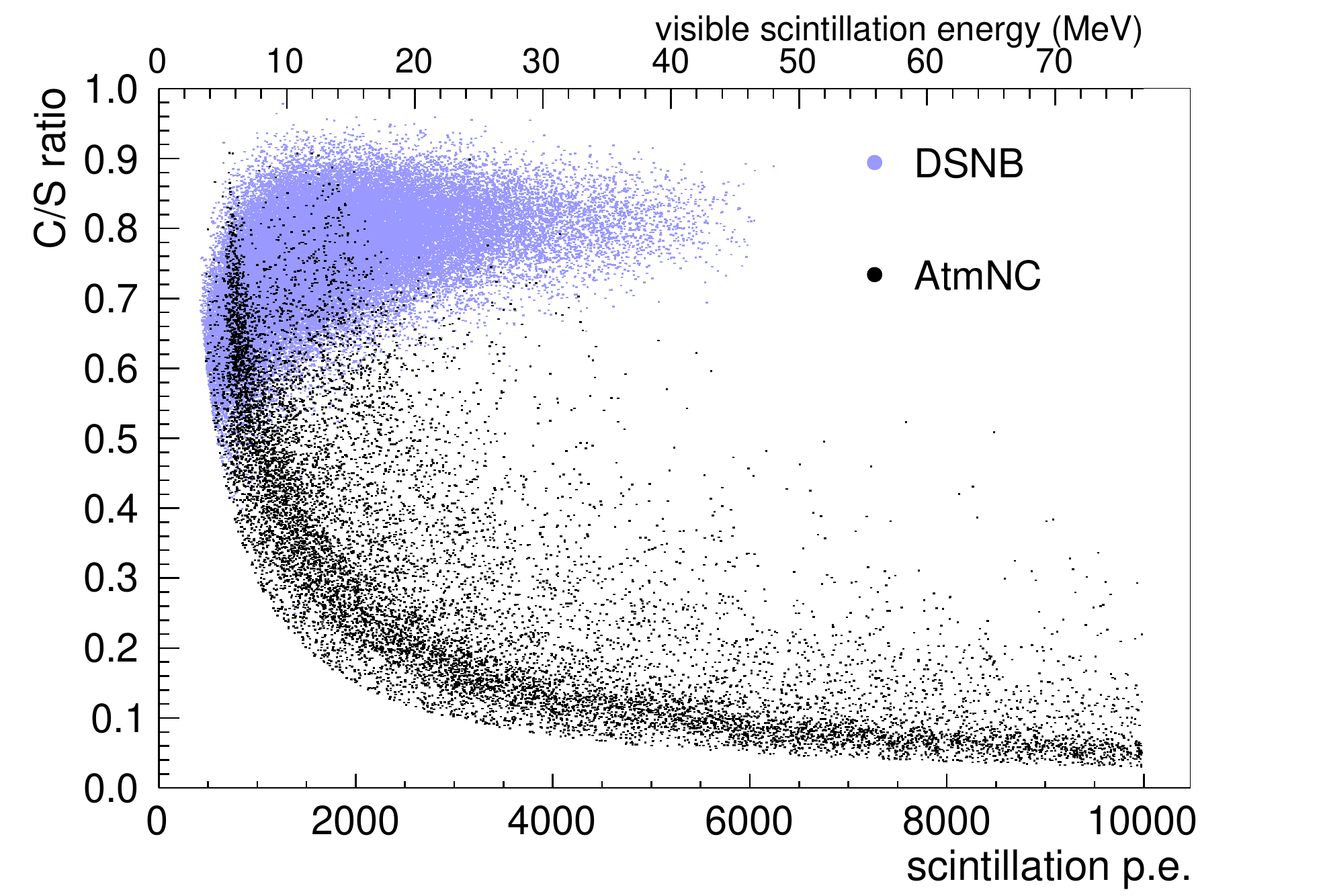}
\includegraphics[angle=270,width=0.49\textwidth]{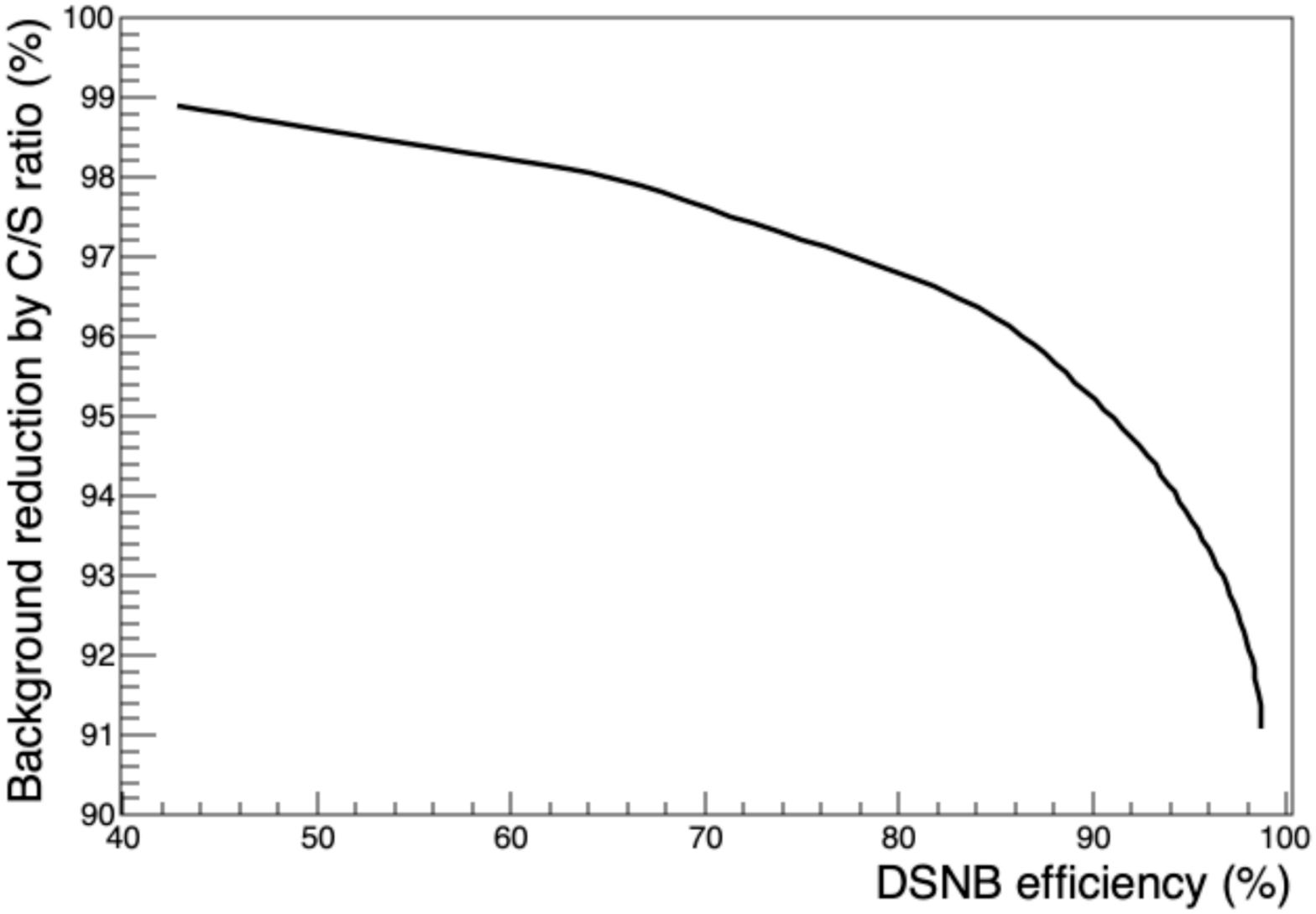}
\caption{{\it Top panel:} The C/S ratio offers a powerful tool for the discrimination of positron-like DSNB ({\it blue}) and hadronic prompt events from atm-NC reactions ({\it black}). While most background events feature no Cherenkov light and, as such, a C/S ratio of 0, some final-state $\gamma$ rays result in a curved band of atm-NC events that leaks slightly into the signal region. {\it Bottom panel:} atm-NC background reduction as function of the DSNB signal efficiency.}
\label{fig:dsnb_csratio}
\end{figure}

The main virtue of \theia lies with the excellent background discrimination capabilities of the WbLS. In the context of the DSNB search, the following have been investigated:
\begin{itemize}
\item A {\it fiducial volume cut} to reject surface background events, especially fast neutrons created by muons in the surrounding rock. In current configuration, \theia-25 will feature 20\,kt of fiducial mass, \theia-100 about 80\,kt.

\item {\it Distance cuts} relative to muon tracks traversing the fiducial volume may be used to veto decays of the $\beta n$-emitter $^9$Li while reducing live exposure by about 1\,\%.

\item {\it Ring counting:} The reconstruction of Cherenkov rings from individual final-state particles provides a  handle to discriminate one-ring positron events from multi-ring atm-NC events. From simulation, we find that about 40\,\% of the latter background events can be discriminated based on this feature\footnote{For this, we require that the subleading ring features at least 20\,\% of the overall Cherenkov photons.}, introducing a negligible loss in signal efficiency.

\item {\it Delayed decays:} In almost 50\,\% of the relevant atm-NC events, the residual nucleus is the $\beta^+$-emitting isotope $^{15}$O with a lifetime of 2.2\,min. Given the low energy threshold, \theia will be able to tag the delayed decay in order to reject half of the atm-NC background without any noticable loss in signal efficiency.

\item {\it Cherenkov-to-scintillation ratio:} Uniquely, a WbLS detector offers the possibility to discriminate events by the magnitude  of the Cherenkov signal accompanying the scintillation light. While $e^\pm$ signals feature a high Cherenkov-to-scintillation (C/S) ratio, that of non-relativistic nuclear recoils is practically zero. As demonstrated by the left panel of Fig.~\ref{fig:dsnb_csratio} that displays the C/S ratios of signal and atm-NC events as a function of visible energy, this parameter provides a powerful tool to reject atm-NC (but also fast neutron) background events. Residual atm-BG events with non-zero C/S values are due to the emission of $\gamma$-rays in nuclear de-excitations of oxygen nuclei. The right panel of Fig.~\ref{fig:dsnb_csratio} shows the relation between signal efficiency and background reduction factor as a function of C/S threshold \cite{Sawatzki:2019}.
\end{itemize}

\subsubsection{Sensitivity to the DSNB signal} 

\begin{table*}[htp!]
\begin{center}
\caption{Rates of DSNB signal and backgrounds within the observation window (8-30\,MeV) for a live exposure of 100\,kt$\cdot$years. While the first column represents the rates before cuts, the following columns apply delayed decay, C/S ratio and ring-counting cuts. The cited fast neutron rate assumes a 2.5\,m fiducial volume cut or presence of corresponding active shielding surrounding the target volume.}
\label{tab:dsnb_rates}
\begin{tabular}{lccccc}
\hline\noalign{\smallskip}
Spectral contribution			& after FV cut & Li veto & delayed decays & single-ring & C/S cut \\
\noalign{\smallskip}\hline\noalign{\smallskip}
DSNB signal				& 25.9	& 25.7	& 25.7	& 24.5	& 24.5\\
Reactor neutrinos			& $-$	& $-$ 	& $-$	& $-$	& $-$ \\
Atmospheric CC			& 2.0		& 2.0		& 2.0		& 1.9		& 1.9 \\
\hline
Atmospheric NC			& 689	& 682 	& 394	& 25.9	& 13.6 \\
$\beta n$-emitters ($^9$Li)	& 55	& $-$	& $-$	& $-$	& $-$ \\
Fast neutrons				& 0.8 	& 0.8  	& 0.8	& $-$	& $-$ \\
\hline
Signal-to-background		& 0.03	& 0.04	& 0.07	& 0.9 	& 1.6  \\
\hline
\noalign{\smallskip}\hline
\end{tabular}
\end{center}
\end{table*}

Table~\ref{tab:dsnb_rates} illustrates the impact of the aforementioned discrimination techniques on the signal and background rates within the observation window. While all background components including the atm-NC events are greatly reduced, the DSNB signal acceptance is hardly affected. The corresponding energy spectra are shown in Fig.~\ref{fig:dsnb_spectrum}b), demonstrating a clear dominance of the signal over the entire energy window.

\theia-25 will require about 6\,years of data taking to achieve a $5\sigma$ discovery of the DSNB signal (assuming standard predictions for flux and spectral energy) \cite{Sawatzki:2019}. Combined with SK+Gd and JUNO, the three detectors will acquire about 5-10 DSNB events per year (with $\sim$40\% of statistics from \theia), so that a spectroscopic analysis of the DSNB based on ${\cal O}(10^2)$ events will become feasible over 10-20 years. At the same time, the C/S signatures of the large atm-NC event sample recorded in WbLS will enable an in-depth study of this most relevant background and will help to reduce the corresponding systematic uncertainties as well for SK-Gd and especially JUNO. 

\theia-100 will take a clear lead in the exploration of the DSNB signal. DSNB discovery is expected in a bit more than 1 year of data taking, and first spectral analyses only a few years later: in the long run, spectroscopy will provide access to the astrophysics of SNe, ranging from the equation of state of neutron stars to the fraction of dark SNe resulting from black-hole collapses \cite{Sawatzki:2019}.

%\subsection{Applied antineutrino physics} This is not very descriptive

\subsection{Geological and reactor neutrino measurements}
%\paragraph{Motivation}
%BRIEF intro to physics motivation, and status of the field: our major competitors
%\paragraph{XX with \theia}
%Electron antineutrinos stream freely from rapidly decaying fission products within nuclear reactors and from long-lived radioactive isotopes and their daughters within Earth \cite{agm15}. Important information about nuclear reactors, Earth, and the properties of neutrinos themselves comes from measuring the rate and energy spectrum of the interactions of these $1-10$ MeV antineutrinos. Detecting antineutrinos from nuclear reactors at short \cite{nucifer15,songs07} and long \cite{nudar13,snif10} distances monitors the operation and identifies the location and power of the reactor with applications for nuclear non-proliferation \cite{adam10}. Such detections also provide fundamental understanding of neutrinos \cite{reines53,reines76,jgl08}. Detecting antineutrinos from the nuclear cascades of thorium-232 and uranium-238 within Earth \cite{kl05} estimates terrestrial radiogenic heating \cite{gando13,agostini15}, leading to a more complete understanding of the composition, structure, and thermal evolution of our planet \cite{dye_etal15}. 

Global antineutrinos emerge from nuclear beta-minus decays, which produce a characteristic energy spectrum in the few-MeV range for each parent isotope. While the mixture of isotopes decaying within a source uniquely determines the energy spectrum of the emitted antineutrinos, neutrino oscillations distort the spectrum of the detected antineutrinos in a pattern determined by the distance from the source. Spectral distortion is pronounced for point-like nuclear power reactors and subtle for diffuse sources within Earth. 

\begin{figure}
    \centering
    \includegraphics[width=0.48\textwidth]{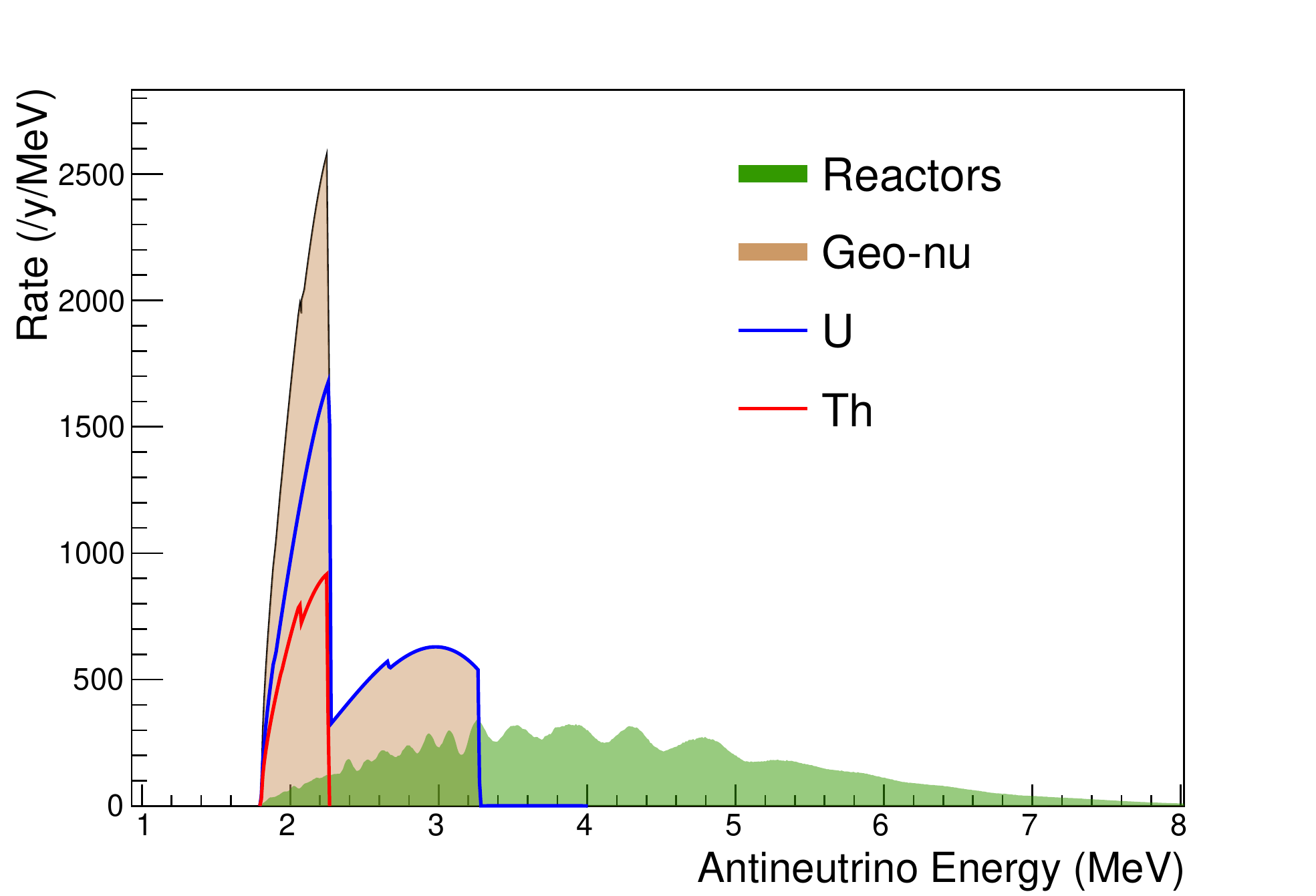}
    \caption{The detected energy spectrum of the predicted rate of antineutrinos from nuclear power reactors and Earth, assuming a $50$ kT water target, located at SURF.}
    \label{fig:theia_reac_geo}
\end{figure}

The rate and energy spectrum of global antineutrino interactions varies dramatically with surface location \cite{agm15}. The following discussion assumes antineutrino interactions by IBD on hydrogen ($E_\nu > 1.8$ MeV) in a water target located at SURF and with a detection efficiency of $90$\%. The realized detection efficiency depends heavily on the actual light production (scintillator loading) and light collection (photosensitive area). The interaction rates given here can be scaled for the different phases of \theia. Fig.~\ref{fig:theia_reac_geo} shows the detected energy spectrum of the predicted rate of antineutrinos from the nuclear power reactors and Earth.

Observations of Earth antineutrinos, or geo-neutrinos, probe the quantities and distributions of terrestrial heat-producing elements uranium and thorium. The quantities of these elements gauge global radiogenic power, offering insights into the origin and thermal history of Earth \cite{dye12}. Spatial distributions reveal the initial partitioning and subsequent transport of these trace elements between metallic core, silicate mantle, and crust types \cite{dye_etal15}. Ongoing observations at underground sites in Japan \cite{kam13} and Italy \cite{bx15} record the energies, but not the directions, of geo-neutrinos from uranium and thorium. Without directions pointing back to source regions, disentangling the signals from various reservoirs requires resolution of differing rates or energy spectra at separate sites. Due to limited statistics and perhaps insufficient geological contrast, the published observations at Japan and Italy do not yet measure distinct rates or energy spectra \cite{dye16}. The large exposure possible with \theia enables these measurements, representing an opportunity to significantly advance observational neutrino geophysics.

The predicted rate of geo-neutrino interactions per kT-year at SURF is $26.5$ ($20.7$ U and $5.8$ Th), which corresponds to a flux of $4.90 \pm 0.13 \times 10^6$ cm$^{-2}$ s$^{-1}$, assuming perfect background suppression, Th/U $=3.9$, and statistical uncertainty only. Systematic uncertainty from incomplete knowledge of the distributions of uranium and thorium abundances is much larger and asymmetric at about the 25\% level. For comparison, reported observed rates of geo-neutrino events from KamLAND and Borexino correspond to $3.4 \pm 0.8 \times 10^6$ cm$^{-2}$ s$^{-1}$ \cite{kam13} and $5.0 \pm 1.3 \times 10^6$ cm$^{-2}$ s$^{-1}$ \cite{bx15}, respectively. While consistent with the Borexino measurement, a measurement at the predicted SURF rate would be almost $2 \sigma$ greater than the KamLAND measurement after an exposure of $50$ kT-y. This would provide the first evidence for surface variation of the geo-neutrino flux. With thousands of geo-neutrino events \theia would precisely measure the uranium and thorium components of the energy spectrum with the potential to test models of differential partitioning and transport of these trace elements between silicate mantle and crust types.

The expected rate of reactor antineutrino interactions at SURF is $\sim 20$ per kT-year. This interaction rate allows \theia to demonstrate techniques for remote reactor discovery \cite{jocher13}, including 
measuring range and direction at distances greater than $1000$ km. A detailed study of the antineutrino capabilities of \theia is in preparation for a separate publication.
% If a reference is needed- 
%https://indico.fnal.gov/event/18662/session/5/contribution/12/material/slides/3.pdf
%This summarizes the potential antineutrino capabilities
In summary, with sufficient detection efficiency and exposure \theia located at SURF has the potential to make significant advances in applied antineutrino physics. \theia could provide the first evidence for surface variation of the geo-neutrino flux and make a precise measurement of the thorium and uranium components of the energy spectrum. \theia could demonstrate basic techniques for remote discovery of nuclear reactors by making antineutrino measurements of range and direction.
%What we bring to the table - pros of \theia design \newline
%Sensitivity estimates with baseline design (one of \theia i--iii)
%\paragraph{Detector Requirements}
%A summary of the impact of different detector choices i.e. what happens if we stray from the relevant baseline

\subsection{Neutrinoless double beta decay}

The \theia search for neutrinoless double beta decay (NLDBD) aims for
sensitivity to the non-degenerate normal hierarchy parameter space
within the canonical framework of light Majorana neutrino exchange and
three-neutrino mixing, at the level of $m_{\beta\beta}\sim5$ meV.
This is achieved through the loading of a very large
mass of a NLDBD candidate isotope into an ultra-pure LS
target, together with coincidence and topological particle identification
techniques.

\subsubsection{Detector configuration}

A search for NLDBD at \theia would involve deploying a balloon or thin vessel containing LS, loaded with the isotope of interest, within the larger WbLS detector. The main advantages of this technique are the higher light yield of pure LS compared to WbLS, the higher radiopurity, and  reduced contamination from external backgrounds. 
The different densities of WbLS and isotopically loaded LS will require
a thoughtful engineering design of the inner containment. The SNO+ experiment
uses a 5~cm think acrylic vessel with an associated rope net to suspend the target LS in a water-filled cavity.   While this approach is
a possibility for the \theia NLDBD phase, a lighter-weight balloon is preferable
due to the simplicity of its deployment and smaller optical impact and
radiological load. Nevertheless, we assume here that, if necessary, the inner
containment could follow the SNO+ design.

%On the other hand, the loading of a balloon or vessel containing a liquid at lower density with respect to the surrounding medium will represent a technical challenge, requiring an anchoring system and selecting the proper thickness to minimise the mechanical stresses. A similar challenge has recently been solved by the SNO+ experiment, where the acrylic vessel containing the liquid scintillator at a density of 0.86 g/cm^{3} is suspended in a cavity full of water [124].
In the following study it is assumed that the double-beta decay isotope under investigation is loaded into a balloon of 8-m radius, filled with ultra-pure LS (LAB + 2 g/l PPO for a density of 0.86 g/cm$^3$). The volume outside the balloon is filled with a 10\%-WbLS (10\% LAB-PPO and 90\% water). We investigate two major loading cases: 3\% enriched Xenon (89.5\% in $^{136}$Xe) and 5\% natural Tellurium (34.1\% in $^{130}$Te). The results presented here are obtaining simulating events in a 20-m fiducial radius and a 40-m height cylinder (50 kT fiducial mass), with a PMT coverage of 90\%. However, due to the long attenuation length in water, this configuration is expected to be equivalent as having the 8-m radius balloon in the Theia-100 detector.

%While the default detector sizes considered in this paper are 25\,kT and 100\,kT, the present sensitivity studies have been obtained with a detector modeled as a cylinder with a 20-m fiducial radius and a 40-m height, for a total fiducial mass of 50\,kT, and with a PMT coverage of 90\%. The double-beta decay isotope under investigation is loaded into a nylon balloon of 435-$\mu$m thickness and 8-m radius, filled with ultra-pure liquid scintillator (LAB + 2\,g/l PPO for a density of 0.86 g/cm$^3$). The volume outside the balloon is filled with a WbLS (10\% LAB-PPO and 90\% water). We investigate two major loading cases: 3\% enriched Xenon (89.5\% in $^{136}$Xe) and 5\% natural Tellurium (34.1\% in $^{130}$Te). Deploying the 8-m radius balloon in the \theia-100 detector configuration, is not expected to change the results obtained in this study. 

The optical properties of the unloaded LAB+PPO cocktail have been measured by the SNO+ collaboration, and are consistent with benchtop measurements.   Those of the WbLS are obtained by weighting contributions of the LAB+PPO and water. As a baseline, an average light yield of 1200 PMT hits per deposited MeV (Nhits/MeV) is assumed (corresponding to about 3\%/$\sqrt{E}$ energy resolution). This value includes the reduction in the light yield due to the addition of the isotope, estimated to be around 30\% at 5\%, or higher, loadings. For the specific case of Xe-loaded scintillator this  light yield is likely an underestimation, as the KamLAND-Zen experiment predicts a reduction of only 15\% at  3\% loading, which at 90\% coverage would correspond to 1530 Nhits/MeV \cite{KZ-2011}. Fig.~\ref{fig:scale}(top panel) shows the impact of  light yield on the \theia sensitivity. 
Simulations are obtained with the Geant4-based RAT-PAC software package~\cite{seib14}; when available radioactive decays are simulated using the Decay0 code \cite{decay0}, otherwise the decay mode is input in RAT-PAC. Reconstructed energy is approximated by assuming the Poisson limit of
photon counting: the true deposited energy, accounting for quenching, is smeared  by a Gaussian resolution function corresponding to the light
yield.

\subsubsection{Backgrounds}

\begin{table*}
\centering
\caption{Dominant background sources expected for the NLDBD search in \theia. The assumed loading is 3\% for Xe, for a $^{136}$Xe mass of 49.5\,t, and 5\% for Te, for a $^{130}$Te mass of  31.4\,t. The events in the ROI/yr are given for a fiducial volume of 7\,m and an asymmetric energy range around the Q-value of the reaction (\textit{see text}). A rejection factor of 92.5\% is applied to $^{10}$C, of 99.9\% to $^{214}$Bi, of 50\% to the balloon backgrounds, and of 50\% to the $^8$B solar neutrinos.}
\label{tab::bckg}
\scalebox{0.9}{
\begin{tabular}{lcccc}
\hline\noalign{\smallskip}
{\bf Source} & {\bf Target level} & {\bf Expected }   & \multicolumn{2}{c}{\bf Events/ROI$\cdot$y} \\
& & {\bf events/y} & 5\% $^{nat}$Te  & 3\% $^{enr}$Xe  \\
\noalign{\smallskip}\hline\noalign{\smallskip}
 $^{10}$C  & & 500 &  2.5       & 2.5\\
$^{8}$B neutrinos (flux from \cite{SNO_3ph}) &  & 2950   & 13.8      & 13.8  \\
$^{130}$I (Te target)  &  & 155  (30 from $^{8}$B) & 8.3 & - \\
$^{136}$Cs ($^\mathrm{enr}$Xe target) & & 478 (68 from $^{8}$B) & -&0.06\\
2$\nu\beta\beta$ (Te, T$_{1/2}$ from \cite{Cuore017}) &   & 1.2$\times$10$^{8}$  & 8.0       &  -\\  
2$\nu\beta\beta$ ($^\mathrm{enr}$Xe, T$_{1/2}$ from \cite{gando16,exo14}) &  &  7.1$\times$10$^{7}$  &   -    & 3.8  \\  
Liquid scintillator & $^{214}$Bi: $10^{-17}$ g$_{U}$/g  & 7300 & 0.4        & 0.4\\
   & $^{208}$Tl: $10^{-17}$ g$_{Th}$/g  & 870  & -& -\\
Balloon  & $^{214}$Bi: $<10^{-12}$ g$_{U}$/g & $<$2$\times$10$^{5}$   & 3.0       & 3.4   \\
 & $^{208}$Tl: $<10^{-12}$ g$_{Th}$/g   & $<$3$\times$10$^{4}$  & 0.03       & 0.02  \\
\noalign{\smallskip}\hline
\end{tabular}}
\end{table*}

The main sources of background included in the this analysis are summarized in Table \ref{tab::bckg} and described below:

\paragraph{Double Beta Decay} Irreducible background from the $2\nu\beta\beta$ decay of $^{130}$Te or $^{136}$Xe. Due to the steeply-falling
spectrum, the number of events  in the region of interest (ROI) depends strongly on the energy resolution.
\paragraph{Cosmogenic Production} These backgrounds are due to activation of nuclei by muons (during data taking) or protons and neutrons (during material production and handling at Earth's surface). The production rates of various radionuclides by cosmogenic neutron and proton spallation reactions in Xe and Te have been investigated in \cite{mei09,baudis15,zhang16,norm05,bard97,wang15,lozza15}. Among the most important nuclides produced there  are $^{60}$Co ($Q=2.8$ MeV, $T_{1/2}=5.27$ y) and $^{110m}$Ag ($Q=3.1$ MeV, $T_{1/2}=250$ d). Mitigation of these background sources requires minimal exposure at sea level, a deep underground cool-down period, chemical purification processes \cite{snop16}, and, to limit  in-situ production during data taking, the use of a water shield. In these studies it is assumed that proper measures are taken to handle the target material, reducing the background to a negligible level.  The most dangerous nuclide for the NLDBD study from in-situ muon induced events is $^{10}$C ($Q=3.65$ MeV, $T_{1/2}=19.3$ s), produced by muon interactions with the carbon atoms of the liquid scintillator. In this study the detector is assumed to be located in the Homestake mine, at a depth of 4300\,m.w.e. The estimated event rate is about 300 events/kt/yr \cite{hagn00} for a muon flux of $4.4\times10^{-9}$ cm$^{-2}$ s$^{-1}$ and an average muon energy of 293\,GeV \cite{mei06}. A reduction of 92.5\% of the $^{10}$C background has been demonstrated by Borexino \cite{bellini13} using a three-fold coincidence technique\cite{galb05}. A machine learning approach, as described in Sec.~\ref{sec:Reconstruction}, could be used in addition to the three-fold-coincidence method to further improve the rejection of $^{10}$C events.
 In the case of Xe-loaded scintillator, another potential background from muon-induced events is $^{137}$Xe. Neutrons produced by cosmic muons in scintillator, once thermalized, could be captured by $^{136}$Xe atoms, emitting a high energy gamma of about 4~MeV and $^{137}$Xe (T$_{1/2}$ = 3.82 min, Q = 4173 $\pm$ 7 keV). At 4300 m.w.e., 165 $^{137}$Xe events/yr are expected (equivalent to 3.3 events/yr/ton of $^{136}$Xe). This number is 30 (18) times smaller than the number of events expected in JUNO (KamLAND-Zen)~\cite{Zhao:2016brs}. Using a rejection method similar to the one applied for $^{10}$C, the overall contribution in the ROI and FV should be less than one event/yr.

\begin{figure}[!h]
\centering
 \includegraphics[width=0.47\textwidth]{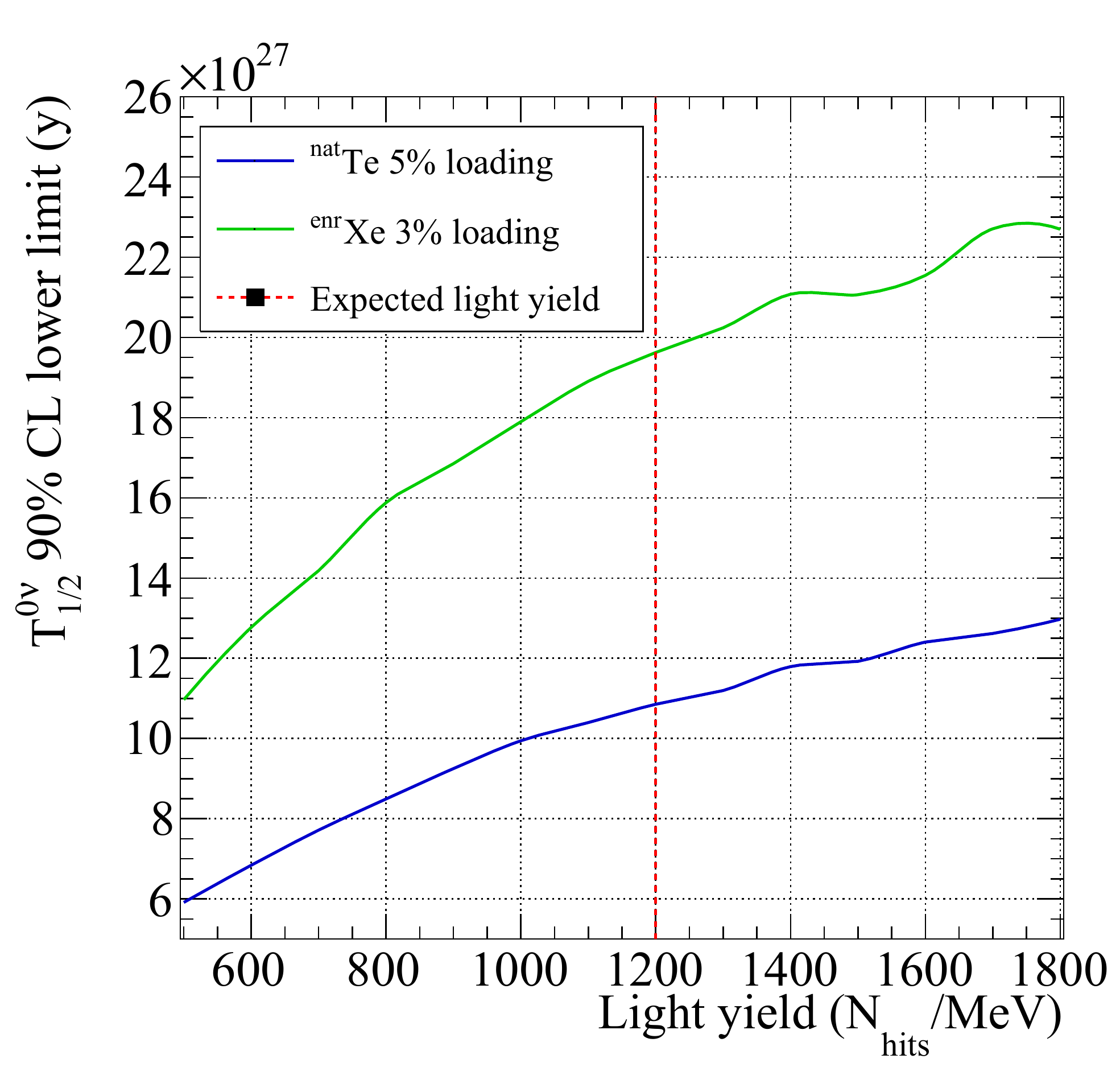}
 \includegraphics[width=0.49\textwidth]{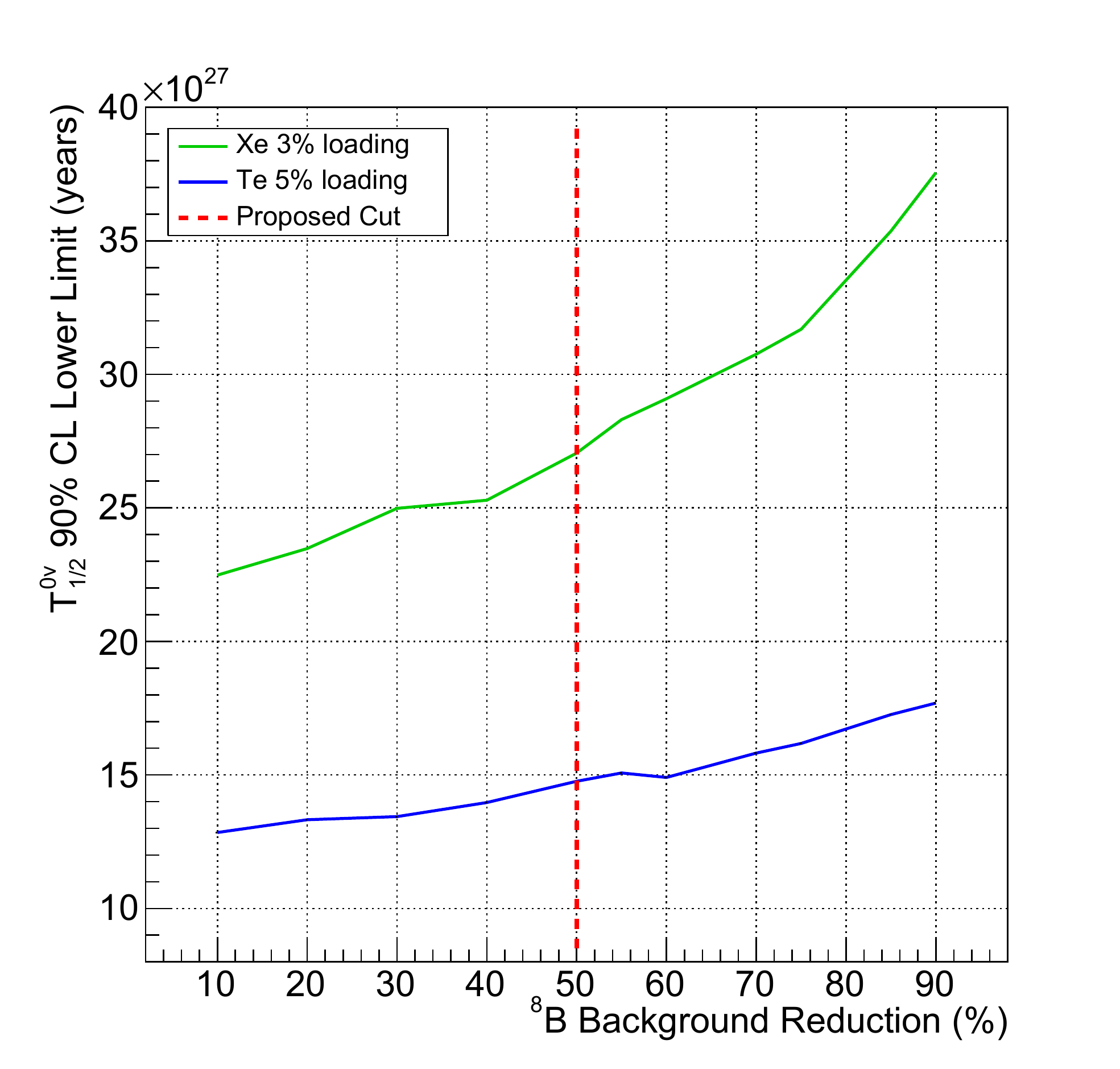}
\caption{Half life sensitivity as a function of key experimental parameters; (top panel) light yield and (bottom panel) $^8$B solar neutrino reduction. The vertical dashed red lines show the values used in the analysis. For the same detector optical properties, a variation in the light yield corresponds to a scaling on the PMT coverage.The plot showing the $^{8}$B reduction as a function of the light yield does not include the signal efficiency of 75\%. }
 \label{fig:scale}
\end{figure}

\paragraph{Solar Neutrinos} $^{8}$B solar neutrino elastic scattering in the target material results in a background that is approximately flat across the NLDBD energy ROI, but that can be rejected using reconstructed event direction. Fig.~\ref{fig:scale}(lower panel) shows how the sensitivity scales with this solar neutrino rejection efficiency. A rejection of at least 50\% is necessary in order to reach the sensitivity goal. Monte Carlo simulations show that, for 2.5\,MeV electrons in water and about 50\% coverage, about 80\% of the $^{8}$B events can be rejected while keeping 75\% of the NLDBD signal (assumed to be isotropic). In the case of \theia, the confounding effect due to the liquid scintillator can be compensated by use of high QE PMTs, together with high coverage. Other options might include the use of slow scintillators. In \cite{elagin19} it is shown that more than 50\% rejection in $^{8}$B can be achieved for an LAB target, retaining more than 70\% of the signal. In the following sensitivity calculations it is assumed a $^{8}$B neutrino rejection of 50\% with 75\% signal efficiency.
Another background induced by solar neutrinos (mainly $^{8}$B and $^{7}$Be) are high Q-value nuclides produced by charged current interaction with $^{130}$Te ($^{130}$I) and $^{136}$Xe ($^{136}$Cs) \cite{eijiri14,eijiri17}. Due to their long half-life, a tagging technique based on a delayed coincidence is expected to have a small efficiency. However, a method as described in Sec.~\ref{sec:Reconstruction} might help in separating multi-gamma events, such as $^{136}$Cs and $^{130}$I decays, from points like events as the neutrinoless double beta decay ones. 
\paragraph{Internal Contamination} $^{214}$Bi ($Q=3.27$ MeV, U-chain) and $^{208}$Tl ($Q=5$ MeV, Th-chain) decays can fall in the NLDBD ROI. The targeted scintillator cocktail purity for the \theia experiment is 10$^{-17}$ g/g in both U and Th. Liquid scintillator purities better than  10$^{-18}$ g/g in U and Th have been obtained in  Phase-II of Borexino \cite{bxo16}, while KamLAND-Zen has reached a cleanliness of the order of  10$^{-16}$ g/g U for  3\% Xe loaded LS \cite{gando13}. The target purity is considered achievable by improving target material purification techniques, i.e the purity grade of the chemicals used to process the tellurium. Delayed coincidence techniques can further reduce the number of $^{214}$Bi decays falling in the ROI. A rejection better than 99.95\% for  $^{214}$BiPo in the ROI has been shown by the KamLAND-Zen experiment \cite{KD-Zen}, while Monte Carlo studies for the SNO+ experiment show that the rejection can be as high as 99.99\% \cite{snop16}. For the aimed target purity, it is required that the $^{214}$Bi is reduced by 99.9\%. Larger reduction factors have a minimum effect on the overall sensitivity.
\paragraph{External Sources} Decays from U- and Th-chain impurities present in the balloon material, the external WbLS, the
shielding water, and in the PMTs also contribute to the background. External background events can be reduced using a fiducial volume cut, and PMT hit-time information. In the following study a rejection factor of 50\% on the top of the fiducial volume is assumed. %However, at the assumed balloon purity, a smaller reduction factor has a small impact on the sensitivity.
%\end{description}

\subsubsection{NLDBD sensitivity: counting analysis}\label{sec::sensitivity}

To estimate the sensitivity, a single-bin counting analysis is employed. Since several backgrounds do not scale with isotope mass (e.g. solar neutrinos
and external $\gamma$ backgrounds), we use the Monte Carlo to evaluate the background expectation, establish a confidence region using the Feldman-Cousins frequentist approach, and derive an expected limit on the NLDBD half-life:
\begin{equation}
\label{eq:sens}
\widehat{T}_{1/2}^{0\nu\beta\beta}(\alpha) = 
\frac{N\cdot \epsilon \cdot t \cdot \ln 2}{\mathrm{FC}(n=b, b; \alpha)}
%\widehat{T}_{1/2}^{0\nu\beta\beta}(\alpha) = \left\langle
%\frac{n\cdot \epsilon \cdot t \cdot \ln 2}{\mathrm{FC}(N, b; \alpha)}
%\right\rangle_{n=\mathrm{Pois}(b)}
\end{equation}
where $N$ is the number of atoms of active NLDBD isotope, $\epsilon$ is the efficiency, $t$ the live time, and $b$ the expected background.
`FC' refers to a Feldman-Cousins interval at confidence level $\alpha$.

\begin{figure*}
\centering
 \includegraphics[width=0.48\textwidth]{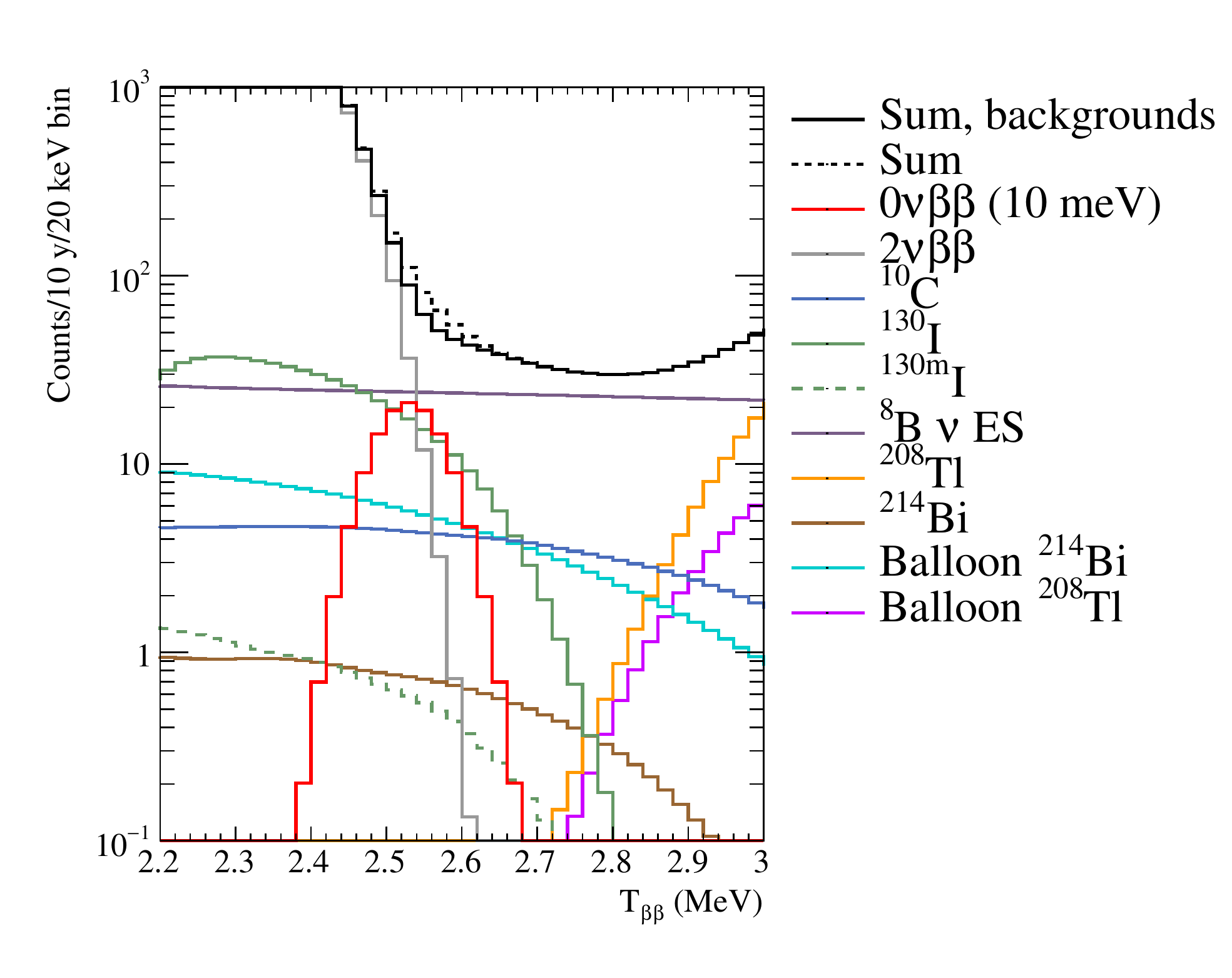}
 \includegraphics[width=0.48\textwidth]{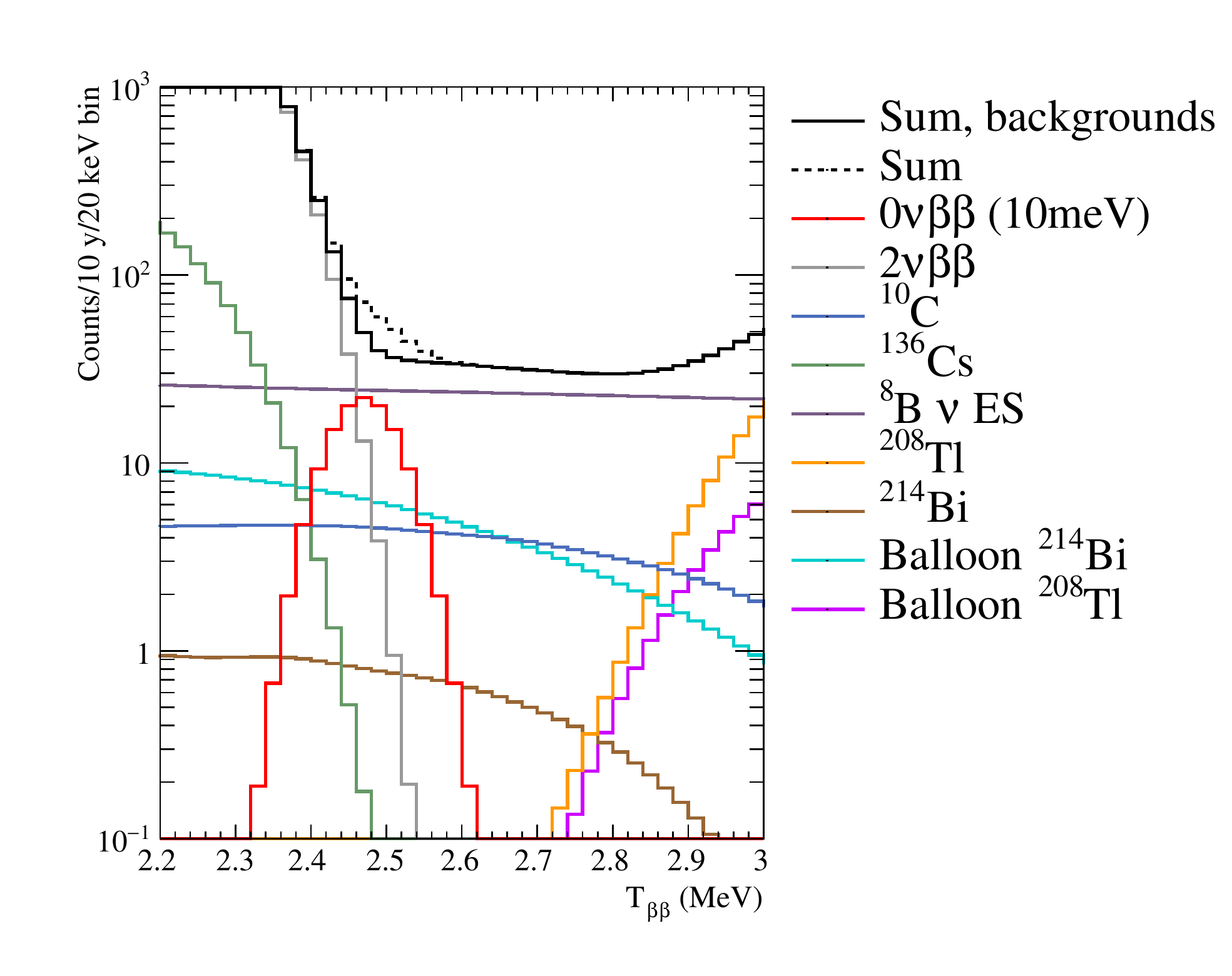}
\caption{Energy spectra near the NLDBD endpoint for events within the 7-m fiducial volume and for 10 years data taking. A rejection factor of 92.5\% is assumed for $^{10}$C, of 99.9\% for $^{214}$Bi, of 50\% for the balloon backgrounds, and 50\% for the $^8$B solar neutrinos.  (Left panel) 5\% $^{nat}$Te loading and (Right panel) 3\% $^{enr}$Xe loading}
\label{fig:spectrum}
\end{figure*}

The expected event rates per year for a $^\mathrm{nat}$Te or $^\mathrm{enr}$Xe loaded \theia detector are given in Table \ref{tab::bckg}, for a fiducial volume radius cut of 7\,m (67\% acceptance) and an asymmetric energy region,  from $-\sigma/2 \to 2\sigma$, to maximize signal acceptance ($\epsilon=66.9$\%) while removing much of the steeply-falling two-neutrino DBD background. Fig.~\ref{fig:spectrum} shows the background spectra near the endpoint in the Te (Fig.~\ref{fig:spectrum}(left panel)) and Xe (Fig.~\ref{fig:spectrum}(right panel)) cases. A 75\% signal efficiency, following the 50\% reduction in the $^{8}$B solar neutrino events, is applied. 

The expected sensitivity (90\% CL) for 10 years of data taking, using phase space factors from \cite{2012PhRvC..85c4316K} and matrix element from \cite{Barea:2013wb} (g$_{A}$=1.269) is:
\begin{eqnarray*}
\mathrm{\bf Te:}~~
  T_{1/2}^{0\nu\beta\beta} > 1.1\times10^{28}~\mathrm{y},~
  m_{\beta\beta} < 6.3~\mathrm{meV}\\
  \mathrm{\bf Xe:}~~
  T_{1/2}^{0\nu\beta\beta} > 2.0\times10^{28}~\mathrm{y},~
  m_{\beta\beta} < 5.6~\mathrm{meV}
\end{eqnarray*}

It should be noted that for the case of Xenon, the use of a more realistic light yield of about 1500\,nhits/MeV, as obtained from  \cite{KZ-2011}, would increase the half-life limit to 2.1$\times 10^{28}$ years, corresponding to $m_{\beta\beta} < 5.4~\mathrm{meV}$. Unfortunately, the required mass of Xe to reach the normal hierarchy is about 10 times the world annual production, which makes the use of Xe likely impractical.

\begin{figure}
\centering
 \includegraphics[width=0.48\textwidth]{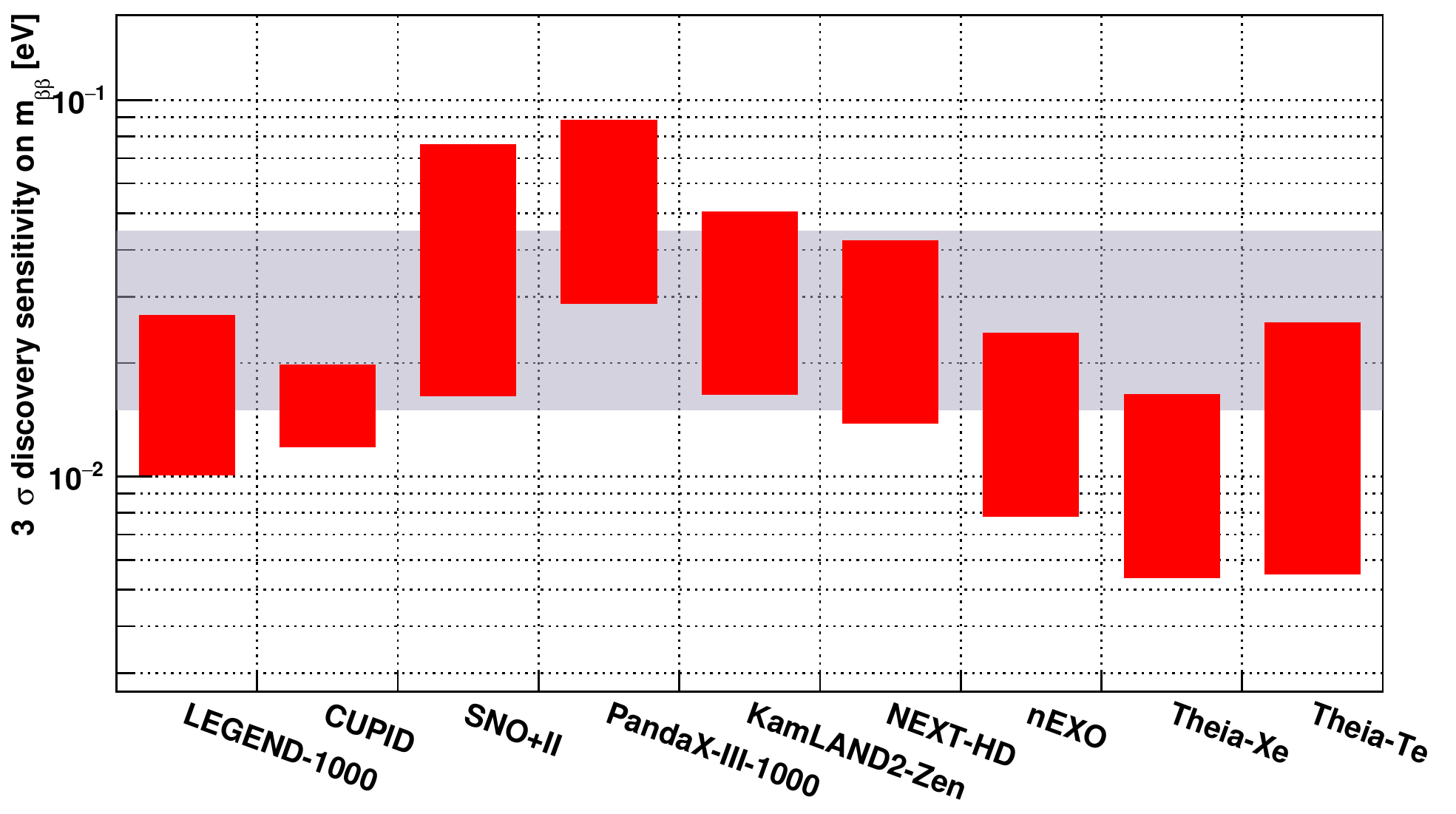}
 \caption{Discovery sensitivity (3$\sigma$) for  proposed future  experiments. The grey shaded region corresponds to the parameter region allowed in the Inverted Hierarchy of the neutrino mass. The red error bars show the $m_{\beta\beta}$ values such that an experiment can make at least a 3$\sigma$ discovery,
within the range of the nuclear matrix elements for a given isotope. The parameters of the other experiments are taken from Refs.~\cite{Agostini:2017jim,Gomez-Cadenas:2019sfa,Galan:2019ake,CUPIDInterestGroup:2019inu,Giuliani:2019uno}.}
\label{fig:dbdcomp}
\end{figure}

A comparison of this sensitivity to other experiments is shown in Fig.~\ref{fig:dbdcomp}.

\noindent
%{\bf TODO:} Comment on prohibitive Xe costs?

\subsubsection{Alternative isotopes}

A few alternative isotopes have been explored, which would be favorable in terms of annual abundance and costs: $^{100}$Mo, $^{82}$Se and $^{150}$Nd. For  these isotopes the main limiting factor is  leakage of the 2$\nu\beta\beta$ into the signal ROI, which is substantially higher than for Te  due to the shorter half-life of the corresponding decay mode. A loading of 2\% for Se and Nd, and of 3\% for Mo, has been chosen based on results of stability tests in table-top experiments, for which the cocktail seems to maintain good stability and optical properties. For Nd and Mo an enrichment factor of 90\% is additionally assumed, resulting in a T$_{1/2}$ limit of 3.6$\times 10^{27}$\,yr ($m_{\beta\beta}$ = 9.1\,meV) and 7.3$\times 10^{27}$\,yr ($m_{\beta\beta}$ = 7.9\,meV), for Nd and Mo respectively.\\
The enrichment option for Se is less promising due to the smaller G$_{0\nu} M^{2}_{0\nu}$ value, and the larger costs (although a larger world annual abundance than Xe is available). The limit in this case is T$_{1/2}$ = 1.6$\times 10^{27}$\,yr ($m_{\beta\beta}$ = 18\,meV).

\subsection{Nucleon decay}
%\paragraph{Introduction}

Many proposed extensions to the Standard Model predict the proton and the
neutron to decay through the introduction of new mediators, whether these be
entirely new gauge fields or supersymmetric partners. In all cases the
lifetime of protons and bounds neutrons must be very long, as modern detectors
have yet to see any sign of such decays. % (and our Universe appears to have an abundance of these particles). 
The subset of theories which assume that the
Standard Model is an effective theory, part of a larger gauge group that is
spontaneously broken at the GUT scale (examples being SUSY and non-SUSY SU(5)
and SO(10)), produce interactions at low-energy which are suppressed by the
energy of the GUT scale $\Lambda$. The effective Lagrangian becomes
\begin{equation}
\mathcal{L}_{eff} = \mathcal{L}(\phi_L) 
    + \sum_{k>4}\frac{C_i}{\Lambda^{k-4}}\mathcal{O}^{(k)}_i(\phi_L),
\end{equation}
where $C_i$ is a dimensionless coefficient and $\mathcal{O}_i^{(k)}$ are local
operators of mass dimension $k$ \cite{Kai2015}. Since $\Lambda$ is large,
operators of increasingly higher dimension are heavily suppressed, with the
lowest order terms (dimension 5) resulting in
\begin{equation}
p\rightarrow \bar{\nu}_iK^+,~ \bar{\nu}_i\pi^+,~ e^+K^0,~ \mu^+K^0,~
    e^+\pi^0,~ \mu^+\pi^0,~e^+\eta,~ \mu^+\eta;~i=e,~\mu,~\tau,
\end{equation}
where the dominant decay mode depends on the particular model of interest. More
exotic modes can also exist, but require a means to suppress the above modes
via some mechanism.  Extra dimensional theories for example can create
constraints on dimension 5 operators such that they are forbidden, allowing for
decays of higher order operators to dominate leading to decays such as $n
\rightarrow 3\nu$ \cite{Mohapatra_6D}. Next generation detectors will be able
to probe deeper into the phase space of such processes, possibly reaching the
sensitivity to measure such a process and give evidence for the type of physics
beyond the Standard Model. The \theia detector has the size and resolution to
contribute to this effort, and in certain modes, provide the dominant
experimental measurement. In the case of modes like $p\rightarrow e^{+}\pi^0$, the efficiency of \theia would be similar to current detectors like Super-Kamiokande and future detectors like Hyper-Kamiokande, and therefore \theia-25 would add only marginally to the DUNE program, and \theia-100 would just add in proportion to the exposure. For modes like $p\rightarrow \overline{\nu} K^{+}$ the contribution to the global sensitivity would be significant, and for modes like $n\rightarrow 3\nu$ \theia would be world-leading. These are discussed in more detail below.

\subsubsection{$p \rightarrow \bar{\nu}K$ and related modes}
\begin{figure}[h]
    \centering
    \includegraphics[width=0.48\textwidth]{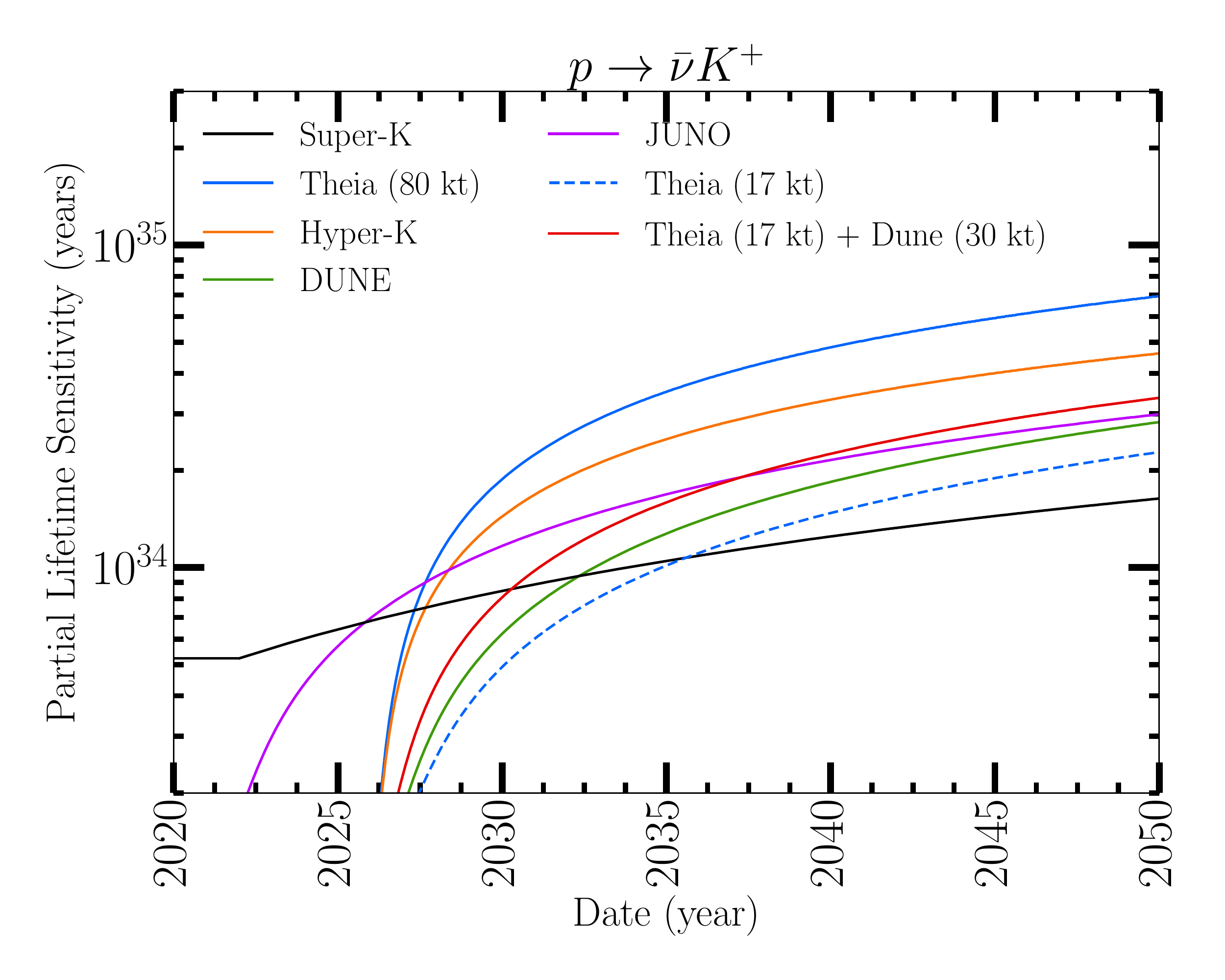}
    \caption{Sensitivity for $p \rightarrow \bar{\nu}K^+$ is highest for \theia, closely followed by the
    Hyper-K detector, whereas JUNO and DUNE will perform similarly.
    }
    \label{fig:nuk_sensitivity}
\end{figure}

Unlike the pion, the kaon is less effected by intranuclear effects, which means
that the primary efficiency will come down to how well any given detector will
be able to distinguish this very specific signature. Water Cherenkov detectors
are able to identify the number of rings corresponding to the kaon and its
decay products:
\begin{itemize}
    \item $K^+ \rightarrow \mu^+ \nu_\mu~~(63.42\%)$
    \item $K^+ \rightarrow \pi^+\pi^0~~(21.13\%)$
    \item $K^+ \rightarrow \pi^+\pi^+\pi^-~~(5.58\%)$
    \item $K^+ \rightarrow \pi^0e^+\nu_e~~(4.87\%)$
    \item $K^+ \rightarrow \pi^+\pi^0\pi^0~~(1.73\%)$
\end{itemize}
whereas a scintillator detector (JUNO) must rely on the timing of the decay to
separate the kaon from its daughters, and a tracking detector (DUNE) can
separate the particles using their tracks. The signal itself has roughly 3
unique energy depositions in each detector: the kinetic energy
deposited by the kaon; the decay products (either a muon or
pions); and the subsequent decay of those muons and pions. JUNO's
primary loss in efficiency comes from the overlap in the kaon energy deposition
and its subsequent decay (using a 7-ns prompt window for the 12-ns kaon
lifetime) \cite{An:2015jdp}. Since the prompt light would make identifying the
Cherenkov ring from the decay product extremely difficult, it is safe to assume
that \theia could do no better in separating these two and would have a similar
efficiency of 55\%. DUNE on the other hand would be able to identify the kaon
track regardless of the decay time, but due to internuclear cascades and Final State Interactions (FSI) is expected to have a signal efficiency
of only 30\%~\cite{dunewineandcheese}. A pure water detector is only sensitive to the
decay products, as the kaon is itself below Cherenkov threshold, making this
particular mode more difficult to measure; however, in lieu of the kinetic energy
deposition from the kaon,  water detectors are sensitive to the deexcitation
$\gamma$s emitted by the daughter nucleus (e.g. $^{16}\mathrm{O} \rightarrow ^{15}\mathrm{N}$.)
Hyper-K  expects an efficiency
of 23\% to see the muon (or charged pion), and the subsequent decay. In all
cases, the primary background is from atmospheric neutrino interactions, with
the expected background contribution given in Table \ref{tab:pdecay_nuk}. The
\theia signal efficiency and backgrounds are estimated to be similar to JUNO for this
mode, since the observation of the kaon energy will dominate deexcitation $\gamma$s
and the dominant selection criterion is inter-event timing.
Fig.~\ref{fig:nuk_sensitivity}
shows the expected sensitivity curves for each of the experiments, including 
\theia-100 (80 kT fiducial volume) and \theia-25 (17 kT fiducial volume). 

%Since an equal size DUNE has  better sensitivity for this mode of nucleon decay, a smaller module \theia in the DUNE cavern would not provide an improvement over a fourth DUNE detector.
\begin{table}[]
    \centering
    \caption{Detection efficiency and background rates for $p \rightarrow
    \bar{\nu}K^+$. \theia and JUNO are assumed to
    have the same relative efficiency, which is dominated by the short lifetime
    of the kaon. Hyper-K efficiency is the lowest due to being unable to detect
    the kaon since it is below the Cherenkov threshold.} 
    \label{tab:pdecay_nuk}
    \begin{tabular}{c|c|c}
\hline\noalign{\smallskip}
        Detector & Efficiency & Bkg [/Mton$\cdot$yr]\\
\noalign{\smallskip}\hline\noalign{\smallskip}
        \theia   & 55\% & 2.5 \\
        Hyper-K & 23\% & 1.6 \\
        DUNE    & 30\% & 1 \\
        JUNO    & 55\% & 2.5 \\
\noalign{\smallskip}\hline
    \end{tabular}
\end{table}

%%%%%%%%%%%%%%%%%%%%%%%%%%%%%%%%%%%%%%%%%%%%%%%%%%%%%%%%%%%%%%%%%%%%%
%%%% n -> 3nu
%%%%%%%%%%%%%%%%%%%%%%%%%%%%%%%%%%%%%%%%%%%%%%%%%%%%%%%%%%%%%%%%%%%%%
\subsubsection{$n \rightarrow 3\nu$ and related modes}
\begin{figure}[h]
    \centering
    \includegraphics[width=0.48\textwidth]{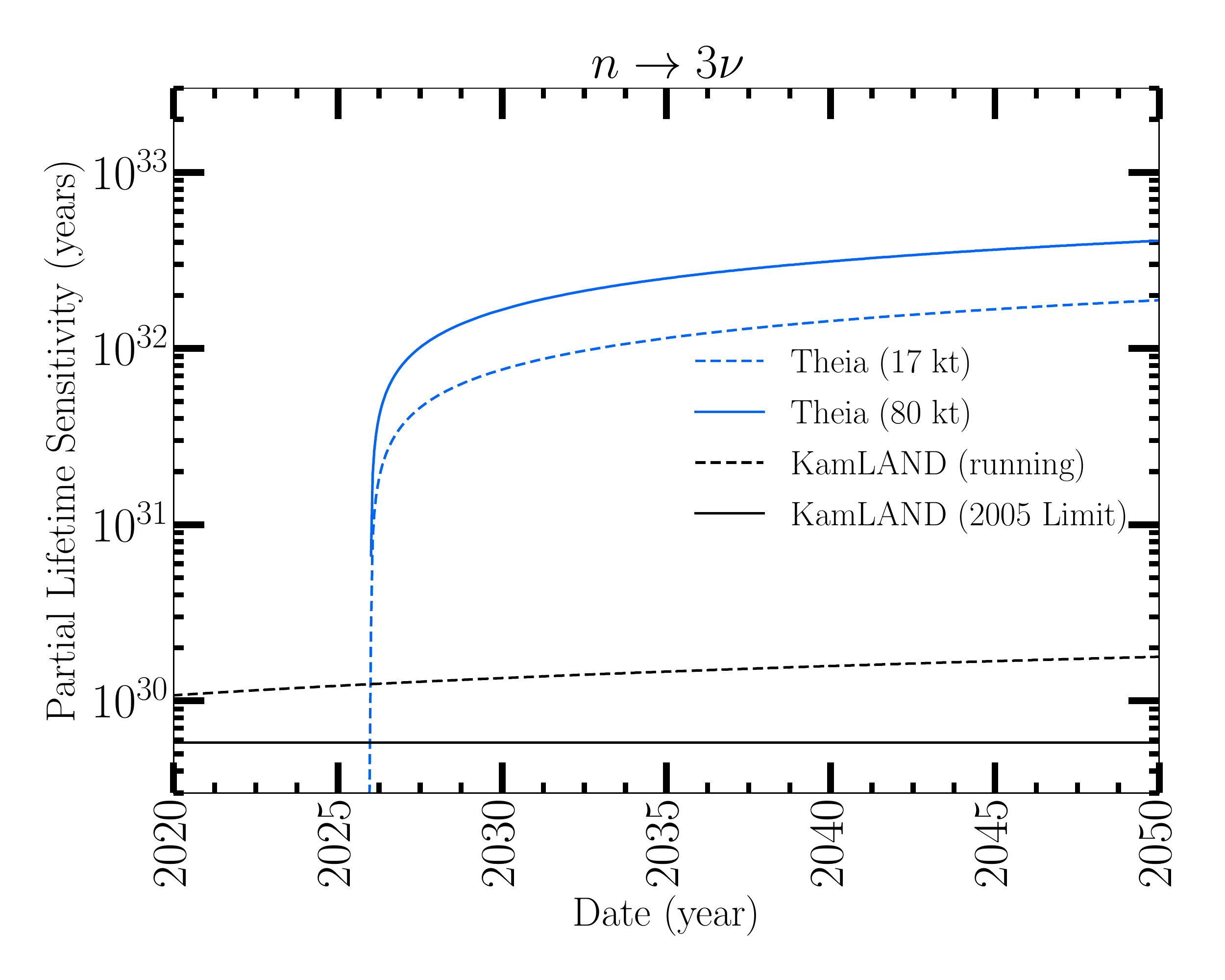}
    \caption{The large size and depth of \theia means that other next generation
    detectors are unlikely to be competitive when looking for very low energy
    modes of nucleon decay.}
    \label{fig:invisible_ndecay}
\end{figure}
A subset of theories predict modes of nucleon decay where the decay products
themselves do not leave a direct visible signature within the detector. It is in this set of potential decay modes that \theia would present a potential increased sensitivity by over two orders of magnitude.  An
example of this would be the decay of a neutron into three neutrinos. For a
bound nucleon this would leave the nucleus in an altered state, which would
have observable deexcitation gamma rays and low-energy emitted nucleons. For
large-scale detectors, the real difference between this decay mode and those
mentioned above is a matter of energy scale and isotope. Water Cherenkov
detectors such as SNO and SNO+ have looked for invisible decay from the oxygen
nucleus, which has a relatively high branching ratio (44 \%) to emit a 6.18-MeV
$\gamma$. The signal is a single event, so the detector is required to have very
low background in order to perform this search. Super-K, for example, is
limited by the production of cosmogenic activation of oxygen to $^{16}$N, which
decays with a 7.13-s half-life, emitting a 6.13-MeV gamma 67 \% of the time.
This limitation also exists for Hyper-K due to its shallow depth, which
means \theia would be the only large-scale water Cherenkov detector available to
look for this decay. Leading limits on invisible neutron and dineutron decay
are set by KamLAND \cite{Araki:2005jt}, and on invisible proton and diproton
decay by SNO+ \cite{Anderson:2018byx}.  Liquid scintillator detectors have both
advantages and disadvantages in the search for invisible neutron decay. While
the branching ratio for carbon is much lower (5.8 \%), the signal itself is a
triple coincidence signal. The primary deexcitation of $^{11}$C emits secondary
particles ($p, n, d, \alpha, \gamma$) providing a secondary signal, which is
followed by the radioactive decay of $^{10}$C---half-life of 19.3~s. KamLAND
identifies no signal that can directly mimic this signal, and so its primary
source of background is from accidental coincidences. Background to the first two components
of the signal is dominated by cosmogenic radioisotopes (particularly
$^{9}$Li), which requires a long cut of 2 s after each muon. Since the KamLAND
muon rate is $\sim 0.34$ Hz, this cut is tolerable; however, JUNO would likely
be insensitive to this decay mode due to its larger size and slightly shallower
depth, which results in a factor of 10 more muons through the detector.

For \theia, at a depth equivalent to DUNE, the primary backgrounds would come
from $^{8}$B solar neutrinos, production of $^{15}$N$^{*}$ through atmospheric
neutrino interactions, and cosmogenic production of $^{16}$N. Compared with the
SNO+ results, \theia would have much less impact from internal
radioactivity and a greater signal efficiency due to enhanced energy
resolution. Direction reconstruction would still play an important role, as the
event direction is used as the primary means for rejecting solar neutrino
events.  The resulting backgrounds would be $\sim 100$/kt$\cdot$y for \theia,
with the expected sensitivity, well above existing limits, shown in Fig.~\ref{fig:invisible_ndecay} for a 17-kt and a 80-kt fiducial volume.

%%%%%%%%%%%%%%%%%%%%%%%%%%%%%%%%%%%%%%%%%%%%%%%%%%%%%%%%%%%%%%%%%%%%%
%%% Conclusion
%%%%%%%%%%%%%%%%%%%%%%%%%%%%%%%%%%%%%%%%%%%%%%%%%%%%%%%%%%%%%%%%%%%%%

%\begin{comment} 
%\subsubsection{Summary}
%In regards to detecting various modes of nucleon decay, \theia would provide the most benefit in the search for invisible decays. All of the other detectors are
%ill-suited for this purpose, and even a smaller  \theia module could provide an order
%of magnitude or two improvement over existing limits. In the case of $p
%\rightarrow e\pi$-like modes, \theia outperforms DUNE for equal size detectors,
%but does not provide a substantial improvement over Super-K with gadolinium
%without a significant mass increase, and cannot compete with the size of
%Hyper-K. Finally the $p \rightarrow \bar{\nu}K$ modes are the most interesting,
%where all of the upcoming experiments are within the same order of magnitude
%(and thus a combined analysis would be most beneficial). In this regime a low
%mass \theia would not be competitive with the currently planned future
%detectors, but a higher mass \theia ($\sim100$ kt) would contribute
%significantly in pushing limits beyond $10^{35}$ years.
%\end{comment}

\section{Conclusions}

The \theia detector design represents an important step forward in the realization of a new kind of large optical detector: a hybrid Cherenkov/scintillation detector. Such an instrument is made possible by the convergence of three technological breakthroughs: (1) the development of ultra-fast and/or chromatically sensitive photosensors, (2) the ability to make novel target materials that produce detectable levels of both kinds of light, and (3) highly sophisticated pattern recognition and data analysis techniques that have moved well beyond the relatively simple methods of a decade ago. In addition, the coming availability of the new Long Baseline Neutrino Facility enables a broad program due to the deep depth and powerful neutrino beam available at the site.

This paper has detailed the exciting possibilities of \theia for new scientific discovery across a broad spectrum of physics, including long baseline neutrino measurements of oscillations and CP violation searches, solar neutrino measurements of unprecedented precision and scope, and the potential to extend the reach of neutrinoless double beta decay searches to the mass scales implied by a Normal Ordering of neutrino masses. In addition, \theia represents a major step forward for advancement in other fields -- ranging from detection of the Diffuse Supernova Background flux to a measurement of geo-neutrinos with unprecedented statistics. 

\section*{Acknowledgments}

The authors would like to thank the SNO+ collaboration for use of their scintillator optical model along with LAB/PPO properties. 

This material is based upon work supported by: the U.S. Department of Energy Office of Science under contract numbers DE-AC02-05CH11231, DE- SC0009999, DE-FG02-88ER40893, and DE-FG02-88ER40479; the U.S. Department of Energy National Nuclear Security Administration through the Nuclear Science and Security Consortium under contract number DE-NA0003180; the U.S. National Science Foundation award numbers 1554875 and 1806440; Funda{\c c}{\~ a}o para a Ci{\^ e}ncia e a Tecnologia (FCT-Portugal); and the Universities of California at Berkeley and Davis.

\bibliographystyle{utphys}
\bibliography{Theia}